# New Equations of State describing both the Dynamic Viscosity and Self-Diffusion Coefficient for Potassium and Thallium in their fluid phases

F. Aitken, F. Volino
Univ. Grenoble Alpes, CNRS, Grenoble-INP, G2ELab, F-38000 Grenoble, France.

**Abstract**. Experimental data on the viscosity and self-diffusion coefficient of two metallic compounds in their fluid phases, i.e. potassium and thallium, are modeled using the translational elastic mode theory which has been successfully applied to the case of water. It is shown that this theory allows the experimental data to be accounted for in accordance with their uncertainties and, above all, it allows the different variations observed between the different authors to be explained. Particularly in the case of thallium, this theory makes it possible to represent viscosity data with much better precision than the so-called reference equation of state. The dilute-gas limit laws connecting various parameters of the theory obtained in the case of water are confirmed here and thus give them a universal character. The elastic mode theory is accompanied by the development of new equations of state, mainly to describe properties along the saturated vapor pressure curve, which greatly extend the temperature range of application of these equations compared to those found in the literature. The whole analysis thus makes it possible to propose precise values of various thermodynamic parameters at the melting and boiling temperature corresponding to atmospheric pressure.

**Keywords**: Shear Elasticity, Viscosity, Self-diffusion coefficient, Density, Liquid, Vapor, Potassium, Thallium, Saturated Vapor Pressure Curve, Equation of State

## 1 Introduction

The properties of fluid metals are important for basic theoretical studies because of the simplicity and ideality of their atomic structure. More specifically, the vapors of the majority of alkali metals and some other metals may be considered as ideal monatomic gases at high temperatures and enough low densities. This is the reason why potassium and thallium gas were used, for example, to test Maxwell's velocity distribution law (Ref. 1).

In particular, alkali metals are studied for applications related to their high latent heat of vaporization, their high thermal conductivity and their high liquid temperature range. For example this makes them attractive as working fluids in turbine power converters (e.g. Ref. 2) and as heat transfer media in nuclear reactors.

The viscosity data for fluids such as alkali metals are also of particular interest. For example, the ablation of meteroids when entering Earth's atmosphere produces metallic vapors, particularly potassium vapors. Modeling the gas flow around a meteroid then requires the knowledge of these metallic vapors transport properties (Ref. 3).

Experimental data on the transport properties of metallic compounds in the liquid phase are mostly confined to the atmospheric isobar while in the gaseous phase they are very limited due to the high temperatures required, low vapor pressures and corrosive nature.

Although limited in number, the study of the transport properties of potassium and thallium are of interest here to test the translational elastic mode theory in fluids that are radically different from water (Ref. 4). The viscosity and self-diffusion coefficients data in these two fluid media present, as in water, different variations and deviations according to the authors that are greater than the experimental uncertainties. Elastic mode theory has made it possible to explain these differences in the case of water by taking into account some geometrical characteristics of the experimental devices. It is thus an important test for the theory. Moreover, dilute-gas limit laws relating several parameters of the theory have been obtained in the case of water with the presumption that these laws are globally universal. This is an important issue that we seek to verify in this paper. To achieve this goal it is necessary to develop new equations of state, mainly to describe properties of potassium and thallium along the saturated vapor pressure curve, which greatly extend the temperature range of application of these equations compared to those found in the literature.

## 2 Summary of the translational elastic mode theory

In this section we will briefly recall the important relations of the elastic mode theory and for more details we refer the reader to Ref. 4.

The starting point of the theory is to model the displacement fluctuations of objects, i.e. a set of $n_B$ molecules or atoms called *basic unit*, on a lattice due to temperature $T$ at thermodynamic equilibrium. We recall that $n_B$ is the number of molecules or atoms in the unit cell of the crystal just below the melting curve. The first fundamental assumption is to consider that the basic unit center of mass displacement $\vec{u}(\vec{r})$ is a Gaussian random variable that can be developed into Fourier series (whose coefficients refer to as *elastic modes*) on the lattice. So, for component $u_x$ of $\vec{u}$, we have:

$$u_x(\vec{r}) = \sum_{\vec{q}} u_x(\vec{q}) e^{i\vec{q}\cdot\vec{r}} = \sum_{\vec{q}} u_{xq} e^{i\vec{q}\cdot\vec{r}} \tag{1}$$

where the amplitudes $u_x(\vec{q})$ are new statistically independent random variables. Each mode is characterized by its wave-vector $\vec{q}$ and its polarization.

The isotropy of reciprocal space is assumed in such a way that the wave-vector moduli $q$ are limited at short length scales by a cut-off wave-vector $q_c$, and towards long length scales by a wave-vector $q_c / N$, where $N/q_c$ represents the fluctuative distance (i.e. the coherence length) compatible with the sample size.

The second fundamental hypothesis consists in introducing an elastic energy functional which combined with the principle of thermal energy equipartition give the following results for the expression of the fluctuation full mean square displacements $< \left| u_x^2 \right| >$ due to transverse modes only:

$$< \left| u_x^2 \right| > = \frac{k_B T q_c}{\pi^2 K} H_N(v) \tag{2}$$

with

$$H_N(v) = \frac{N^{v-1} - 1}{v - 1} \tag{3}$$

where $K$ is a shear elastic constant and exponent $\nu$ is a thermodynamic function determined by the ratio $T/T_t$ , $T_t$ being a temperature associated with the glass transition that occurs when $T_t$ = $T$. As long as one is limited to studying the medium properties in the disordered phase (i.e. for $T > T_t$), then one has:

$$\nu - 1 = \left(1 - \frac{T_t}{T}\right)^{\frac{1}{4}} \tag{4}$$

We have seen in Ref. 4 that the parameters $K$, $N$ and $q_c$ depend on temperature and density of the medium through the reduced functions $K^*$ , $f_N$ and $f_{q_c}$ such that:

$$K^*(\rho) = K(\rho)/K_0 \text{ with } K_0 = \frac{12\pi^2}{n_B}\frac{R_g T_c \rho_c}{M} \tag{5}$$

$$N - 1 = d_N \frac{q_{c0,\text{crit}}}{2\pi} \text{ with } d_N = f_N(\rho,T)\, d \tag{6}$$

$$q_c(\rho,T) = f_{q_c}(\rho,T) \times q_{c0}(\rho) \text{ with } q_{c0}(\rho) = \left(\frac{6\pi^2 \rho\, \mathfrak{N}_a}{M\, n_B}\right)^{1/3} \tag{7}$$

where $R_g$ is the perfect gas constant, ($T_c$, $\rho_c$) are the critical parameters of the fluid (*i.e.* absolute temperature and density respectively), $M$ is the molar mass of the fluid, $q_{c0,\text{crit}} = q_{c0}(\rho_c)$ and $\mathfrak{N}_a$ is the Avogadro number.

The distance $d$, called the dissipative distance, is associated with system out-of-equilibrium and represents the characteristic region of the sample volume where the velocity gradient is important. It also represents a characteristic geometric distance of the system when it is associated with $d_N$.

Putting the system out-of-equilibrium by an external disturbance makes it possible to define transport coefficients namely the self-diffusion coefficient $D_t$ and the dynamic viscosity $\eta$ such as:

$$D_t = \frac{3k_B\, T\, q_c\, H_N(\nu)}{2\pi^2 d\sqrt{K\rho}} \tag{8}$$

$$\eta(\rho,T) = \eta_l(\rho,T) + \eta_{Knu}(\rho,T) = \frac{d}{H_N(\nu)}\sqrt{\rho K} + \tilde{\rho}_{Knu}\sqrt{\frac{R_g T}{M}}\frac{2\pi}{q_{c0,\text{crit}}} \tag{9}$$

where $\tilde{\rho}_{Knu} = \rho_{Knu}\, \delta\, q_{c0,\text{crit}}/(2\pi)$ . The parameter $\rho_{Knu}$ represents the density of the released gas due to the shear stresses action on the system and $\delta$ is a characteristic distance which in the case of a Poiseuille type flow can be identified with the dissipative distance $d$.

To determine the self-diffusion coefficient and the viscosity, the scaling parameters $K_0$ and $q_{c0,\text{crit}}$ must first be determined and then the functions $K^*(\rho)$, $f_N(\rho,T)$, $\tilde{\rho}_{Knu}(\rho,T)$ and $f_{q_c}(\rho,T)$. The first three functions can be obtained by analyzing the viscosity data while the fourth one is obtained by analyzing the self-diffusion coefficient data.

The aim of the paper is therefore to determine the different parameters and functions to represent the self-diffusion coefficient and the dynamic viscosity with Eq. (8) and Eq. (9), for the two fluid media: potassium and thallium. We will begin by studying potassium because there are much more data for potassium than for thallium. Thus the thallium analysis will be copied on the potassium analysis.

## 3 Application to potassium fluid

The scaling $K_0$ of the shear elastic constant $K$ is determined from the critical parameters of the fluid as well as the crystal structure through the value of the parameter $n_B$. The values of these different parameters are grouped in Table 1.

| $M$ (g/mole) | $n_B$ | $T_c$ (K) | $P_c$ (MPa) | $\rho_c$ (g/cm$^3$) | $\mathcal{V}_c$ (Å$^3$) | $z_c$ |
|---|---|---|---|---|---|---|
| 39.083 | 2 | 2223 | 16.212 | 0.194 | 334.669 | 0.176 |

Table 1. Characteristic parameters of fluid potassium. The value of $M$ is from Ref. 5 and the critical parameters are from Ref. 6. Below the melting line, solid potassium is body-centred cubic with 2 atoms per unit cell (Ref. 7) thus the value of $n_B$.

The following values are then deduced for the scalings $K_0$ = 5.4309 GPa and $q_{c0,\text{crit}} = 4.45593\times10^{7}\ \text{cm}^{-1}$.

*A Mathematica application with the potassium equations of state can be freely downloaded by following* Ref. 8.

### 3.1. Density equation of state in the gaseous phase

In order to be able to analyze the viscosity and self-diffusion coefficient data in the gaseous phase, it is necessary to know the equations of state in this phase and especially the one describing the density.

Ewing *et al.* (Refs. 9 and 10) derived a state equation of the compressibility factor $z(T, V)$ valid between 1140 K and 1660 K in the gaseous phase. In order to cover the range of potassium viscosity data in the gaseous phase and to be consistent with the state equations determined to describe the Saturated Vapor Pressure curve (SVP), we have added to the data of Ewing *et al.* some points corresponding to the density on SVP (see section 3.2). Then the coefficients in the expression of Ewing *et al.* have been recalculated using more practical units. Thus the expression of the compressibility factor is now written:

$$z(T,V)=\frac{PV}{R_gT}=1+\frac{B}{V}+\frac{C}{V^2}+\frac{D}{V^3} \tag{10}$$

with

$$\begin{cases}\log_{10}|B|=-2.01657+\dfrac{2982.8}{T}+\log_{10}(T),\ B<0\\ \log_{10}(C)=2.66215+\dfrac{5710.87}{T},\ C>0\\ \log_{10}|D|=3.42866+\dfrac{9196.84}{T},\ D<0\end{cases}$$

where $T$ is expressed in Kelvin and $V$ in cm$^3$/mole. This new equation of state is valid between 1000 K and 1700 K and for pressures varying between 0.1 MPa. and 2.6 MPa. Fig. 1 shows that the deviation of densities calculated from Eq. (10) is entirely within ±0.7% but most of the points lie within ±0.5%. This deviation is a little higher than the one given by Ewing *et al.* which is ±0.4% but it allows a small extension in temperature. Fig. 1 does not show the points added on SVP. These points are treated separately in the section 3.2 (see Fig. 3).

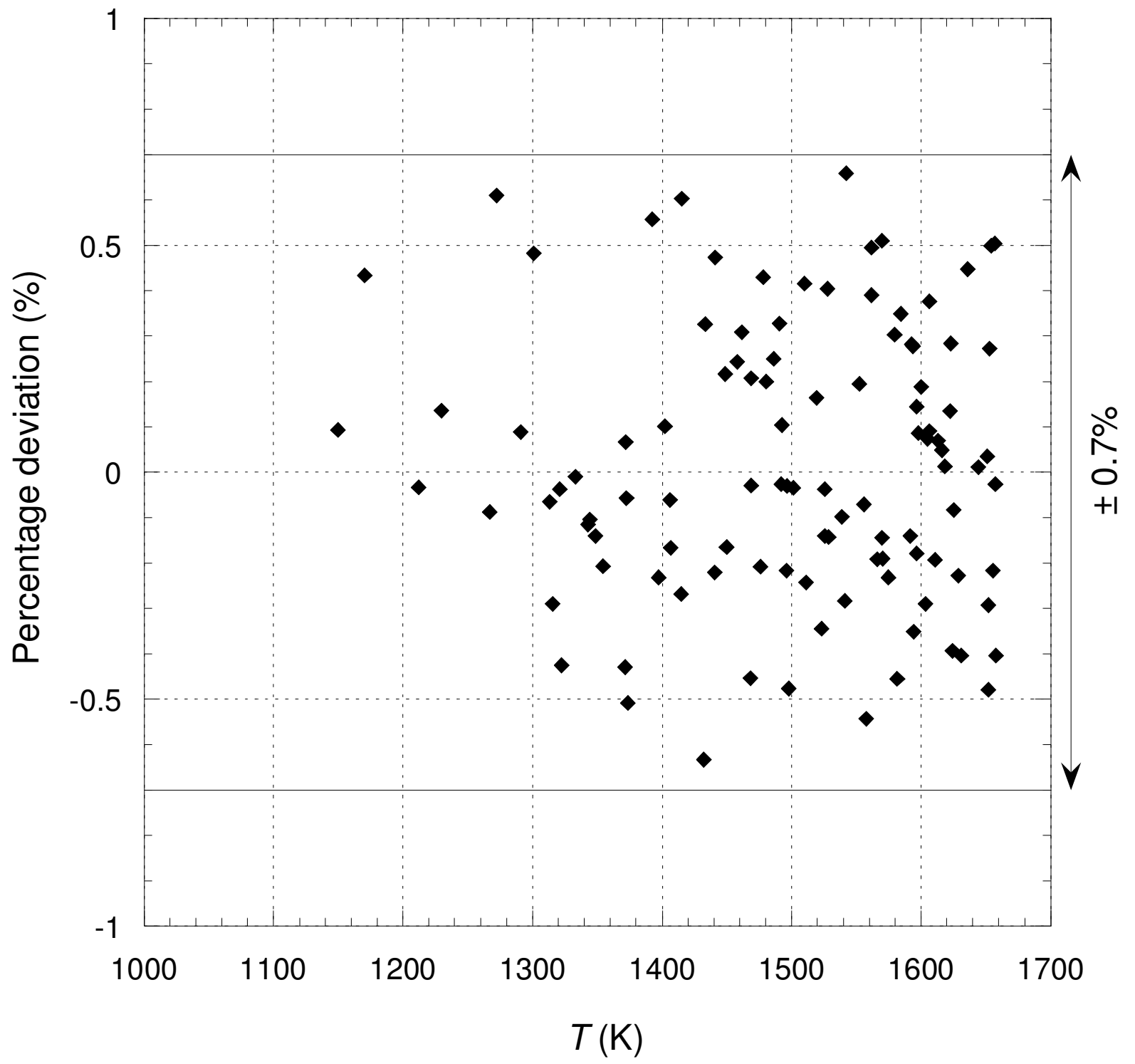

Fig. 1. Deviation plot for the potassium vapor densities, i.e. $100\left(\rho_{\mathrm{exp}}-\rho_{\mathrm{calc}}\right)/\rho_{\mathrm{calc}}$ where $\rho_{\mathrm{exp}}$ corresponds to the data from Ref. 9 (i.e. Table 3) and $\rho_{\mathrm{calc}}$ is determined from Eq. (10) of the present modeling.

It can be noted that Vargaftik *et al.* (Ref. 11) developed a state equation for the compressibility factor in potassium vapor valid between 1075 K and 2150 K and for pressures varying between 0.1 MPa and 10 MPa. This equation covers a pressure and temperature range greater than Eq. (10). However, this equation is expressed as a function of the reduced parameters $\rho/\rho_c$ and $T/T_c$ . But the values of $\rho_c$ and $T_c$ are not given in the paper which is a handicap for the use of this equation. In the range of temperature and pressure corresponding to Eq. (10), the authors mentioned that the deviation of their equation with some experimental data varies between -0.7% and +1.3%. The two descriptions are thus quite equivalent in their overlapping range.

## 3.2. Thermodynamic properties along the saturated vapor pressure curve

To determine some parameters as well as to perform some calculations, it is useful to know the evolution of the pressure and of the liquid and vapor densities along SVP. This will allow us, among other things, to determine the evolution of the latent heat of vaporization and thus put it into perspective with the value of $K_0$ as we have seen in Appendix B of Ref. 4.

In this section we will establish new state equations in order to take into account the maximum of experimental data but also in order to extend the description of the SVP curve to the largest temperature range, i.e. from melting temperature at atmospheric pressure $T_{\text{m.p.}}$ = 336.65 K to the critical temperature $T_c$.

To begin, the liquid density data from Ref. 6 have been modeled so as to connect to the critical point using the following relationship:

$$\frac{\rho_{\sigma,\text{Liq}}}{\rho_c} = 1 + 0.67096\,\theta_r^{\;6.1442} + 3.3663\,\theta_r^{\;0.54393} \qquad (11)$$

where $\theta_r = 1 - T/T_c$.

Fig. 2 shows that the deviation between Eq. (11) and the data is less than ±1% with the exception of the point corresponding to the melting temperature at atmospheric pressure, i.e. $T_{\text{m.p.}}$. This point at $T_{\text{m.p.}}$ must correspond to a density lower than that corresponding to atmospheric pressure, which is not the case according to the experimental value. It is therefore necessary to deviate from this value. However, the deviation is not exaggerated if we refer to authors' comment of Ref. 6:

> "For the liquid densities the standard deviation is ±0.015 in the range from 0.931 to 1.200 g/cc, which is an error of about ±1.5%. For the vapor densities, the standard deviation is 0.0015 in the range from 0.0075 to 0.1190 g/cc, which is an error of about ±20% in the lower end of the range to ±2% in the higher end of the range."

Ewing *et al.* (Ref. 9) used to describe the liquid density on SVP their Eq. (9) which describes the liquid density along the atmospheric isobar. It is therefore not surprising that this equation represents rather poorly the data of Dillon *et al.* on SVP as shown in Fig. 2.

Caldwell *et al.* (Ref. 12) also developed a state equation to represent the liquid density on SVP based on the data from Ewing *et al.* (Ref. 9). Fig. 2 shows that Eq. (5) from Caldwell *et al.* is identical to that of Ewing *et al.* in the overlap temperature range and therefore does not reproduce very well the data, even if the deviation is of the same order as that obtained with Eq. (11). Above 1600 K the two equations diverge strongly. However, Eq. (11) is by construction compatible with the chosen value of $\rho_c$. It is therefore coherent to retain only Eq. (11) in the following.

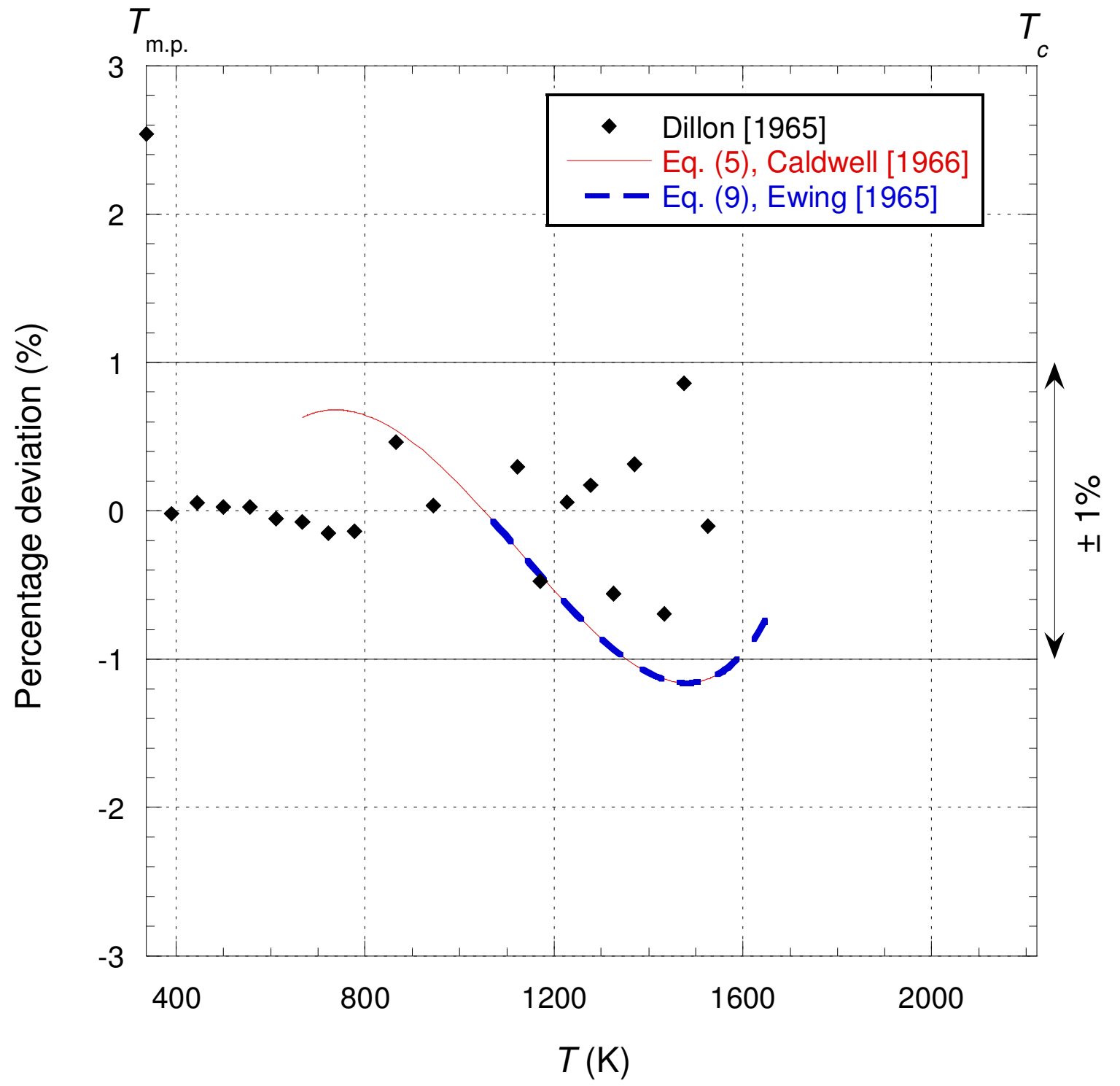

Fig. 2. Deviation plot for the liquid densities of potassium along the saturated vapour pressure curve, i.e. $100\left(\rho_{\exp}-\rho_{\text{calc}}\right)/\rho_{\text{calc}}$ where $\rho_{\exp}$ corresponds to the data from Dillon *et al.* (Ref. 6) and $\rho_{\text{calc}}$ is determined from Eq. (11) of the present modeling. The red curve represents the deviation between Eq. (5) from Caldwell *et al.* (Ref. 12) and Eq. (11). The dashed blue curve represents the deviation between Eq. (9) from Ewing *et al.* (Ref. 9) and Eq. (11). $T_{\text{m.p.}}$ represents the melting temperature for the atmospheric pressure and $T_c$ represents the critical temperature.

To determine an expression for the gas density on SVP that covers the temperature range from $T_{\text{m.p.}}$ to $T_{\text{c}}$, we have considered the data from Table 4 of the report of Dillon *et al.* from 1965 (Ref. 6) with those from Table 4 of Ref. 9. One must also take into account Eq. (10) which leads to describe a part of SVP in its restricted temperature range. The expression thus obtained is as follows:

$$\log_{10}\left(\frac{\rho_{\sigma,\text{Vap}}}{\rho_c}\right)=-85.2049\,\theta_r^{9.74181}-1.74903\,\theta_r^{0.388134}-0.140485\,\theta_r^{4/3}-7.06168\,\theta_r^{9/3}+97.3286\,\theta_r\exp\left(-\left(\frac{T+502.412}{598.176}\right)^{1.9}\right) \quad (12)$$

where $\theta_r=1-T/T_c$ and $T$ is expressed in Kelvin.

Fig. 3 shows that the deviation of Eq. (12) from the data is less than ±2% overall. This overall deviation with the data of Dillon *et al.* (Ref. 6) is largely consistent with the experimental error we cited earlier concerning the liquid density on SVP. Ewing *et al.* (Ref. 9) describe their data in their small temperature range with an expression such that the average deviation is ±0.34%. With the exception of two points, the deviation with the data from Ewing *et al.* is less than ±0.5%, but it can be observed that this deviation is necessary to obtain a description consistent with Eq. (10). Moreover, a deviation of ±0.5% is consistent with the deviation corresponding to Eq. (10) which confirms the present modeling. The confirmation of the present modeling also comes from the model of Caldwell *et al.* (Ref. 12) which, in the temperature range 1100 K to 1600 K, is practically identical to Eq. (12). On the other hand, it can be observed that for temperatures below 1100 K the model of Caldwell *et al.* is not consistent with the data.

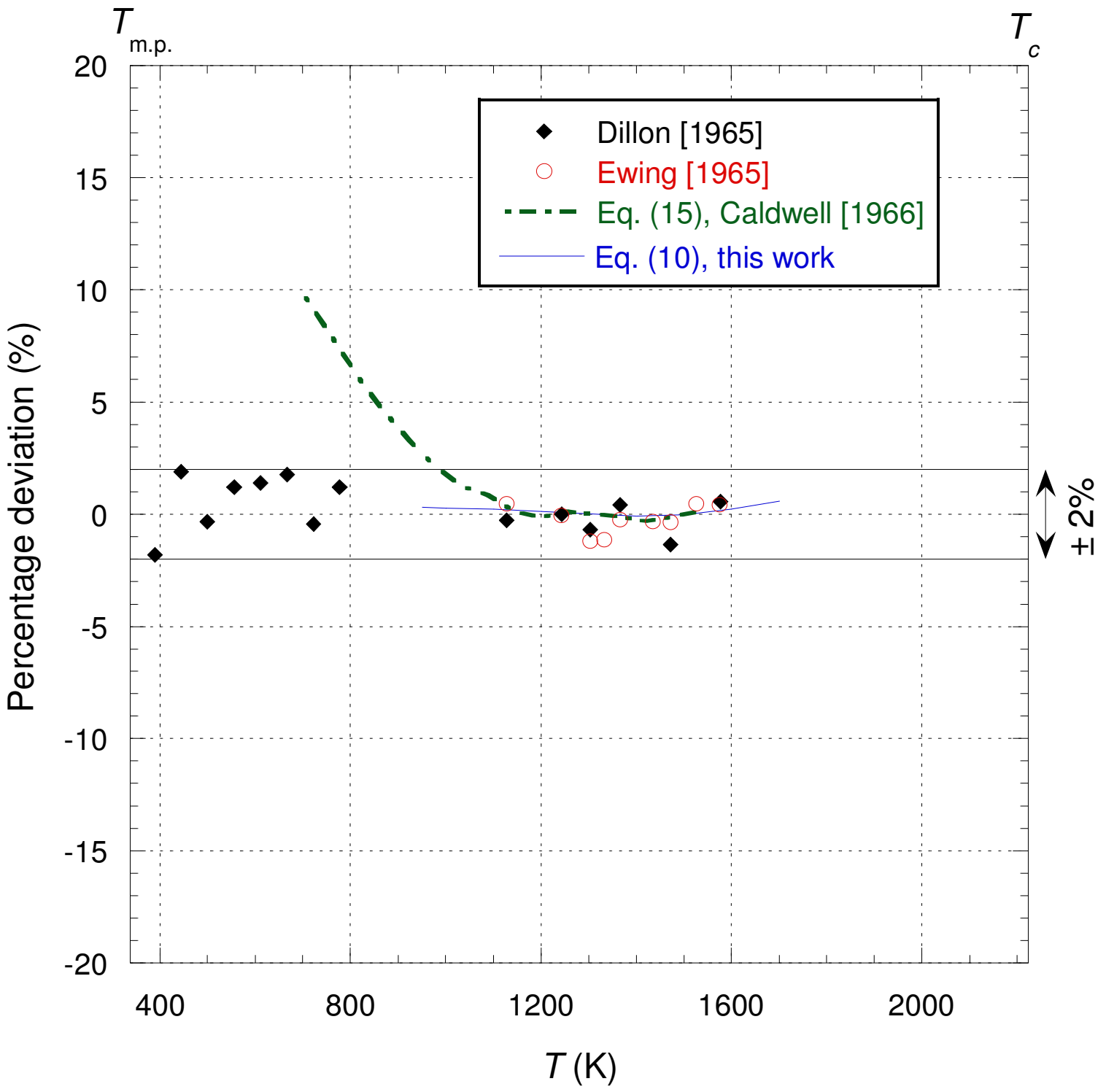

Fig. 3. Deviation plot for the gas densities of potassium along the saturated vapor pressure curve, i.e. $100\left(\rho_{\text{exp}} - \rho_{\text{calc}}\right)/\rho_{\text{calc}}$ where $\rho_{\text{exp}}$ corresponds to the data from Dillon *et al.* (Ref. 6) and Ewing *et al.* (Ref. 9) and $\rho_{\text{calc}}$ is determined from Eq. (12) of the present modeling. The green dot-dashed curve represents the deviation between Eq. (15) of Caldwell *et al.* (Ref. 12) and Eq. (12). The blue curve represents the deviation between Eq. (10) along SVP and Eq. (12). $T_{\text{m.p.}}$ represents the melting temperature for the atmospheric pressure and $T_c$ represents the critical temperature.

In order to be able to determine the density on SVP from Eq. (10), it is also necessary to know the relation $P_\sigma(T)$, i.e. the vapor pressure versus temperature. To determine this relationship we used together the data from Ewing *et al.* (Table 6 from Ref. 9) and Hicks (Table VI from Ref. 13). The data in Table 3 of Dillon's 1965 report (Ref. 6) were discarded because they are largely shifted with the other two.
Ewing *et al.* proposed two equations to represent their data in such a way that:

> "The average deviation of the observed vapor-pressure data in Table 6 from corresponding values computed from Eq. (3) is ±0.31%, and from Eq. (4) is ±0.32%."

Fig. 4a shows the deviation of the vapor pressure data of Ewing *et al.* with their Eq. (4) and it can be observed a maximum deviation of about ±1.2%. Nevertheless, this Eq. (4) is interesting if it can be extrapolated to lower temperatures. However, Fig. 4b shows that if Hicks' data agree with Eq. (4) for temperatures higher than $0.4T_c$, this is no longer the case for lower temperatures. It is therefore necessary to determine a new expression of $P_\sigma(T)$ to cover the entire SVP curve.

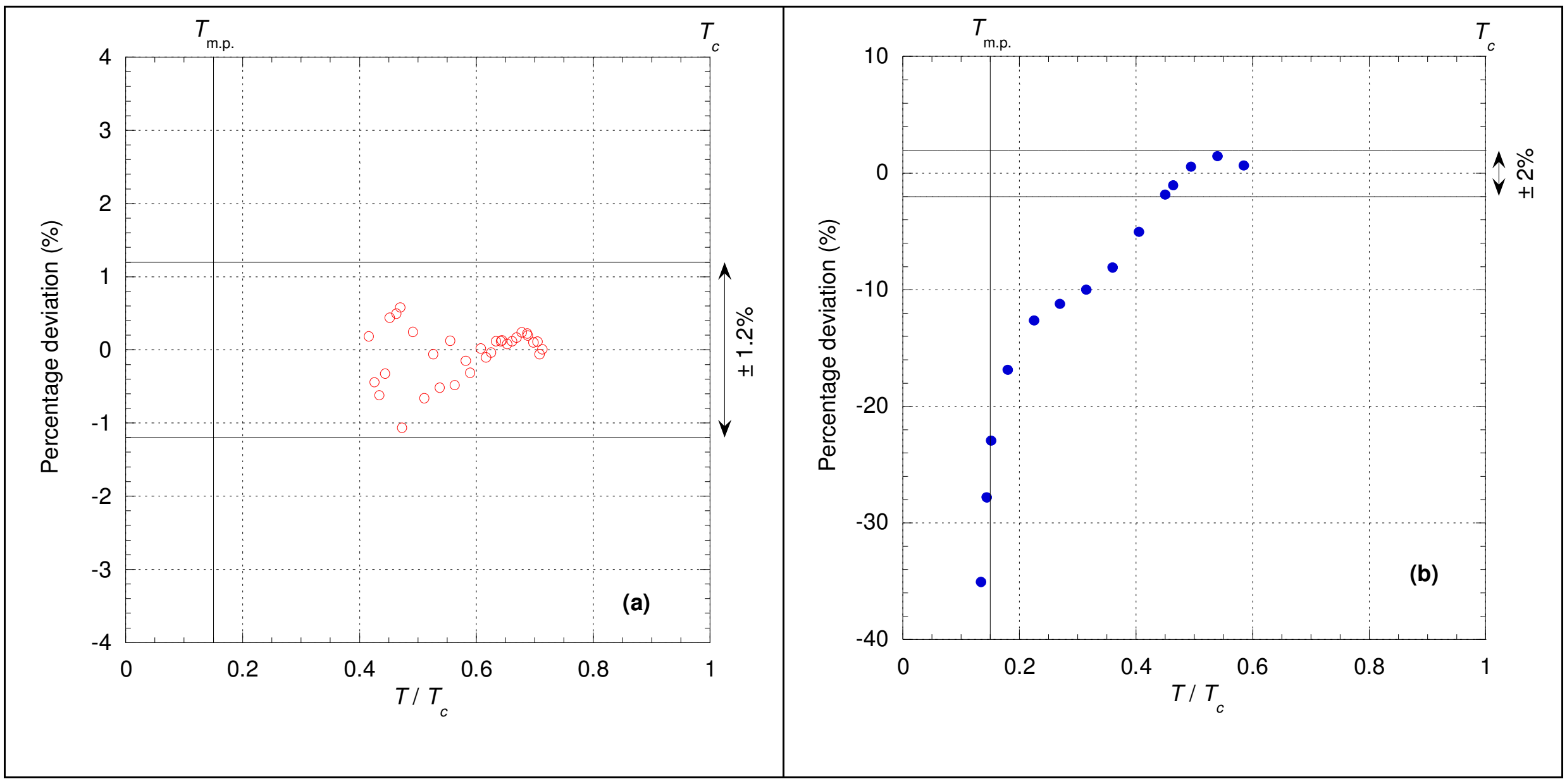

Fig. 4. Deviation plot for the pressure along the saturated vapor pressure curve of potassium, i.e. $100\left(P_{\text{exp}} - P_{\text{calc}}\right)/P_{\text{calc}}$ where $P_{\text{calc}}$ is determined from Eq. (4) of Ref. 9: (a) $P_{\text{exp}}$ corresponds to vapor pressure data from Table 6 of Ref. 9; (b) $P_{\text{exp}}$ corresponds to vapor pressure data from Table VI of Ref. 13. $T_{\text{m.p.}}$ represents the melting temperature for the atmospheric pressure and $T_c$ represents the critical temperature.

As an alternative to Eq. (4) from Ewing *et al.* one can look at the results of Eq. (1) by Caldwell *et al.* (Ref. 12). Fig. 5 shows that the model of Caldwell *et al.* does not provide a better description than Eq. (4) of Ewing *et al.* and therefore cannot be considered for use.

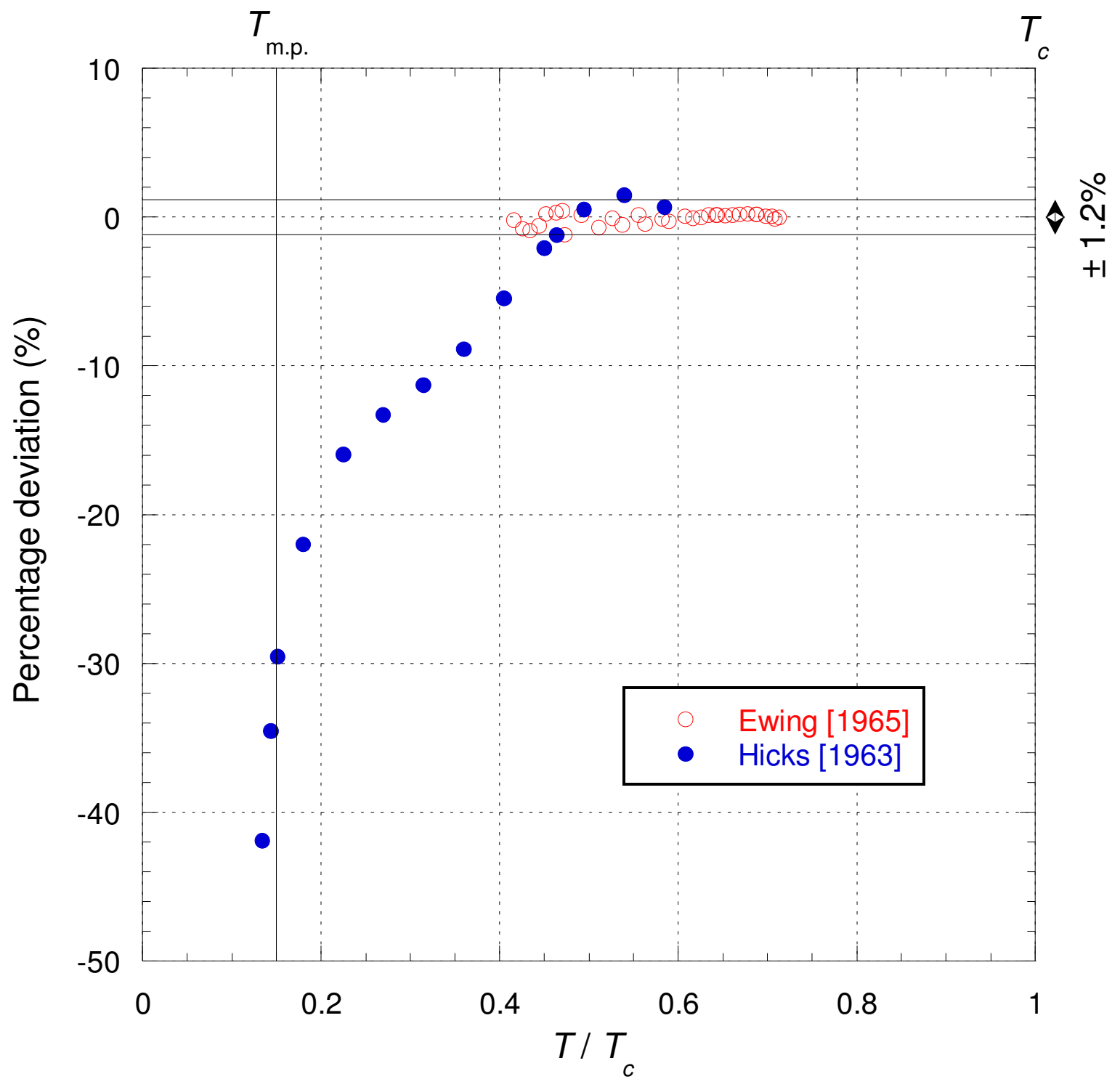

Fig. 5. Deviation plot for the pressure along the saturated vapor pressure curve of potassium, i.e. $100\left(P_{\text{exp}}-P_{\text{calc}}\right)/P_{\text{calc}}$ where $P_{\text{exp}}$ corresponds to vapor pressure data from Table 6 of Ref. 9 and from Table VI of Ref. 13 and $P_{\text{calc}}$ is determined from Eq. (1) of Ref. 12. $T_{\text{m.p.}}$ represents the melting temperature for the atmospheric pressure and $T_c$ represents the critical temperature.

To take into account Hicks data (Ref. 13), it is therefore necessary to find a new equation that covers at least the temperature range between $T_{\text{m.p.}}$ and $T_c$. Thus the expression of $P_\sigma(T)$ that accounts for both Ewing *et al.* and Hicks data is as follows:

$$\log_{10}\left(\frac{P_\sigma}{P_c}\right)=-(-0.097263+44.002)-\frac{0.097263}{T_r^{\ 1.9396}}+44.002\,T_r-42.095\,T_r^{\ 0.51788}\ln(T_r) \qquad (13)$$

where $T_r = T/T_c$ .

Fig. 6 shows that the maximum deviation is now ±3%, which may seem high, but it can be observed that the data from Hicks and Ewing *et al.* in their overlapping area are in opposition (i.e. they are shifted on both sides of the zero) and therefore it is not possible to make a better representation if we want to get through the middle of the points. That said, it can be observed that the majority of points correspond to a deviation of ±1.2%, which is consistent with the deviations obtained from Eq. (4) of Ewing *et al.* (Ref. 9) in its region of validity.

Freyland *et al.* (Ref. 14) also performed pressure measurements on SVP. These authors considered that:

> “The absolute error for the temperature measurement was estimated as about 0.5% and for pressure measurement 1% which was established by calibration.”

Fig. 6 shows that the data of Freyland *et al.* are not compatible with those of Hicks and Ewing *et al.* with the uncertainty defined by the authors for the two points around 0.6 $T_c$. On the other hand, it is observed that the present modeling is compatible with the data of Freyland *et al.* at higher temperatures if a deviation of ±1.2% is accepted. This deviation value can be considered as compatible with the data since the deviation is fairly well centered on zero percent.
It can be concluded that Eq. (13) allows a consistent description of the data sets between $T_{\mathrm{m.p.}}$ and $T_c$.

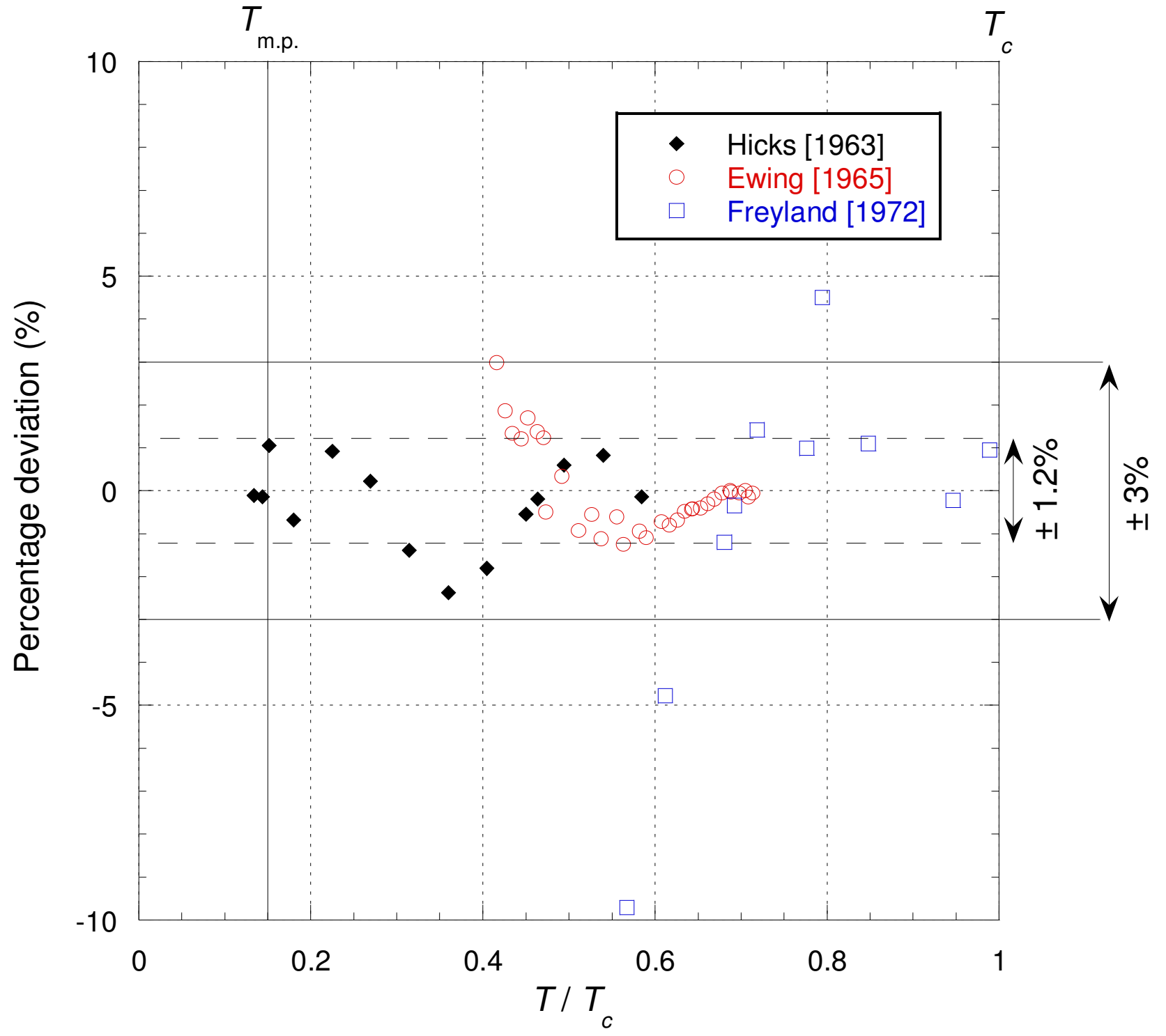

Fig. 6. Deviation plot for the pressure along the saturated vapor pressure curve of potassium, i.e. $100\left(P_{\mathrm{exp}} - P_{\mathrm{calc}}\right)/P_{\mathrm{calc}}$ where $P_{\mathrm{exp}}$ corresponds to vapor pressure data from Table 6 of Ref. 9, from Table VI of Ref. 13 and from Table 1 of Ref. 14, and $P_{\mathrm{calc}}$ is determined from Eq. (13) of the present modeling. $T_{\mathrm{m.p.}}$ represents the melting temperature for the atmospheric pressure and $T_c$ represents the critical temperature.

From Eq. (11), Eq. (12) and Eq. (13) the latent heat of vaporization $L_v$ can be calculated using the Clapeyron equation. Fig. 7a shows the function $L_v(T)$ in comparison with the value of the scaling $K_0$ of the shear elastic constant $K$. We observe that the variation of $L_v(T)$ near the melting temperature decreases exactly as observed in the case of helium (see Appendix B of Ref. 4). This is therefore both a new result obtained due to the accuracy of Eqs. (11)-(13) but also a behavior expected for many other fluids. Fig. 7b shows a comparison between Eq. (15) of Ewing *et al.* (Ref. 9) and Eq. (17) from Caldwell *et al.* (Ref. 12) with the present modeling. First, it can be observed that the model of Caldwell *et al.* represents an extension of the modeling of Ewing *et al.* The present modeling shows slightly different variations with temperature but the deviation with Ewing *et al.* modeling as well as with Caldwell *et al.* modeling does not exceed 2%. However, it can be noted that the model of Ewing *et al.* tends to plunge too quickly at high temperatures which seems difficult to be compatible with the zero value at $T_c$.

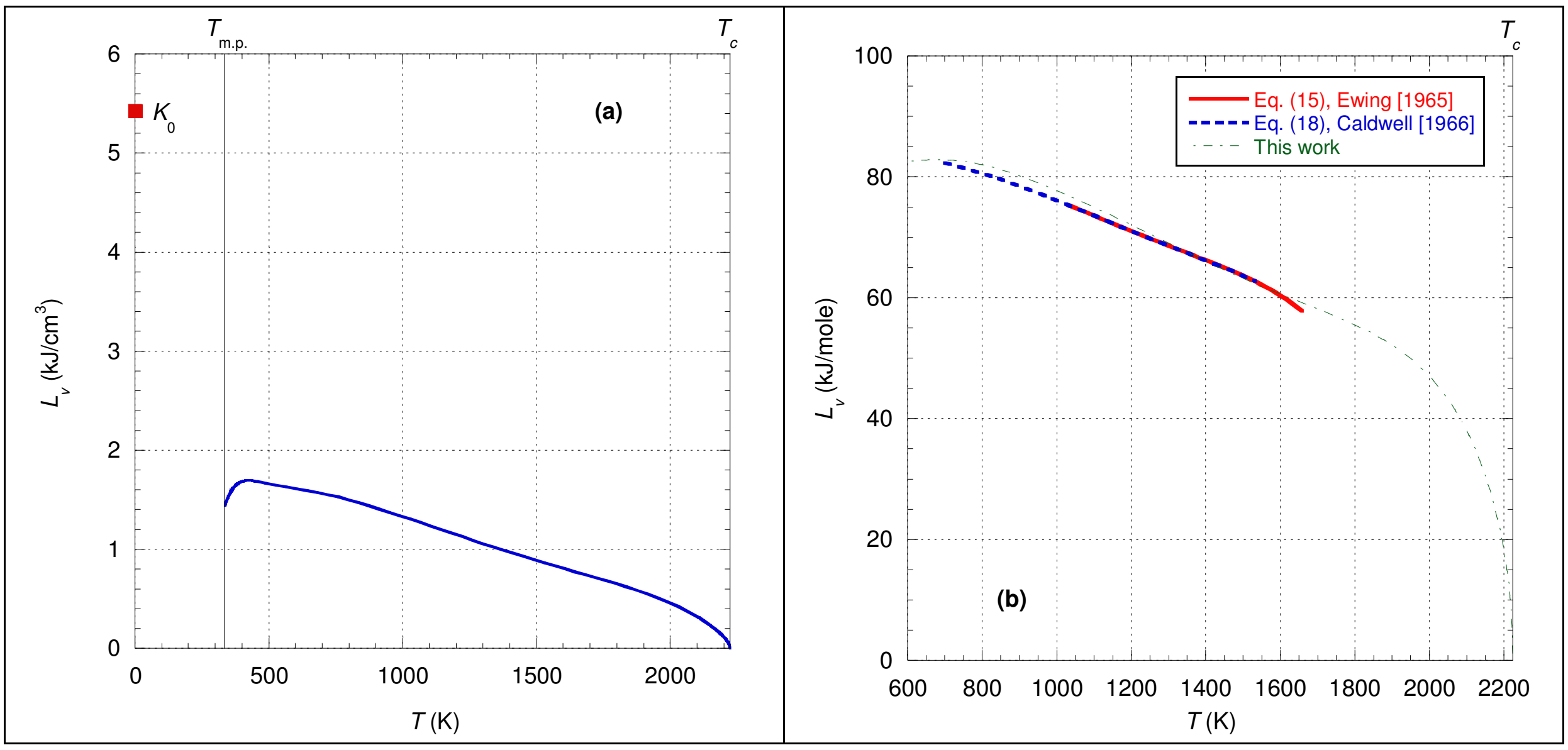

Fig. 7. Latent heat of vaporization of potassium as function of temperature: (a) calculated from Clapeyron equation (blue curve); (b) Eq. (15) from Ref. 9 (red curve) and Eq. (17) from Ref. 12 (dashed blue curve) versus the present modeling (dot-dashed green curve). $T_{\mathrm{m.p.}}$ represents the melting point temperature for the atmospheric pressure and $T_c$ represents the critical temperature.

### 3.3. Density equation of state along the atmospheric isobar

Most of the data in the liquid phase are found along the atmospheric isobar which thus extends from $T_{\mathrm{m.p.}}$ (i.e. the melting temperature, see Table 4) to $T_b$ (i.e. the boiling temperature, see Table 3). To be able to analyze the data, it is therefore necessary to have a good representation of the density variation along the atmospheric isobar. Ewing *et al.* (Ref. 9) proposed a third degree polynomial to represent the variation of density as a function of temperature, knowing that the variation is mainly linear (i.e. the non-linear terms are very small). In order to obtain an equation with more practical units, we recalculated a polynomial equation considering the data of Hagen (Ref. 15), Rinck (Ref. 16) and Stokes (Ref. 17), the latter not having been considered by Ewing *et al.* because they are slightly posterior. To these, we have added those given by Ewing *et al.* (Ref. 18) in his paper with the viscosity data in order to be consistent with the analysis of these data. Given the very low non-linearity of density variation with temperature, the data set is enough to cover the temperature range of the atmospheric isobar in the liquid phase. We have thus determined the following new polynomial expression:

$$\rho_{1\,\mathrm{atm.}}(T) = 0.93502 - 3.8621\times10^{-4}\,T + 2.6342\times10^{-7}T^2 - 1.4093\times10^{-10}\,T^3 \qquad (14)$$

where $T$ is expressed in Kelvin and $\rho_{1\,\mathrm{atm.}}$ is obtained in g/cm$^3$.

Fig. 8 shows that the maximum deviation is obtained with Rinck's data (except for two points corresponding to his test n°1 which are obviously erroneous). For the other data sets the maximum deviation is less than ±0.2%. This deviation value is also the mean deviation value indicated by Ewing *et al.* (Ref. 9) between their Eq (9) and the data. Fig. 8 shows, however, that Eq. (9) of Ewing *et al.* is shifted on average by -0.2% with Eq. (14) of the present modeling.

A point of consistency is to recover as closely as possible the value of the boiling point density deduced from Eq. (12) and Eq. (13). Eq. (14) leads to a deviation of +0.036% while Eq. (9) of Ewing *et al.* leads to a deviation of +0.070%. Therefore, in order to ensure consistency, the use of Eq. (14) of the present modeling should be preferred.

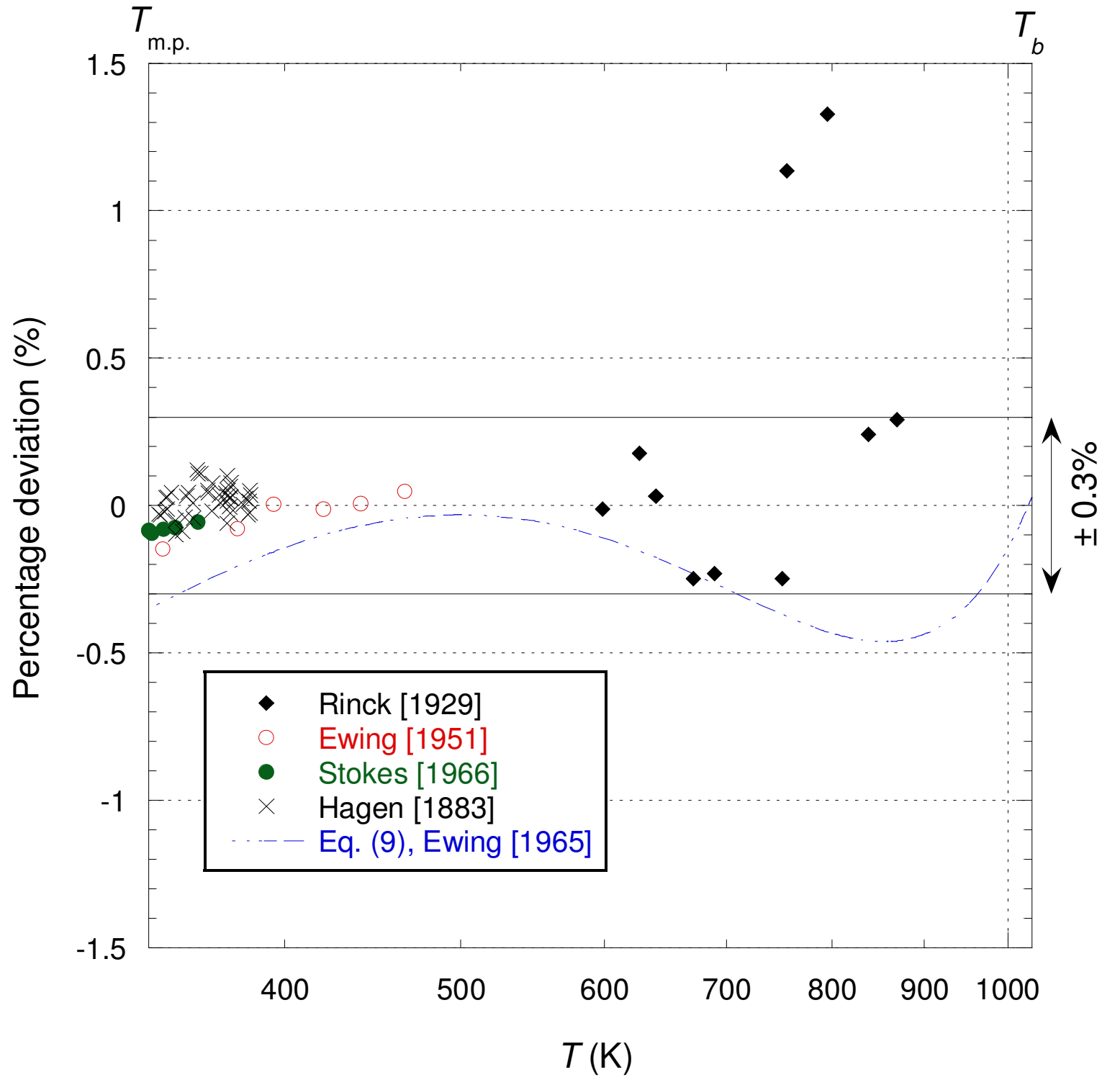

Fig. 8. Deviation plot for the liquid density of potassium along the atmospheric isobar, i.e. $100\left(\rho_{\text{exp}}-\rho_{\text{calc}}\right)/\rho_{\text{calc}}$ where $\rho_{\text{calc}}$ is determined from Eq. (14) and $\rho_{\text{exp}}$ corresponds to the data from Ref. 15 to 18. The dot-dashed blue curve represents the deviation between Eq. (9) of Ewing *et al.* (Ref. 9) and Eq. (14) of the present modeling. $T_{\text{m.p.}}$ represents the melting temperature and $T_b$ the boiling temperature for the atmospheric pressure.

### 3.4. Viscosity and self-diffusion coefficient in the gaseous phase

According to Eq. (9), the viscosity is the sum of a liquid-like term $\eta_l$ and a gas-like term $\eta_{Knu}$ related to the gas released by the shear stresses. It appeared that in the liquid phase of water, the term $\eta_{Knu}$ is small and can be neglected as a first approximation. On the other hand, in the gaseous phase, it is the term $\eta_{Knu}$ that produces the main part of the viscosity variations while the liquid term $\eta_l$ is almost constant and equal to its limit value $\eta_{l0}=\pi\hbar/\mathcal{V}_{\text{mol}}$ where $\hbar$ is the reduced Planck constant and $\mathcal{V}_{\text{mol}}=\sqrt{\mathcal{V}_{\text{c}}\mathcal{V}_0}$ represents the geometric mean of the two characteristic molecular volumes of the medium i.e. the critical molecular volume $\mathcal{V}_{\text{c}}$ and the molecular volume at zero temperature $\mathcal{V}_0$. It is therefore essential to determine the volume $\mathcal{V}_0$. This value can be determined from Eq. (11) by extrapolating the liquid density to $T$ = 0 K. Then it is deduced that $\rho_{\sigma,\text{Liq}}(0\,\text{K})=0.97723\,\text{g/cm}^3$ i.e. $\mathcal{V}_0=66.438\,\text{Å}^3$. As a result, the dilute-gas limit value of the liquid-like term is: $\eta_{l0}=\lim_{\rho\to 0}\eta_l=2.2218\times10^{-3}$ mPa.s.

On the other hand, it has been shown for water (Ref. 4) that in the dilute gas limit, the elastic shear constant varies proportionally to the cube of the density such that $K^*_{\lim}(\rho) = c_{K0}\left(\frac{\rho}{\rho_c}\right)^3$ and the function $f_N$ (which defines the fluctuation distance $d_N$ such that $d_N = f_N(\rho,T)d$ ) varies proportionally to the square of the density such that $f_{N,\lim}(\rho) = c_{N0}\left(\frac{\rho}{\rho_c}\right)^2$. These relations imply that the expression for the viscosity limit of the liquid-like term is $\eta_{l0} = 2\pi \frac{\sqrt{c_{K0}}}{c_{N0}} \frac{\sqrt{K_0 \rho_c}}{q_{c0,\text{crit}}}$.

We will admit that these relationships are still valid. This last relation with the limit viscosity $\eta_{l0}$ will allow us to determine one of the two constants $c_{K0}$ or $c_{N0}$. As the experimental devices are similar to those used to determine the viscosity of water, the fluctuative distance $d_N$ must also have similar values, therefore the coefficient $c_{N0}$ is fixed to the value of water, i.e. $c_{N0} = 6.563$ and it is deduced that $c_{K0} = 1.015 \times 10^{-4}$. For the gaseous phase, the expression of the liquid-like term is thus completely determined such that: $\eta_l(T, \rho \le \rho_{\sigma,\text{Vap}}) = \frac{d}{H_N(v)}\sqrt{\rho\, K^*_{\lim}(\rho) K_0}$ with $N - 1 = f_{N,\lim} d \frac{q_{c0,\text{crit}}}{2\pi}$.

The analysis of the viscosity data from Lee *et al.* (Ref. 19) leads to the following expression for the density of the released gas in the temperature range from 950 K to 1500 K:

$$\tilde{\rho}_{Knu}(T, \rho \le \rho_{\sigma,\text{Vap}}) = \tilde{\rho}_{Knu,0}(\rho)\exp\left(-\left(\frac{T_{Knu}}{T}\right)^{\gamma_{Knu}(\rho)}\right) \qquad (15)$$

with $T_{Knu} = 1050\,\text{K}$ and

$$\begin{cases} \tilde{\rho}_{Knu,0}(\rho) = 0.063039 + (0.025174 - 0.063039)\,\text{erf}\left((\rho/0.010419)^{1.08}\right) \\ \gamma_{Knu}(\rho) = 1 + (7-1)\,\text{erf}\left((\rho/0.01519)^{3/2}\right) \end{cases}$$

where $\text{erf}(\bullet)$ represents the error function. In Eq. (15), $\rho$ must be expressed in g/cm$^3$ and $T$ in Kelvin. It can be noted that Eq. (15) and its constituent parameters have variations similar to those of water in the gaseous phase (see Ref. 4).

Hence, in the gaseous phase, for the temperature range under consideration, the viscosity can be simply determined by using Eq. (9) where $\tilde{\rho}_{Knu}(\rho,T)$ is given by Eq. (15) and $N - 1 = f_{N,\lim}(\rho)\, d \frac{q_{c0,\text{crit}}}{2\pi}$.

We compare below the viscosity data in Fig. 5 of Ref. 19 with the present modeling along the different isobars. In view of the wide dispersion of the data, it is not relevant to plot this comparison as a deviation curve. Fig. 9 shows that present modeling provides a good representation of the data whose deviations are comparable to the curves drawn by Lee *et al.* It even appears that on the isobar corresponding to 1.06 atm., the present modeling provides a better representation of the data than the best of the correlations proposed by Lee *et al.*

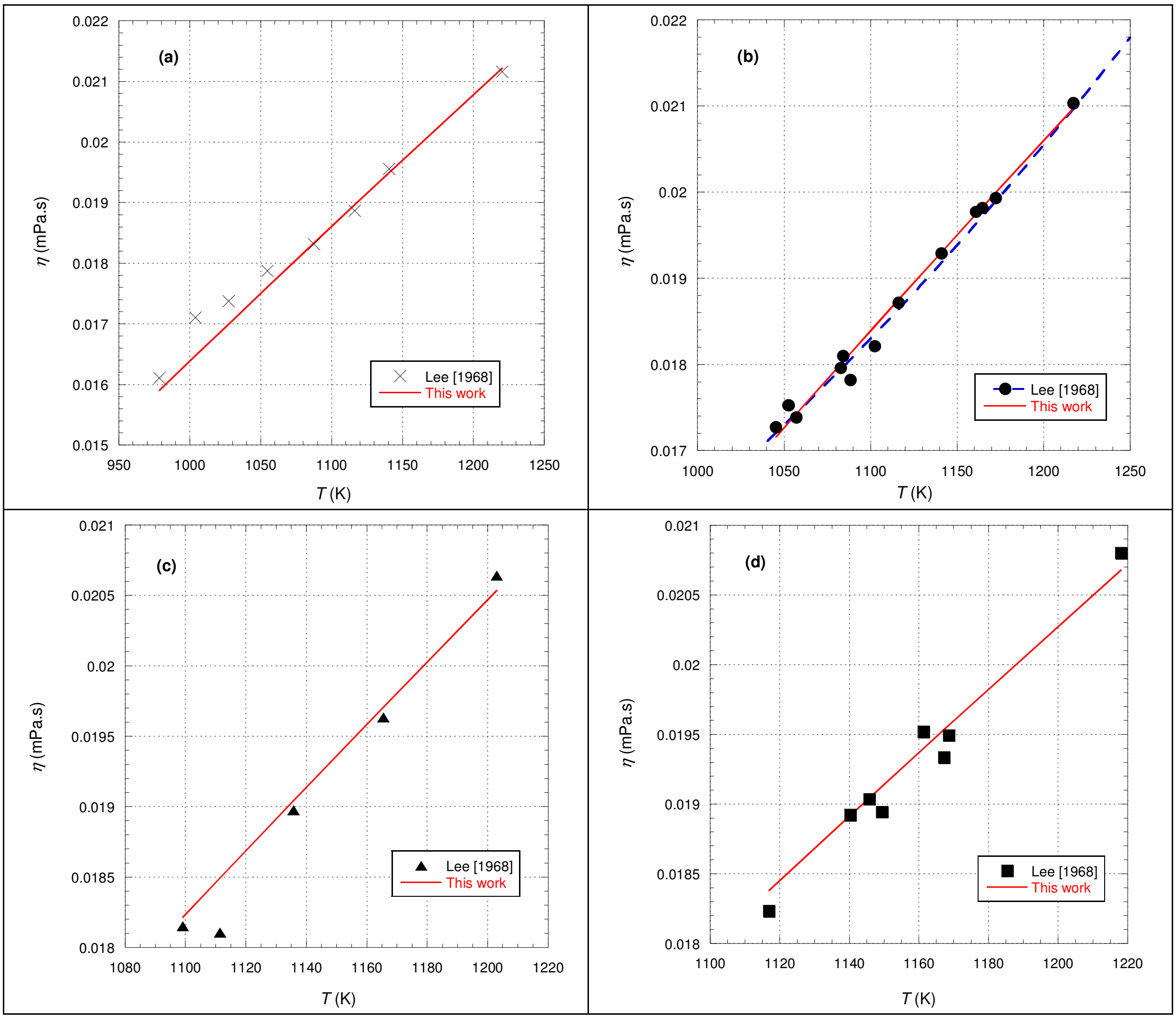

Fig. 9. Comparison of viscosity data (i.e. black points) from Ref. 19 with the present modeling (i.e. red curves) along different isobars in the gaseous phase of potassium versus the temperature: (a) 0.55 atm.; (b) 1.06 atm.; the blue dashed curve represents the best correlation given by Lee *et al.* for the atmospheric pressure (c) 1.44 atm.; (d) 2.02 atm.

Lee *et al.* proposed a table (i.e. Table 3) with viscosity values calculated and extrapolated for higher pressures (i.e. 0.506 MPa). Fig. 10 shows that the overall deviation is included within ±1.5%. It can be seen that the largest deviation is observed for the points extrapolated on SVP but this deviation is of the same order of magnitude as for the previously processed raw data. It can be concluded that the present modeling is also compatible with these calculated data.

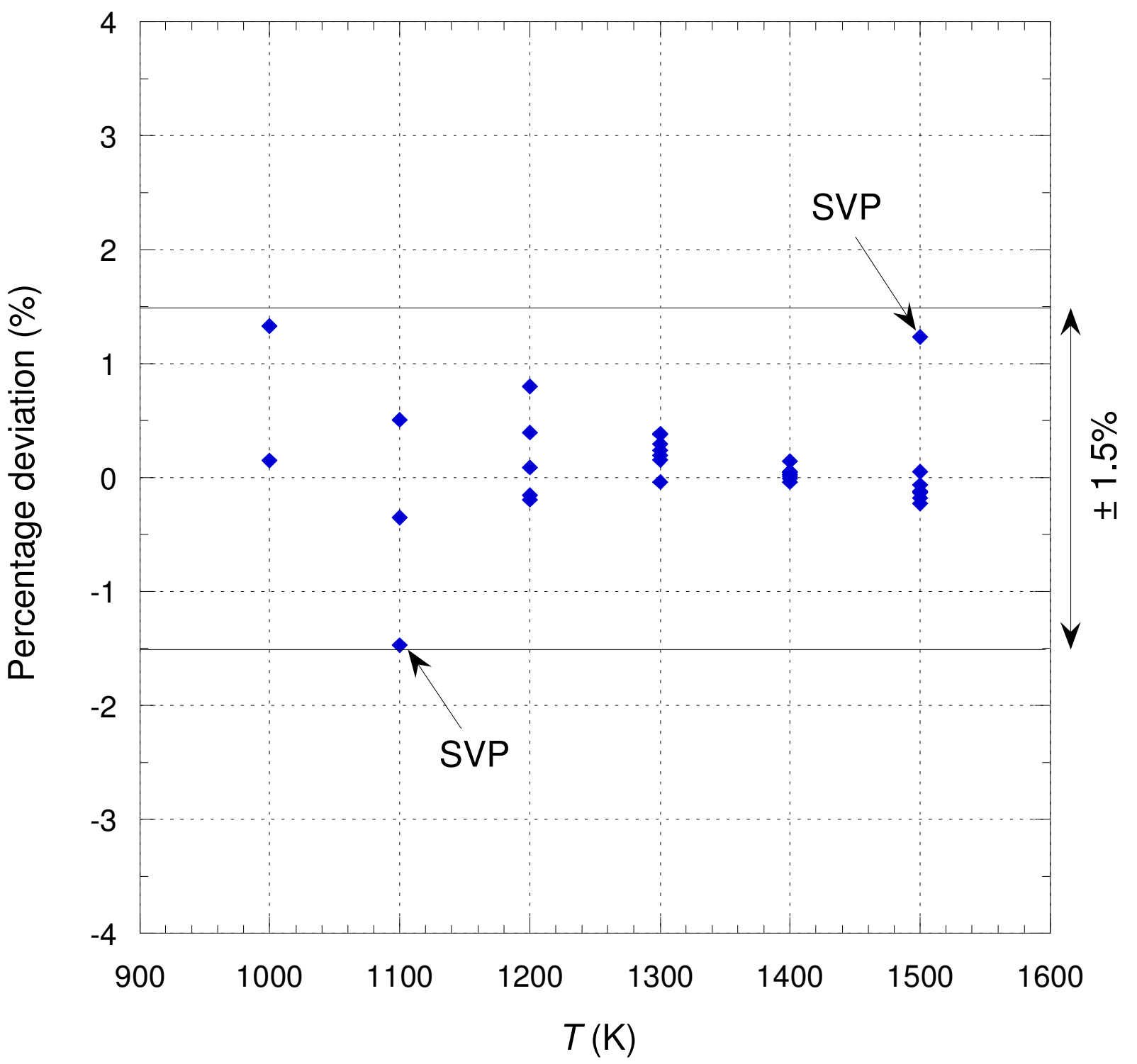

Fig. 10. Deviation plot for the viscosity data from Table 3 of Ref. 19 with the present modeling, i.e. $100\left(\eta_{\mathrm{Lee}} - \eta_{\mathrm{Eq.(A.4)}}\right)/\eta_{\mathrm{Eq.(A.4)}}$ . The points indicated by an arrow with "SVP" correspond to the points on the saturated vapor pressure curve at the corresponding temperature.

Fig. 11 shows the viscosity evolution determined by the present modeling in the temperature range 900 K to 1600 K: we observe that the viscosity at high temperature is about ten times the limit value $\eta_{l0}$ but we can see that the value converges well towards the limit value $\eta_{l0}$ as the temperature decreases and the density increases.

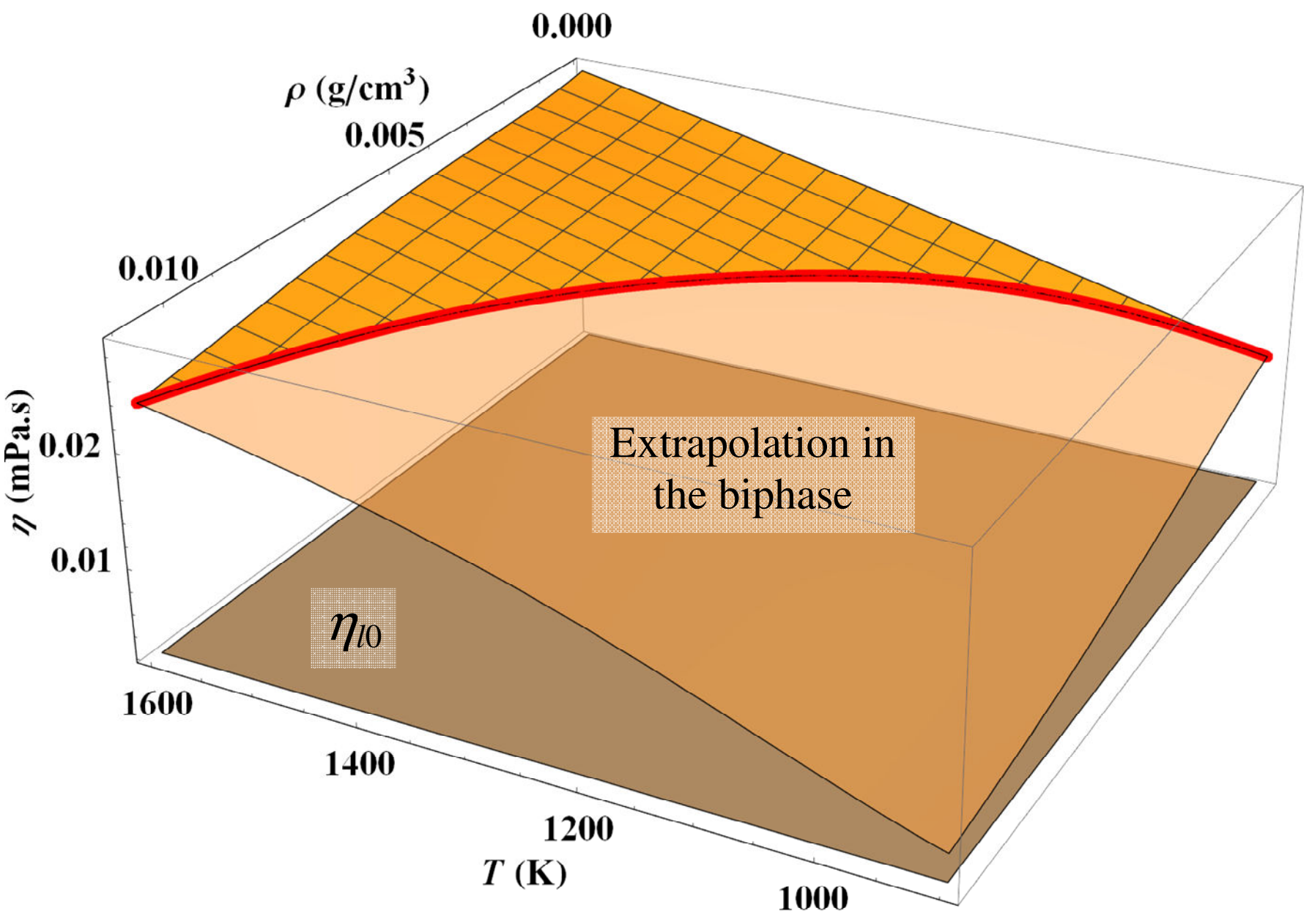

Fig. 11. Viscosity as function of density and temperature in the gaseous phase of potassium. The thick red curve represents the gaseous states along SVP from 900 K to 1600 K. The horizontal plane represents the dilute-gas limit value of the liquid-like term.

Stefanov *et al.* (Ref. 20) also made viscosity measurements over a wider range of pressure and temperature than Lee *et al.* Fig. 12 shows the data of Lee *et al.* and Stefanov *et al.* for two isobars close to atmospheric pressure. It appears that the data are shifted with respect to each other, which could be explained by the fact that $d_N$ is lower in the experiment of Stefanov *et al.* than in that of Lee *et al.*, but in the absence of details on Stefanov *et al.*'s experiment, it is not possible here to make a more detailed analysis of these data.

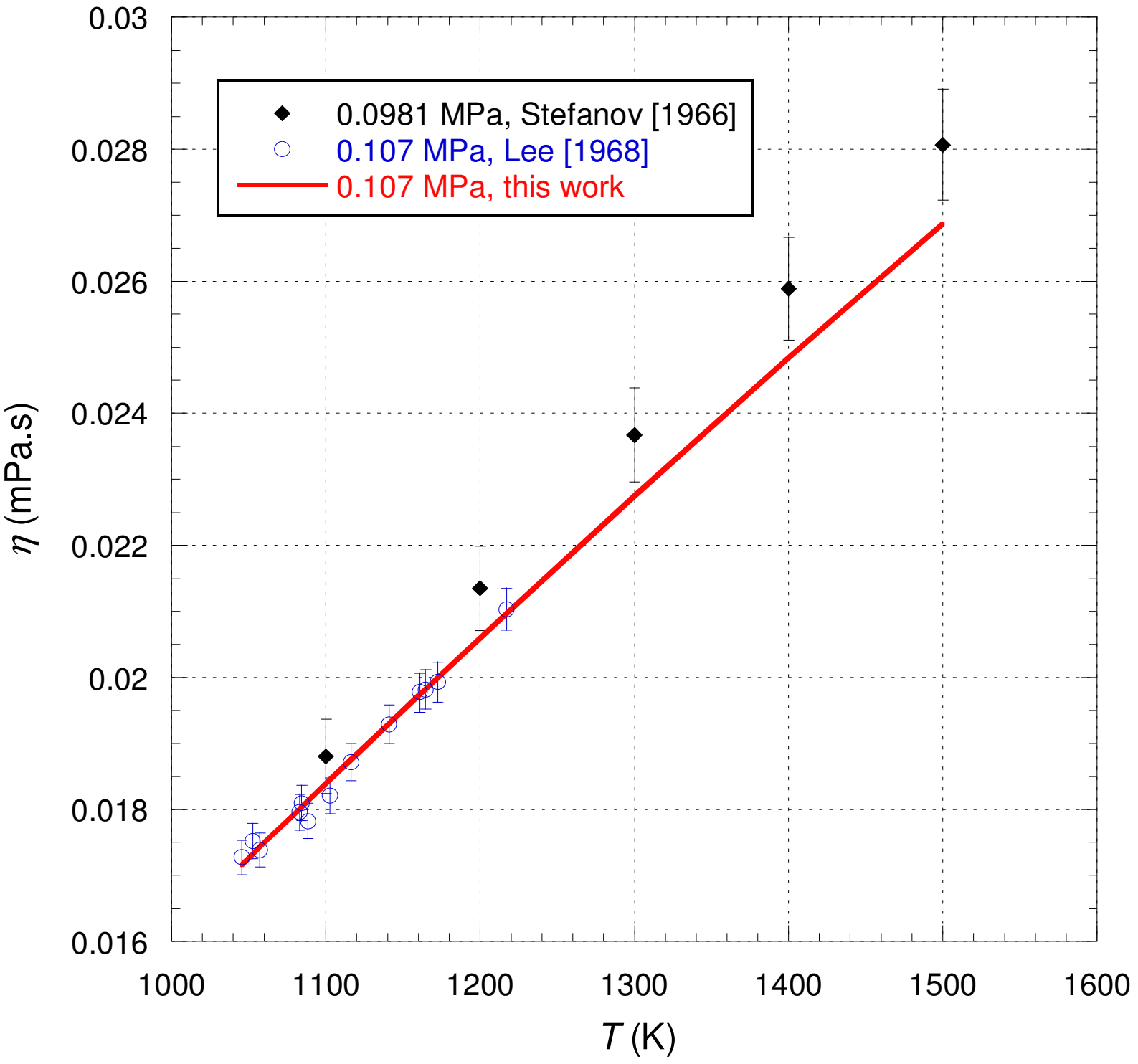

Fig. 12. Viscosity data along two isobars close to atmospheric pressure in the gaseous phase of potassium: the black diamonds correspond to the data of Stefanov *et al.* (Ref. 20) and the blue circles correspond to the data of Lee *et al.* (Ref. 19). The red curve corresponds to the present modeling calculated along the isobar equal to 0.107 MPa.

In the gas phase there are no data, *a priori*, about the self-diffusion coefficient of potassium. However we have shown in Ref. 4 that in water, the self-diffusion coefficient becomes equal to the thermal diffusion coefficient $D_{th}$ as soon as the density is sufficiently low compared to the critical density. Moreover, the coefficient $D_{th}$ that must be considered is the one determined from the isochoric heat capacity $C_V$, i.e.

$$D_{th} = \frac{\lambda}{\rho C_V} \tag{16}$$

where $\lambda$ represents the thermal conductivity.

Caldwell *et al.* (Ref. 12) have determined some equations of state to describe the evolutions of $\lambda$ and of the isobaric heat capacity $C_P$ along SVP. It has been seen previously that the equations of Caldwell *et al.* do not give a very good representation of the experimental data below 1000 K. Fig. 13 shows that this is still the case with thermal conductivity. Thus one obtains an average deviation of 25% with Briggs experimental data (Ref. 21), while the latter are given to have an uncertainty of ±10%. It is also observed that the model of Caldwell *et al.*

is quite different from the data of Stefanov *et al.* (Ref. 20) along SVP. But the temperature range of Stefanov *et al.* data is too small to do without the Caldwell *et al.* modeling.
From derivatives of Eq. (10) and well-known thermodynamic formula, it is possible to deduce corresponding $C_V$ values and then to calculate $D_{th}$ values. It should be considered that the uncertainty corresponding to the determined values of $D_{th}$ from Caldwell *et al.* modeling can be between 10% and 25%.

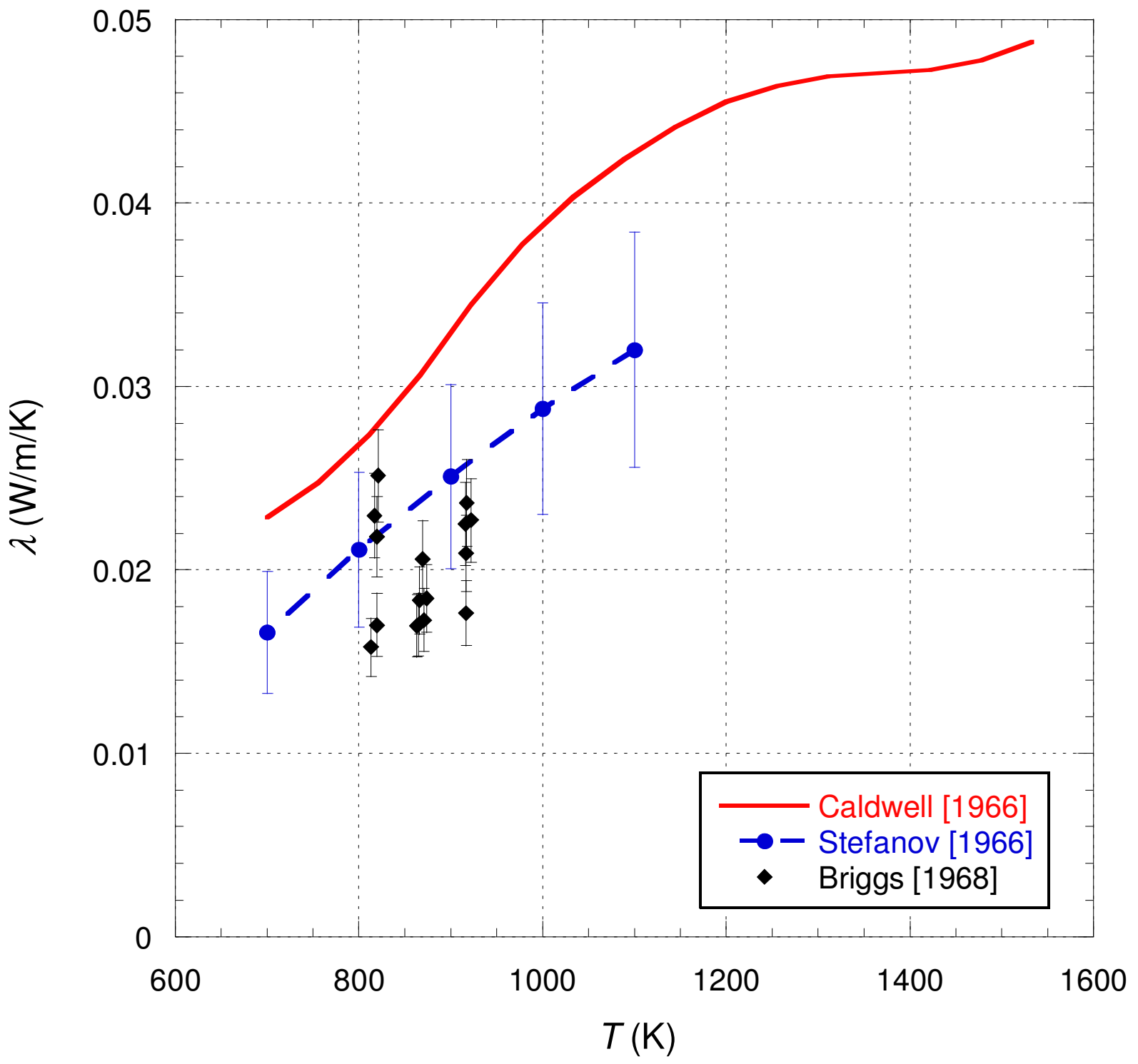

Fig. 13. Thermal conductivity as a function of temperature along SVP in the gaseous phase of potassium. The red curve represents the model of Caldwell *et al.* (Ref. 12) while the black diamonds represent Briggs' experimental points (Ref. 21, Table VI).

Caldwell *et al.* model should be considered as providing plausible variations of some parameters but not absolute values. To obtain absolute values, one must consider the data of Gerasimov *et al.* (Ref. 22) who have measured the thermal conductivity of potassium vapor at very small pressures (between 787 Pa and 147 Pa) and for a temperature range between 1189 K and 1931 K. This temperature range coincides in part with that of Caldwell *et al.* modeling, which allows for a possible connection between the two approaches, theoretical and experimental. Gerasimov *et al.* considered that it is a monoatomic potassium vapor, therefore we will simply transform these values of thermal conductivity into thermal diffusivity by assuming that the vapor density $\rho$ is deduced from $P$ and $T$ by the perfect gases law and that the isochoric heat capacity $C_V$ is equal to $\frac{3}{2}R_g$.

Assuming that the function $f_{q_c,\mathrm{Gas}}(\rho,T)$ has the same structure as the corresponding function in the case of water, namely

$$f_{q_c,\mathrm{Gas}}(\rho,T)=1+\alpha_{q_c,\mathrm{Gas}}(T)\exp\left(-\frac{\rho}{\rho_{q_c}(T)}\right)\left(1-\frac{\rho}{\rho_{q_c}(T)}\right)\left(\frac{\rho_c}{\rho}\right)^{\frac{4}{3}} \tag{17}$$

then the identification of $D_t$ with $D_{th}$ leads to the expression of the parameter $\alpha_{q_c,\mathrm{Gas}}(T)$ such that:

$$\alpha_{q_c,\mathrm{Gas}}(T) = 0.043734\exp\left(-\left(\frac{T}{823.188}\right)^{6}\right) + 0.01715\exp\left(-\left(\frac{T}{1181.5}\right)^{10}\right) + 0.05266\exp\left(-\left(\frac{T}{2125.25}\right)^{22}\right) \quad (18)$$

Eq. (17) is valid as long as $\rho < \rho_{q_c}$ and $\rho_{q_c}(T)$ identifies with $\rho_{\sigma,\mathrm{Liq}}(T)$ as long as $T \le T_c$. Fig. 14 shows the comparison between the data of Gerasimov *et al.* transformed into thermal diffusivity and the present modeling. It is observed that the least well reproduced isobar is the one corresponding to the lowest pressure taking into account error bars of 6% if one follows the authors recommendation:

> "Although the experimental points are scattered with RMS deviation of 6.9 % a statistical analysis permits one to reduce the overall 95% confidence error to 6%."

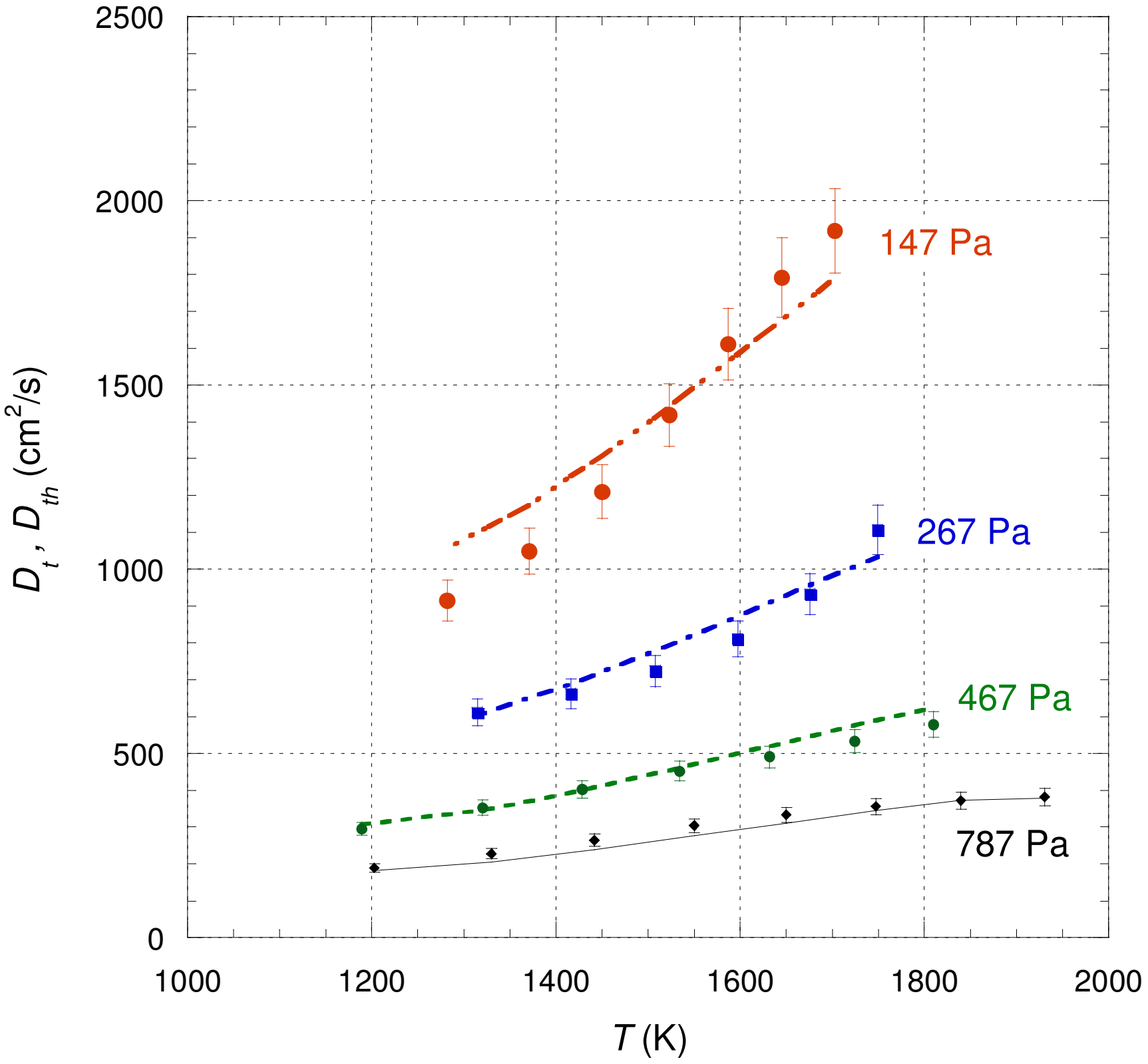

Fig. 14. Comparison of thermal diffusivity deduced from Gerasimov *et al.* data (Ref. 22) with the present modeling (color curves) along different isobars in the gaseous phase of potassium.

The deviation obtained may seem important but Fig. 15 shows that this deviation is almost identical to that obtained between Eq. (5) of the authors and their data. We can therefore deduce that the representation of these data by the present modeling is in conformity with the deviation implicitly admitted by the authors.

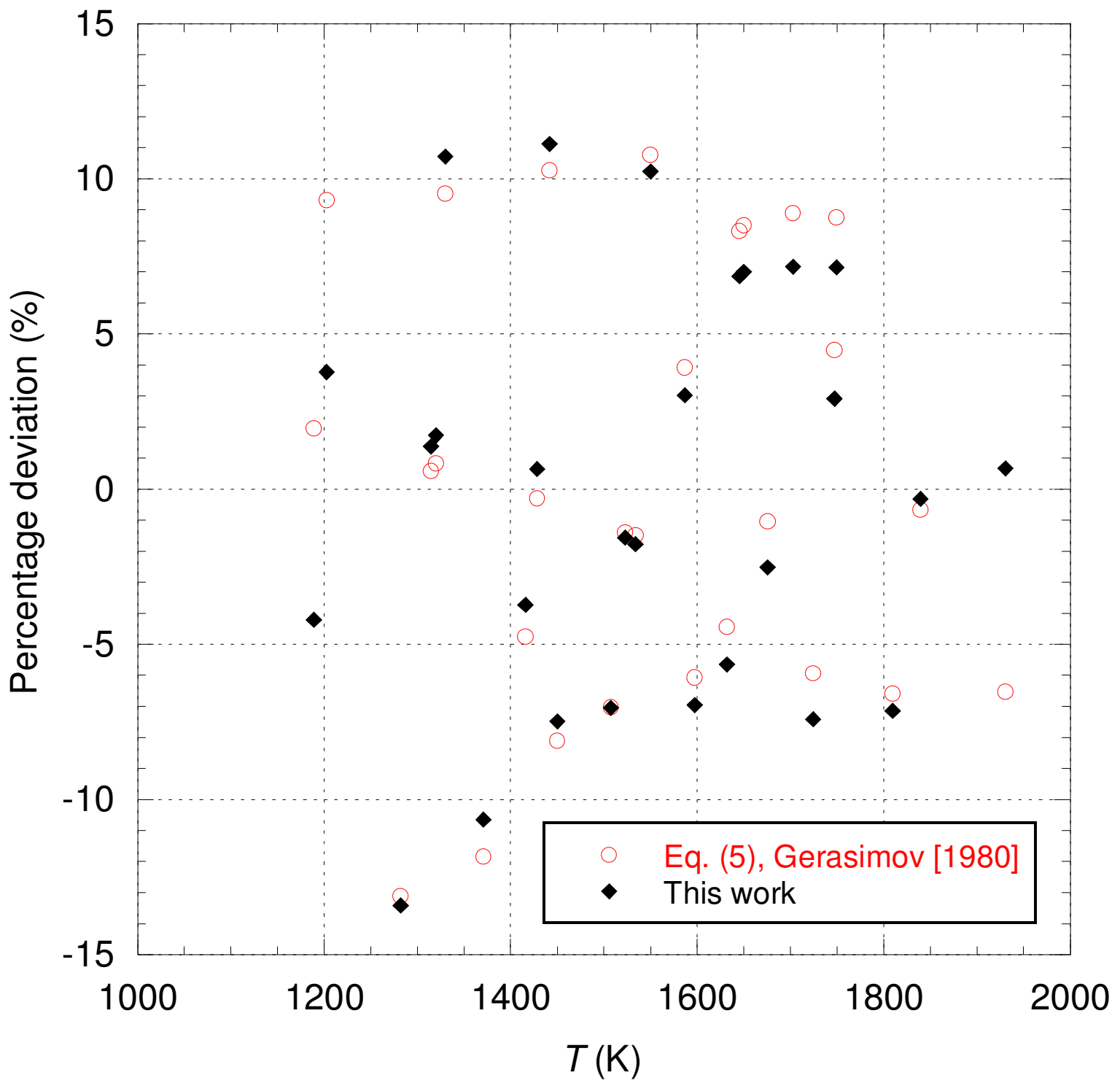

Fig. 15. The red circles correspond to the deviation of the potassium thermal conductivity data from Gerasimov *et al.* (Ref. 22) with their Eq. (5) (i.e. $100\left(\lambda_{\text{exp}} - \lambda_{\text{calc}}\right)/\lambda_{\text{calc}}$ ) and the black diamonds correspond to the deviation of the potassium thermal diffusivity data deduced from Ref. 22 with the present modeling (i.e. $100\left(D_{th,\text{exp}} - D_{t,\text{calc}}\right)/D_{t,\text{calc}}$ ).

Stefanov *et al.* (Ref. 20) made thermal conductivity measurements also along different isobars. But the lowest pressure of the isobars is higher than those of Gerasimov *et al.* Therefore the data of Stefanov *et al.* do not correspond to the monoatomic vapor of potassium. This said, by treating the isobar of smaller pressure as if it were monoatomic vapor, Fig. 16 shows that the data of Stefanov *et al.* are well reproduced by the present modeling taking into account the error bars.

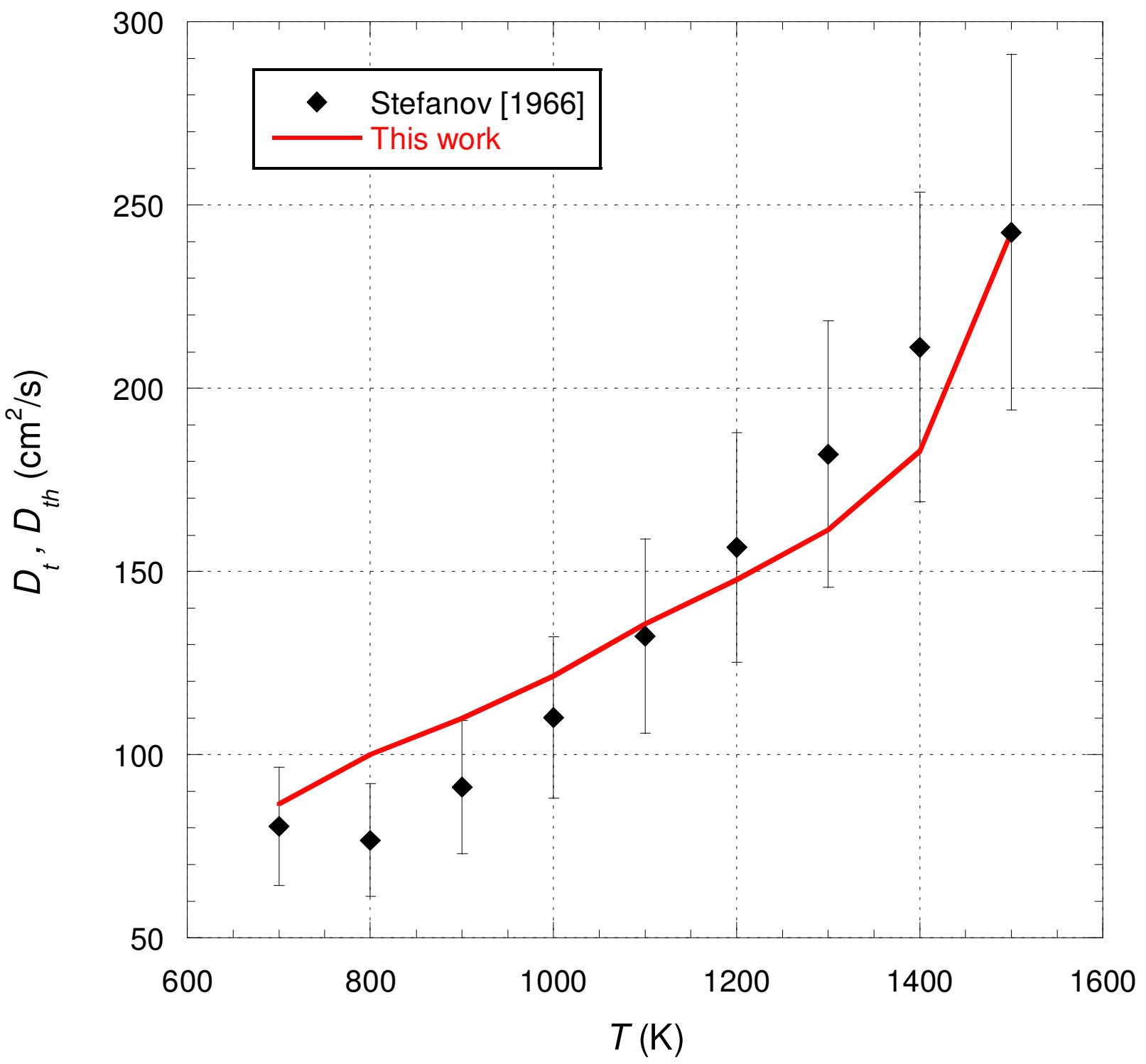

Fig. 16. Comparison of potassium thermal diffusivity deduced from Stefanov *et al.* data (Ref. 20) with the present modeling (red curve) along the isobar equal to 981 Pa.

Fig. 17 shows finally the self-diffusion coefficient evolution determined by the present modeling in the temperature range 700 K to 1900 K: it is observed that it varies little with temperature along isochors.

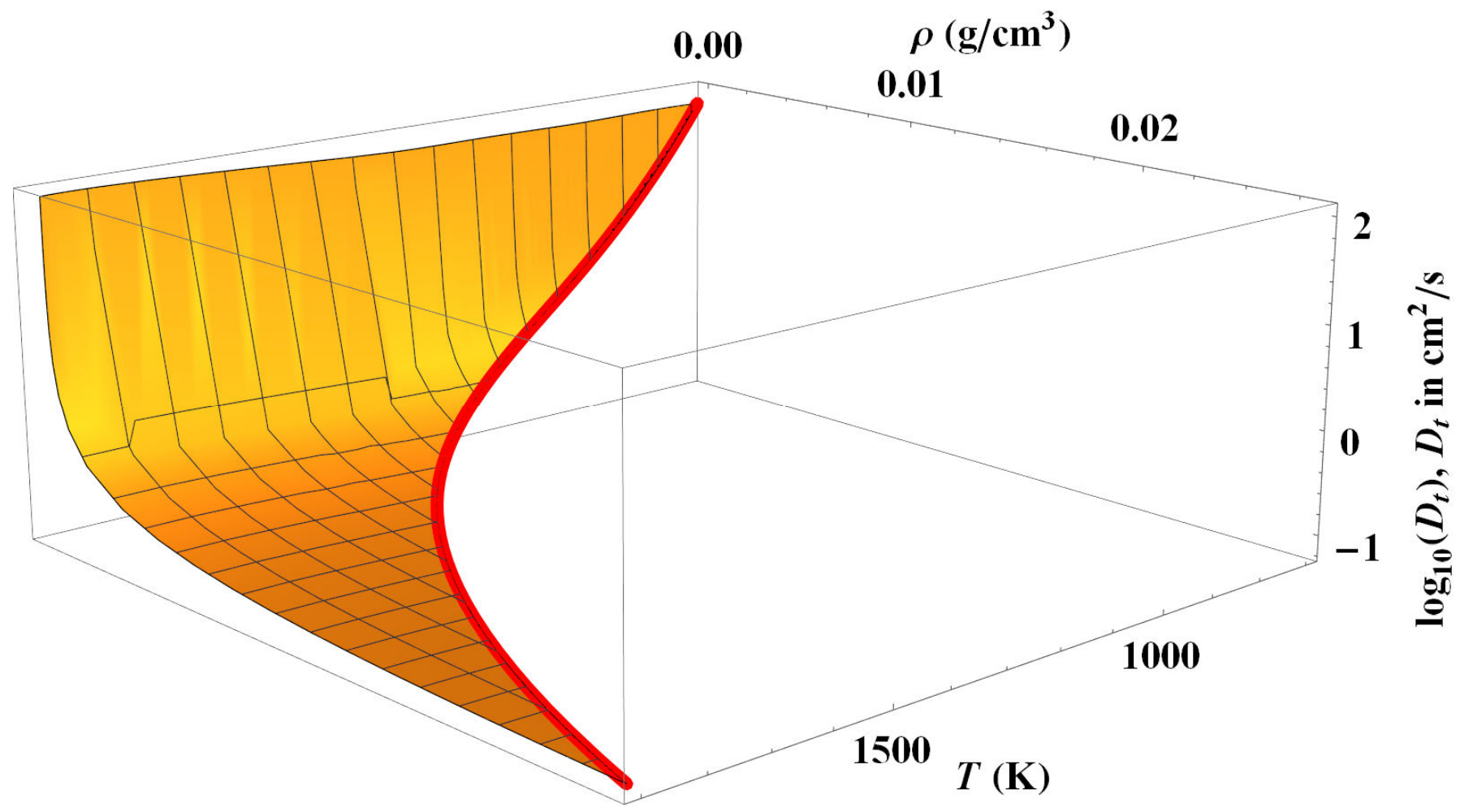

Fig. 17. Evolution of the logarithm of the self-diffusion coefficient as function of density and temperature in the gaseous phase of potassium. The thick red curve represents the gaseous states along SVP from 700 K to 1900 K.

### 3.5. Viscosity and self-diffusion coefficient of potassium in the liquid phase

In the liquid phase, one can find different datasets of viscosity measurement in the literature, but only two datasets have a fairly wide temperature range along the atmospheric isobar (i.e. Refs. 18 and 23). It is these two datasets that we will analyze here to which we have added the few data points of Ref. 24 obtained by the same authors of Ref. 18. Let us note however that another dataset (Ref. 25) exists corresponding to viscosity variations along a quasi-isochor with a very wide temperature range but the lack of information on the experimental set-up does not allow a relevant analysis.
Fig. 18 shows that the data sets have quite different variations and seem to deviate more and more with increasing temperature, as Ewing *et al.* wrote in 1951 (Ref. 18):

> "the values for potassium, though coinciding with those by Chiong at 70°, diverge as the temperature is increased and differ by as much as 10% at 200°."

then they updated their commentary in 1954 (Ref. 24) by writing:

> "the values for potassium, though coinciding with those by Chiong at 70°, diverge at higher temperatures and differ by as much as 14% at 350°."

The discrepancies found between the authors make Lemmon *et al.* (Ref. 25) say that:

> "The viscosity of liquid potassium has been studied by various workers; however; the agreement among the several sets of data was, in general, out side of the indicated precision of the individual sets."

Indeed, Ewing *et al.* considered in 1951 that:

> "the over-all accuracy of the viscosity results will be stated as ±0.8%."

while Chiong (Ref. 23) considers that its data can be represented with a maximum uncertainty of ±0.7%.

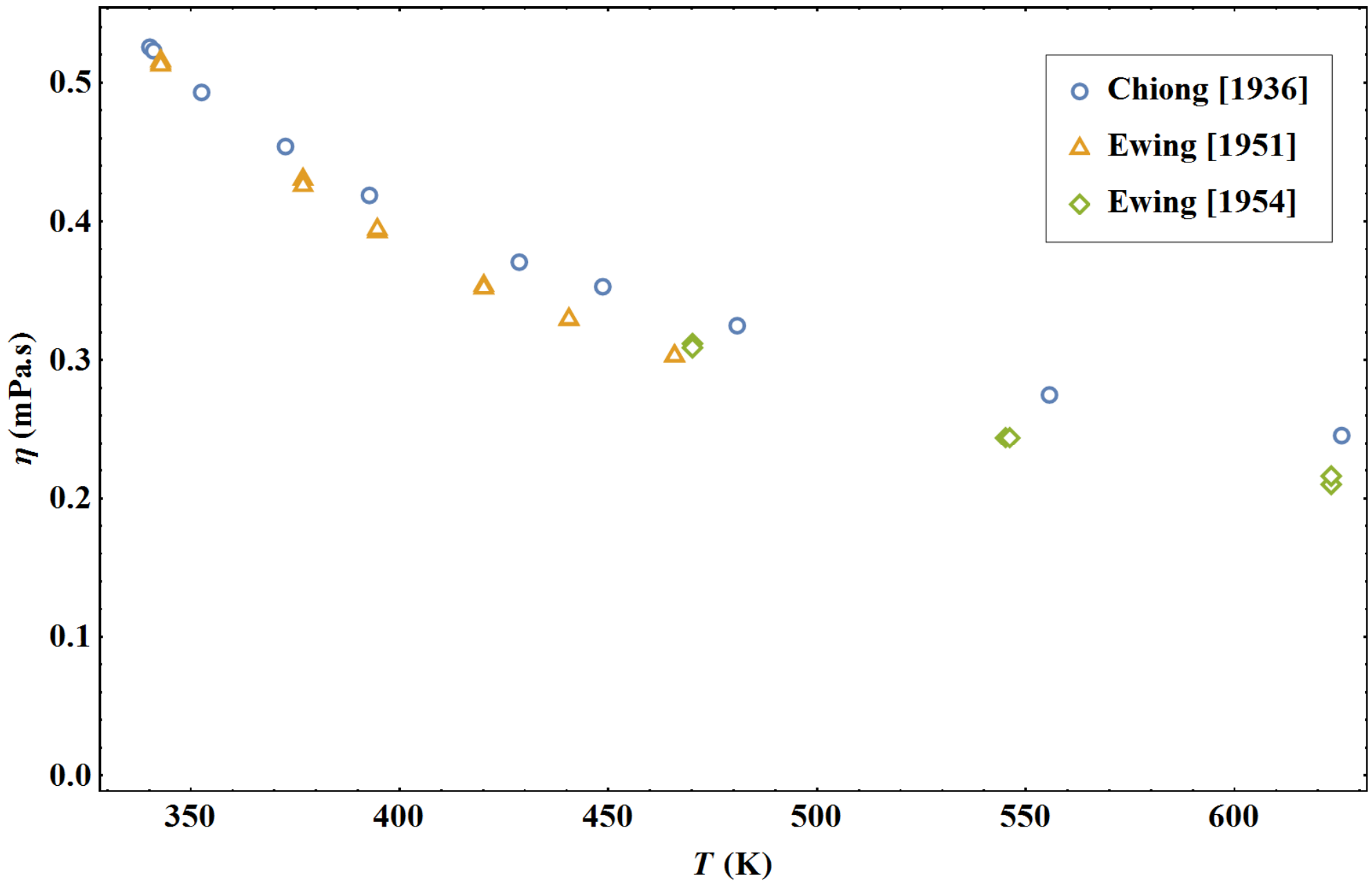

Fig. 18. Potassium viscosity data versus temperature along the atmospheric isobar from Chiong (Ref. 23), Ewing *et al.* (Refs. 18 and 24).

To analyze such deviations, it is necessary to be able to determine the parameters related to the geometrical characteristics of the experiments. These parameters are not clearly given in the references but it turns out that the different experimental devices have been calibrated using water as the test fluid along the atmospheric isobar. For Chiong device, the data on water can be found in Ref. 26. We recall that for water modeling, the dissipative distance $d$ is set by default to the value $d$ = 0.01 cm and the fluctuative distance $d_N$ is calculated such that $d_N = f_N(\rho,T)d$ . These distances are multiplied by the coefficients $C_d$ and $C_N$ to take into account the specific geometrical characteristics of the experimental devices.

Chiong's experiment consists in using oscillating spheres with a diameter varying approximately between 4 cm and 5 cm. Fig. 19a shows that the water viscosity data can be reproduced using the model of Ref. 4 with a maximum deviation of ±1% in accordance with the deviation that the authors found in Table 2 of Ref. 26. The value of $d_N$ varies between 1.35 cm and 1.5 cm and is perfectly compatible with the spheres radii. This said, the deviation could be reduced if we could have analyzed the points corresponding to each sphere diameter separately.

In 1951, Ewing *et al.* (Ref. 18) determined viscosities using an Ostwald type viscometer having a

> "measuring bulb (30 ml. with a mean diameter of 5 cm.) and a receiving bulb with an average diameter of 6 cm."

but there is no information concerning the capillary used. Fig. 19b shows that the water viscosity data can be reproduced using the model of Ref. 4 with a maximum deviation of ±0.12% in accordance with what the authors have written:

> "The last column in the table gives the water viscosity as calculated from these empirical constants. The maximum deviation of any value, so calculated, from the corresponding NBS value was only 0.12%."

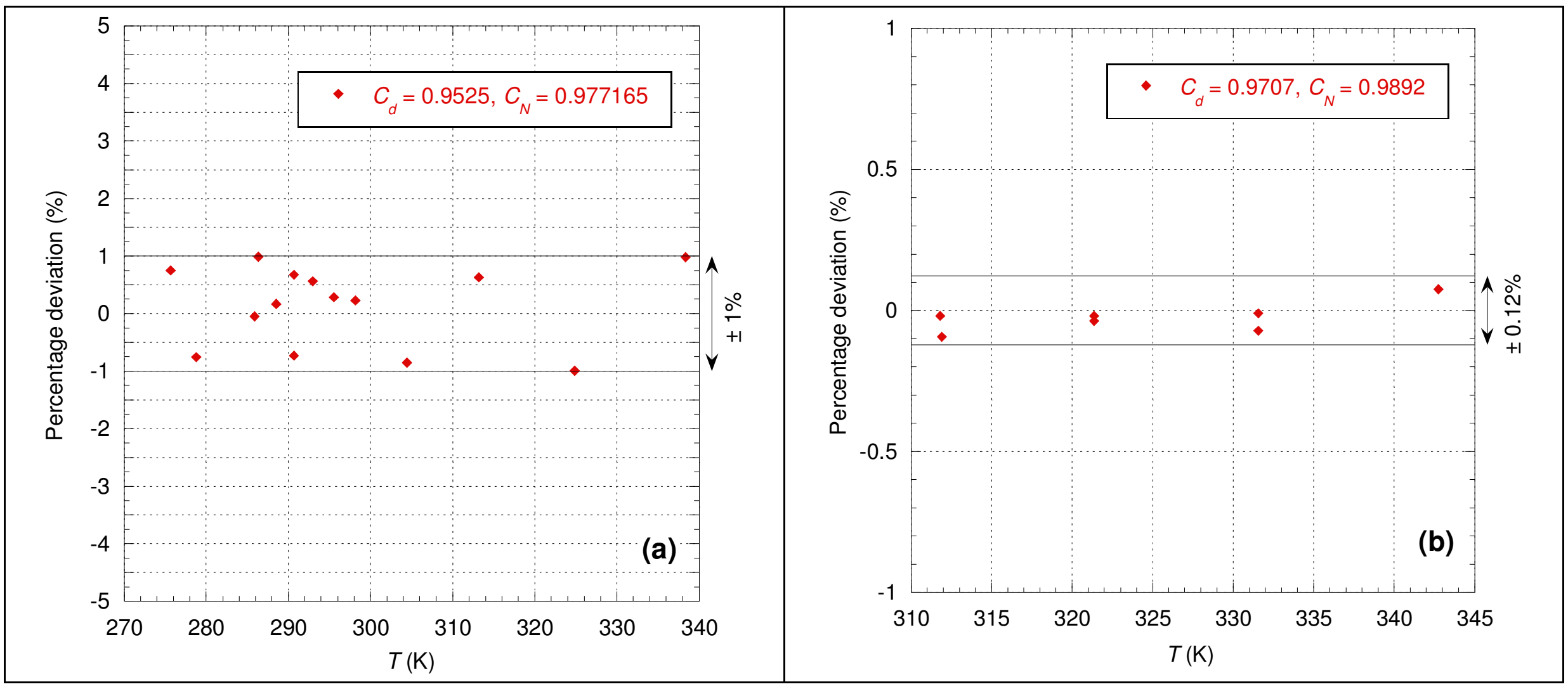

Fig. 19.Deviation plot for the viscosity data of liquid water along the atmospheric isobar, i.e. $100(\eta_{\rm exp} - \eta_{\rm calc})/\eta_{\rm calc}$ where $\eta_{\rm calc}$ is calculated from the model developed in Ref. **4**: (a) $\eta_{\rm exp}$ corresponds to the data from Ref. 26; (b) $\eta_{\rm exp}$ corresponds to the data from Ref. 18. $C_d$ and $C_N$ represent the coefficients that multiply the values of $d$ and $d_N$, respectively.

To analyze the potassium data, the same $d$ values as those obtained for water in each of the experimental devices are then imposed. The single data along the atmospheric isobar does not allow to deduce a general law of evolution of the parameters $K^*(\rho)=K/K_0$ and $f_N(\rho,T)$. It has been shown in Ref. 4 that $K^*$ varies little with temperature on the atmospheric isobar of liquid water and its value remains close to 1. Consequently we will admit by simplification that $K^*$ can be considered as constant along the atmospheric isobar for potassium liquid. In the same way, $f_N$ can be considered as a constant along the atmospheric isobar, in first approximation. The densities of potassium being comparable to those of water, we set $f_N$ to a mean value determined from its expression for water.

The knowledge of data at atmospheric pressure alone does not also make it possible to generalize Eq. (15) to the liquid densities. One must admit a simplified expression which depends only on $T$. In the case of water, it has been shown on the atmospheric isobar that the variation of density $\tilde{\rho}_{Knu}(T,1\,\text{atm.})$ had a bell shape with a maximum in the vicinity of the maximum density of water. In other words, in normal liquid water, $\tilde{\rho}_{Knu}(T,1\,\text{atm.})$ decreases quasi-exponentially as $T$ increases along the atmospheric isobar. Therefore, it is assumed for liquid potassium at atmospheric pressure that $\tilde{\rho}_{Knu}$ can be written in the following form:

$$\tilde{\rho}_{Knu}(T \geq T_{\text{m.p.}},1\,\text{atm.}) = (\tilde{\rho}_{\text{m.p.}} - \tilde{\rho}_\infty)\exp\left(-\left(\frac{T-T_{\text{m.p.}}}{\Delta T_{Knu}}\right)^{\gamma_{Knu,1\,\text{atm.}}}\right) + \tilde{\rho}_\infty \qquad (19)$$

where $\tilde{\rho}_{\text{m.p.}}$ and $\tilde{\rho}_\infty$ are two constant values of density which respectively fix the value of $\tilde{\rho}_{Knu}(T,1\,\text{atm.})$ at the melting temperature $T_{\text{m.p.}}$ and infinite temperature. Thus the expression of the viscosity of liquid potassium along the atmospheric isobar is determined by using Eq. (9) where $\rho_{1\,\text{atm.}}(T)$ is given by Eq. (14) and $\tilde{\rho}_{Knu}(T,1\,\text{atm.})$ by Eq. (19).

The analysis of the viscosity data in potassium leads to set the values given in Table 2. The parameters $\Delta T_{Knu}$ and $\gamma_{Knu,1\,\text{atm.}}$ have not been set to a precise value because we will see that this value depends on the analyzed data.

| $d$ (cm) | $K^*$ | $f_N$ | $\tilde{\rho}_{\text{m.p.}}$ (g/cm$^3$) | $\tilde{\rho}_\infty$ (g/cm$^3$) |
|---|---|---|---|---|
| 0.01 | 0.225 | 158.45 | 1.318 | 0.262 |

Table 2. Characteristic parameters to describe the viscosity of potassium along the atmospheric isobar.

We will start by analyzing the oldest data, i.e. Chiong's data (Ref. 23). Fig. 20 shows that Chiong's data can be represented with the expected experimental uncertainty with the exception of the point at $T$ = 337.5 K which was also discarded by Chiong himself from his analysis. Although the uncertainty obtained is comparable to that of Andrade's formula (i.e. Eq. (3) of Ref. 23), it can however be noted that the deviation corresponding to the present modeling is better centered on the zero value so it provides a slightly better description.

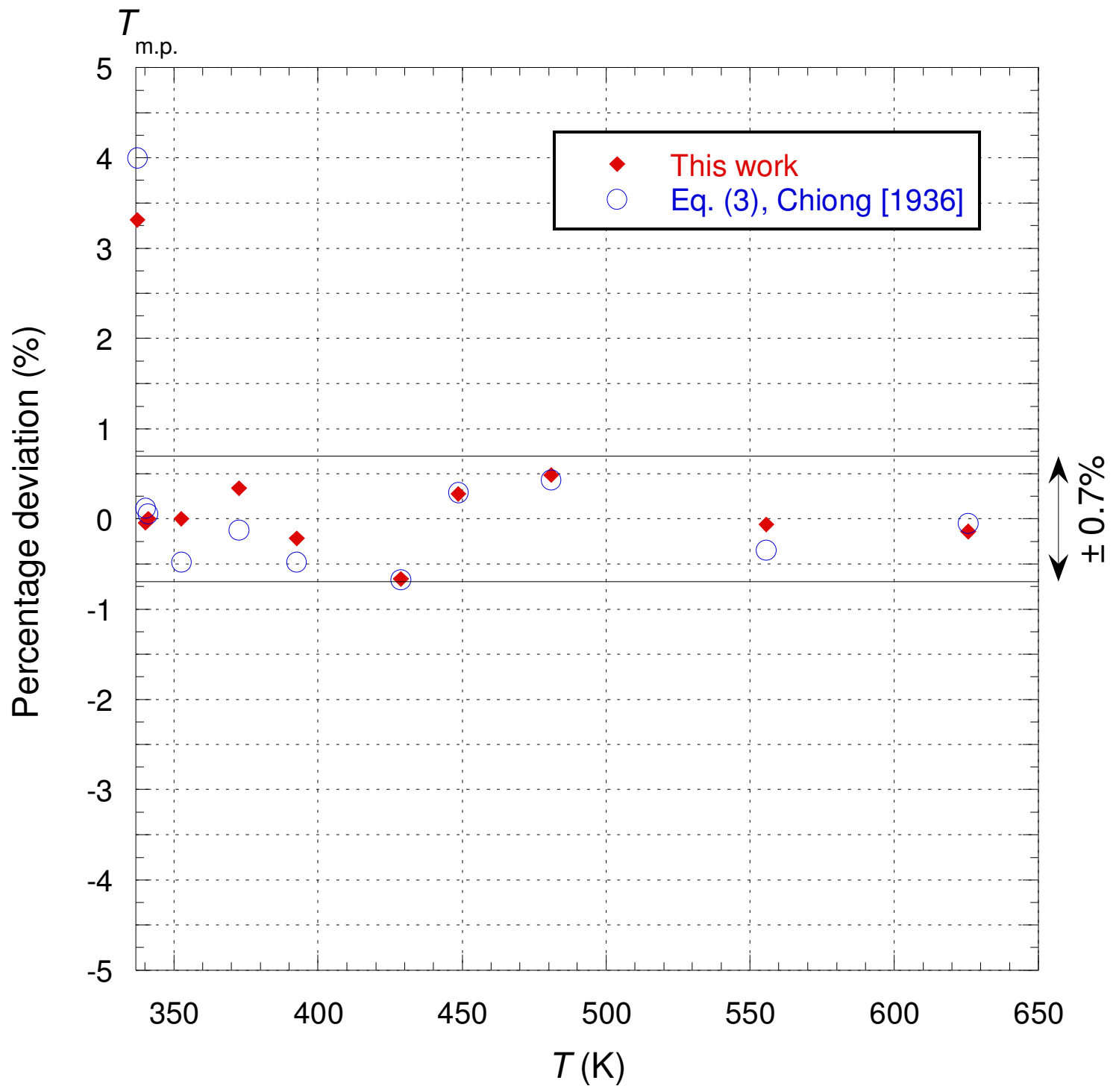

Fig. 20. Deviation plot for the viscosity data of liquid potassium along the atmospheric isobar, i.e. $100(\eta_{\text{exp}} - \eta_{\text{calc}})/\eta_{\text{calc}}$ where $\eta_{\text{exp}}$ are from Table II of Ref. 23. $\eta_{\text{calc}}$ represents either Eq. (3) of Ref. 23 or Eq. (9) with $C_d = 0.9525$, $C_N = 0.986928$, $\Delta T_{Knu} = 148.192$ K and $\gamma_{Knu,1\,\text{atm.}} = 0.867196$. $T_{\text{m.p}}$ represents the melting temperature for the atmospheric pressure.

Ewing *et al.* (Refs. 18 and 24) conducted two sets of experiments, in 1951 and 1954, and considered that the results could be combined in such a way that

> "a composite set of viscosity-composition isotherms can be readily drawn for the full temperature range, from which viscosity-temperature relationships for any composition can be interpolated."

However, as the authors wrote in 1954,

> "Viscosity coefficients to 200° for sodium, potassium and their alloys were measured by the present authors in a modified Ostwald viscometer of glass. A larger capillary type viscometer of nickel has been used to extend these measurements to 700°."

two different viscometers have been used, therefore within the framework of the present modeling these two datasets must be analyzed separately.

The 1951 experiment corresponds to the same viscometer as the data analyzed for water in Fig. 19b. By considering the same value of *d* as for water data analysis, Fig. 21 shows that the present modeling can represent the data with the expected experimental uncertainty. As for Chiong, it can be noted that the deviation of the present modeling is slightly better centered on the zero percent value than the Andrade's equation.

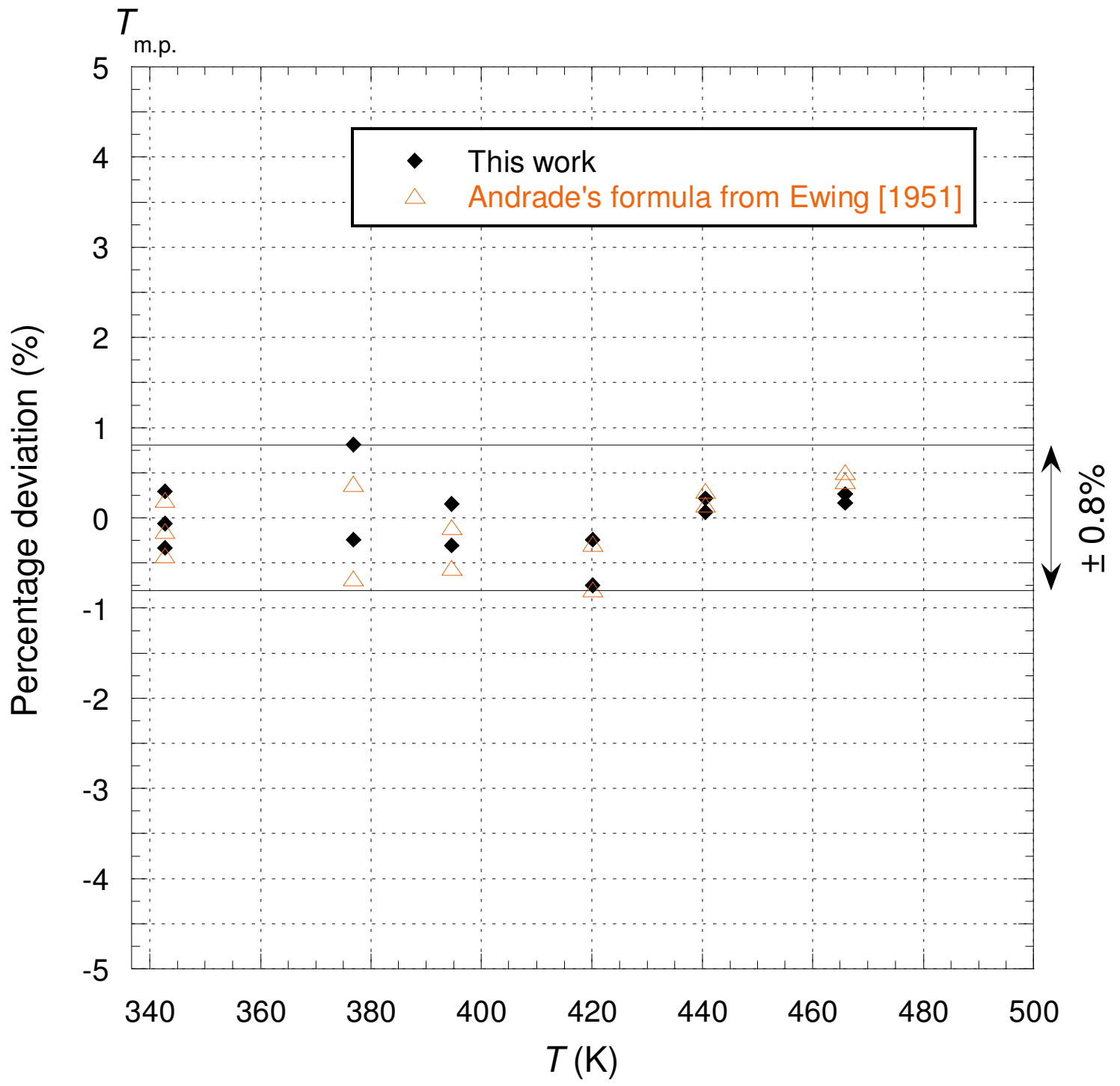

Fig. 21. Deviation plot for the viscosity data of liquid potassium along the atmospheric isobar, i.e. $100\left(\eta_{\text{exp}}-\eta_{\text{calc}}\right)/\eta_{\text{calc}}$ where $\eta_{\text{exp}}$ are from Table VI of Ref. 18. $\eta_{\text{calc}}$ represents either the Andrade's equation with the coefficients of Table VII of Ref. 18 or Eq. (9) with $C_d$ = 0.9707, $C_N$ = 0.993346, $\Delta T_{Knu}$ = 118.729 K and $\gamma_{Knu,1\,\text{atm.}} = 0.885317$ . $T_{\text{m.p.}}$ represents the melting temperature for the atmospheric pressure.

Concerning the experimental set-up of 1954, the authors indicated:

> "The viscometer consisted essentially of two cylindrical 3-liter tanks which were connected by a long capillary. [...] A smaller capillary (0.159 cm. diameter) of the same length [i.e. 520 cm] was substituted for the potassium experiments."

This information is used to set the value of $d$ to the radius of the capillary. Concerning the experimental uncertainty, the authors wrote:

> "Coefficients estimated in this manner should have an error of ±2% below 200° [i.e. 473.15 K] and above 200° a graded error from ±2% to ± 10% at the highest temperature."

Fig. 22 shows that the present modeling allows the data to be reproduced in accordance with the estimated uncertainty while Andrade's equation does not agree with the authors' recommendation below 473.15 K. It is further observed that Andrade's equation does not have the right variation while the deviation of the present modeling is quite centered on zero percent. The parameters $d$ and $C_N$ lead to a value of $d_N$ = 9.28 cm which is perfectly compatible with the large volumes of liquid involved in this viscometer.

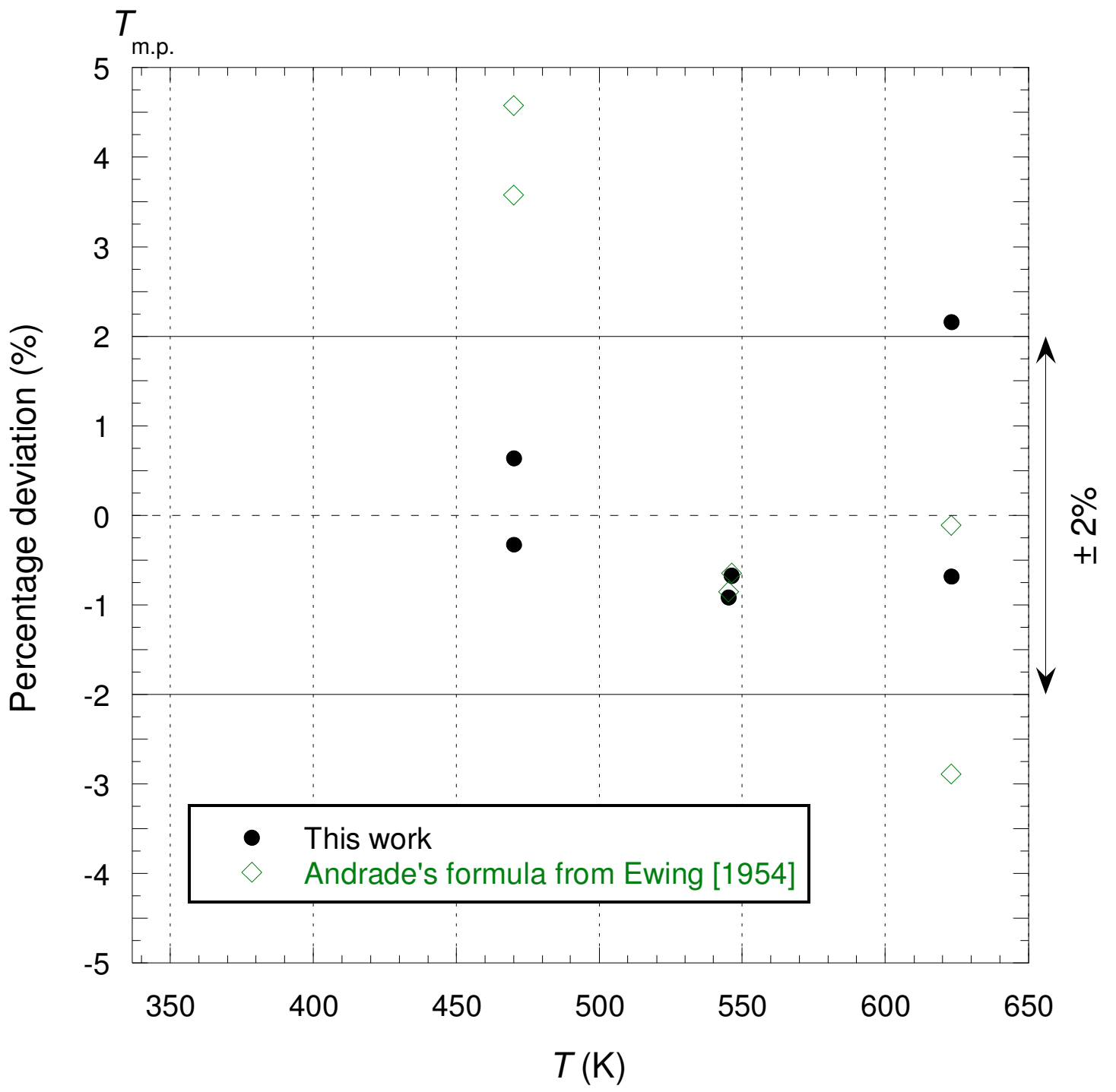

Fig. 22. Deviation plot for the viscosity data of liquid potassium along the atmospheric isobar, i.e. $100\left(\eta_{\exp}-\eta_{\mathrm{calc}}\right)/\eta_{\mathrm{calc}}$ where $\eta_{\exp}$ are from Table V of Ref. 24. $\eta_{\mathrm{calc}}$ represents either the Andrade's formula with the coefficients of Table VI of Ref. 24 or Eq. (9) with $C_d$ = 7.95, $C_N$ = 0.737039, $\Delta T_{Knu}$ = 118.729 K and $\gamma_{Knu,1\,\mathrm{atm.}} = 1.01458$. $T_{\mathrm{m.p.}}$ represents the melting temperature for the atmospheric pressure.

It appeared that in order to analyze the data of the different authors, the density of the released gas must be assumed to be slightly different in each of the experiments as shown in Fig. 23. This reflects the fact that the potassium liquids used in the different experiments are slightly different, i.e. they do not have the same purity. Indeed Chiong wrote:

> "The potassium used was supplied by Schering-Kahlbaum as specially pure, and was said to contain only a slight trace of iron. Probably the trace of iron was reduced after the distillation. Thus, from the point of view of viscosity measurement, where small contaminations produce only a small effect, these metals may be said to be satisfactorily pure."

while Ewing *et al.* wrote in 1954:

> "Within the limits of the experimental method of analysis, the pure metals analyzed to 100.0% purity."

In Ref. 4 it was shown that impurities have an impact on the value of the shear elastic constant *K* but here it is not possible to see this effect given the too limited number of data.
We observe in Fig. 23 that the values of $\tilde{\rho}_{Knu}(T,1\,\mathrm{atm.})$ are ten times higher than in water on the atmospheric isobar but if we determine the "real" value of the density then it is found that $\rho_{Knu}(T,1\,\mathrm{atm.})$ varies between $1.85\times10^{-5}$ g/cm$^3$ and $4.02\times10^{-6}$ g/cm$^3$ considering $\delta = d$, which corresponds to the same order of magnitude as for water.

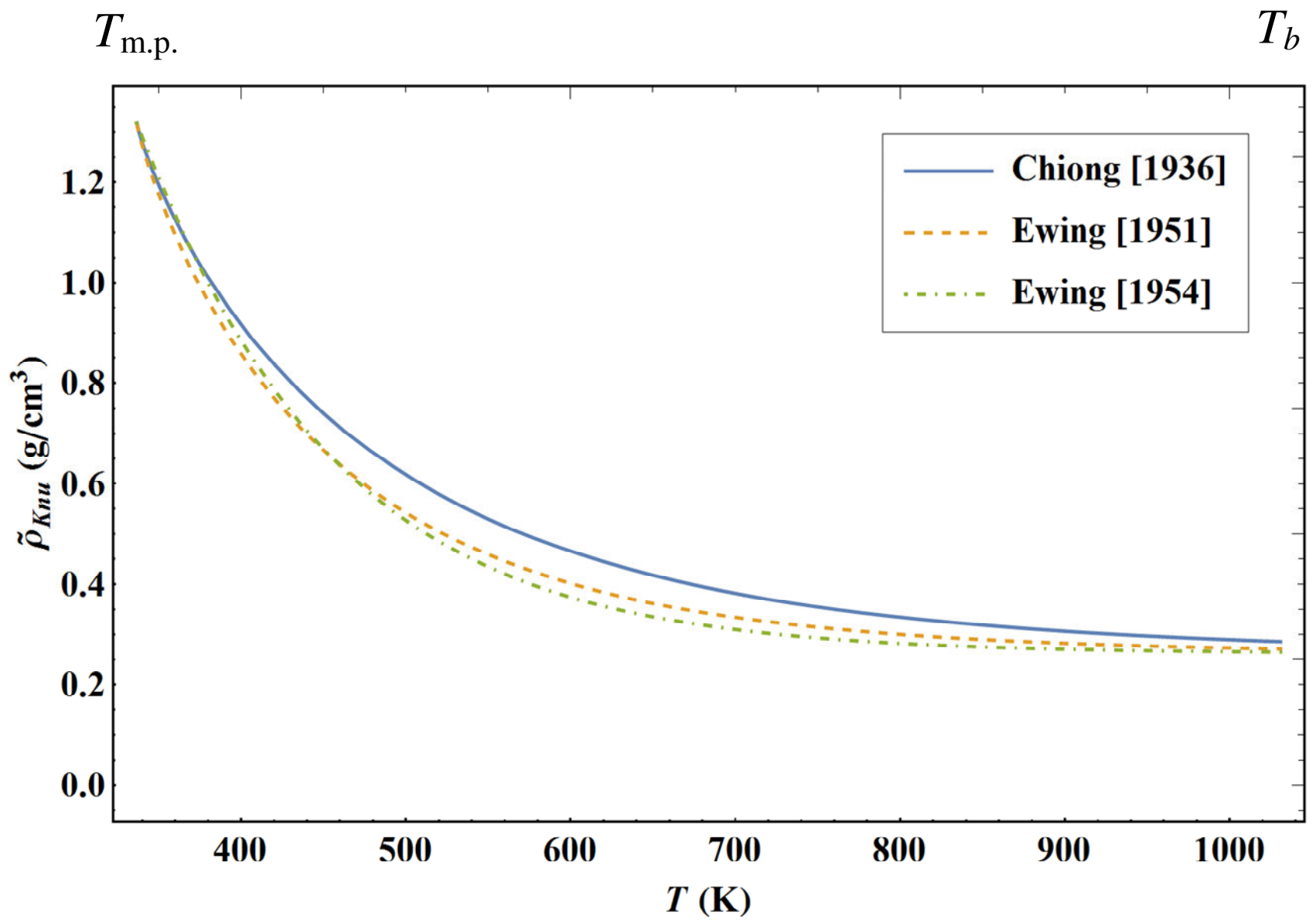

Fig. 23. Representation of Eq. (19) with the parameters used in the present modeling corresponding to the different experiments analyzed (Refs. 18 to 24) for liquid potassium. $T_{\text{m.p.}}$ represents the melting temperature and $T_b$ the boiling temperature for the atmospheric pressure.

Finally, the present modeling allows to calculate the viscosity of potassium in the gaseous phase and along the atmospheric isobar. It is not possible to define a general law of evolution of the functions $K^*(\rho)$ and $f_N(\rho)$, but a partial picture of this evolution can be obtained as we can see on Fig. 24: the evolution of these two functions is unknown between $\rho_{\max} = 0.0076\,\text{g/cm}^3$ and $\rho_{1\text{atm.}}(T_{\text{m.p.}})$ but it is still possible to imagine a function that links the two evolution curves. As for water, it can be seen that the deviation of the reduced elastic shear constant $K^*(\rho)$ from its limiting law is quite large, while $f_N(\rho)$ deviates little from its limiting law as the density increases.

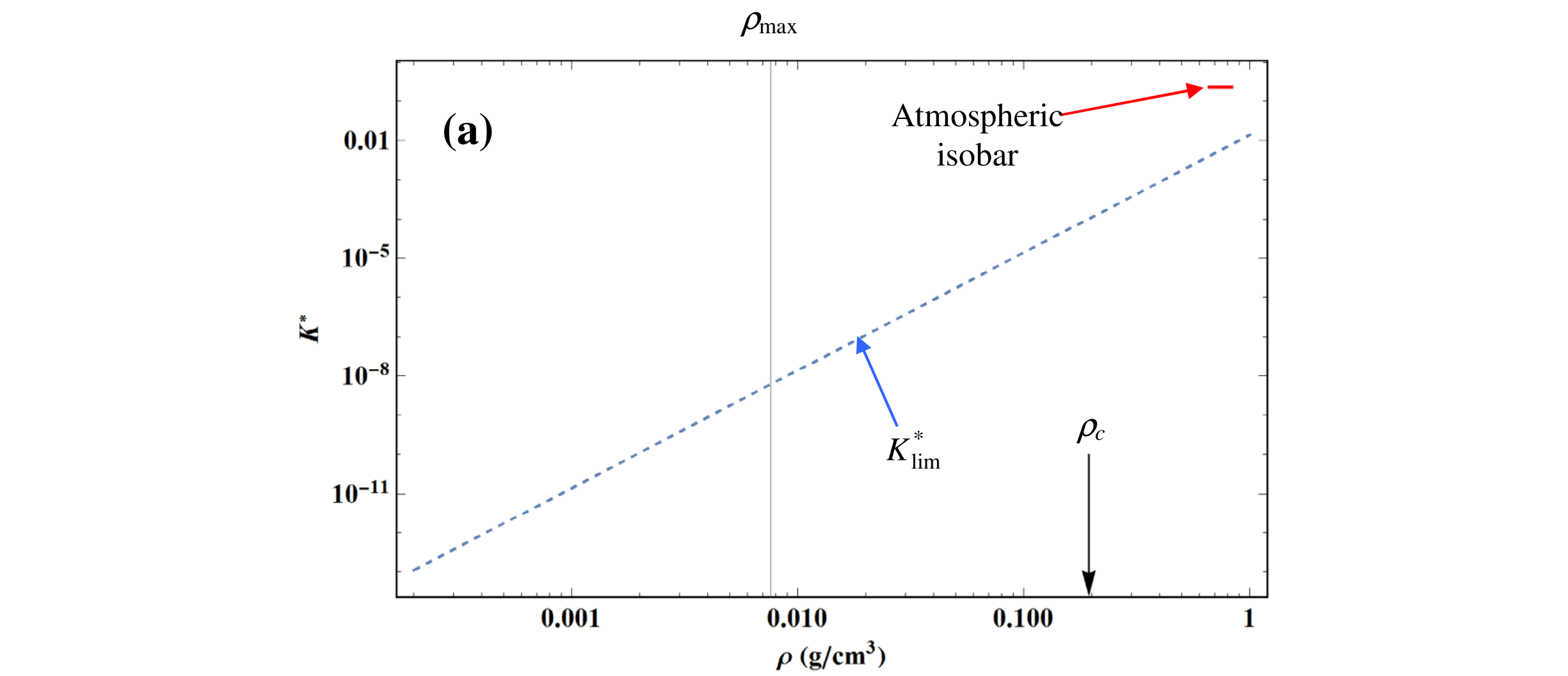

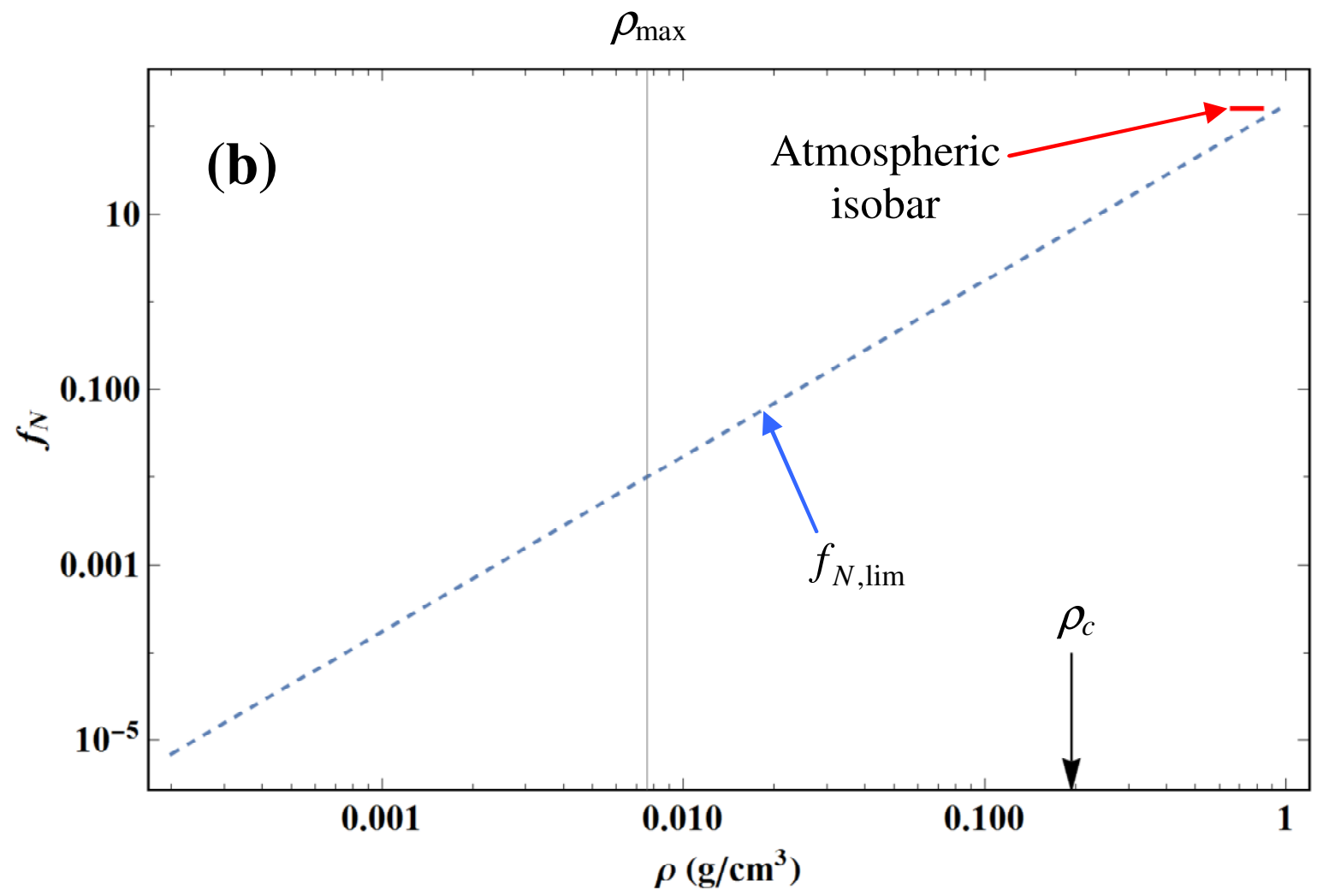

Fig. 24. Logarithmic plot of an overview for fluid potassium of the evolution versus density of **(a)** the reduced shear elastic constant and **(b)** the reduced fluctuative distance. $\rho_{max} = 0.0076\,g/cm^3$ represents the maximum density of Lee *et al.*'s dataset (Ref. 19) in the gaseous phase.

To conclude this analysis of potassium viscosity data, Table 3 regroups precise values, calculated with the different equations of the present modeling, of different properties at the boiling point and Table 4 regroups precise values of different properties at the melting temperature $T_{m.p.}$.

| $T_b$ (K) | $\rho_{\sigma,Liq}$ (g/cm$^3$) | $\rho_{\sigma,Vap}$ (g/cm$^3$) | $\eta_{\sigma,Liq}$ (mPa.s) | $\eta_{\sigma,Vap}$ (mPa.s) | $L_v$ (kJ/mole) |
|---|---|---|---|---|---|
| 1030.793 | 0.662175 | 0.000500 | 0.202665 | 0.016855 | 76.93 |

Table 3. Characteristic parameters of fluid potassium at the boiling point (i.e. for a pressure of 1 atm.). The viscosity value of $\eta_{\sigma,Liq}$ is determined by considering Chiong's experimental parameters (i.e. $C_d$ = 0.9525, $C_N$ = 0.986928, $\Delta T_{Knu}$ = 148.192 K and $\gamma_{Knu,1\,atm.} = 0.867196$). The values of these parameters are defined with an uncertainty of $\pm 1\%$.

| $T_{m.p.}$ (K) | $\rho_{m.p.,Liq}$ (g/cm$^3$) | $\eta_{m.p.,Liq}$ (mPa.s) | $P_\sigma$ (Pa) | $L_v$ (kJ/mole) |
|---|---|---|---|---|
| 336.65 | 0.82948 | 0.539208 | $1.20\times10^{-4}$ | 67.6051 |

Table 4. Characteristic parameters of liquid potassium at the melting temperature. The viscosity value of $\eta_{\sigma,Liq}$ is determined by considering Chiong's experimental parameters (i.e. $C_d$ = 0.9525, $C_N$ = 0.986928, $\Delta T_{Knu}$ = 148.192 K and $\gamma_{Knu,1\,atm.} = 0.867196$). The values of these parameters are defined with an uncertainty of $\pm 1\%$.

To complete this section we will analyze the self-diffusion coefficient data determined by Hsieh *et al.* (Ref. 27) along the atmospheric isobar. These data were obtained by measuring the diffusion of a tracer which is the radioactive isotope $^{42}$K. As indicated by the authors, the dispersion of the data is of the order of 10% so showing a deviation plot does not provide much information.

In Ref. 4, it was shown that the self-diffusion coefficient $D_t$ is expressed by a single term (whatever the phase) corresponding to Eq. (8). The coefficient $D_t$ depends on the cut-off wave-vector modulus $q_c(\rho,T)=f_{q_c}(\rho,T)\,q_{c0}(\rho)$ contrary to viscosity and at this stage of the analysis only the function $f_{q_c}(\rho,T)$ remains to be determined.

It has been shown that in normal water along the atmospheric isobar the function $f_{q_c}(\rho,T)$ remains almost constant around the unit value but the extension then following SVP shows that $f_{q_c}(\rho,T)$ is a monotonically increasing function with increasing temperature. As for viscosity, one cannot deduce here the general law of $f_{q_c}(\rho,T)$ but only a valid expression along the atmospheric isobar. The analysis of the data leads to the following expression:

$$f_{q_c}(T,1\,\text{atm.})=1+\left(\sqrt[3]{n_B}-1\right)\frac{\left(\dfrac{T-T_{\text{m.p.}}}{407.187}\right)^2}{1+\left(\dfrac{T-636.78}{407.187}\right)^2} \tag{20}$$

This function varies in a very similar way to the function $f_{q_c}(\rho,T)$ for water. The writing here of this function in terms of $n_B$ is equivalent to saying that $q_c$ varies in such a way that $n_B$ goes from the value of 2 to the value of 1 when $T$ tends towards infinity (i.e. the size of the unit cell decreases when the temperature increases). Thus along the atmospheric isobar $D_t$ is determined by Eq. (8) where $\rho_{1\text{atm.}}(T)$ is given by Eq. (14), $f_{q_c}(T,1\,\text{atm.})$ by Eq. (20) and the values of $d$, $K^*$ and $f_N$ are given in Table 2.

It was also shown in water that tracers diffusion is accompanied by an increase of the reduced elastic constant $K^*$ and a decrease of the cut-off wave-vector modulus $q_c$ (i.e. an increase in the size of the objects) with respect to the pure product, i.e. tracers generally form clusters. So in order to analyze Hsieh *et al.*'s data correctly, these effects must be taken into account. The analysis of the data shows in fact that to reproduce them correctly one must decrease $q_c$ by a factor $C_{q_c}=0.793522$ and increase $K^*$ by a factor $C_{K^*}=1.21108$ in accordance with what can be expected with the tracer used.

By imposing values for $C_d$ and $C_N$ that correspond to the measuring tubes dimensions used by Hsieh *et al.* (i.e. 1 mm in diameter for 10 cm in length), Fig. 25 shows that the present modeling can finally reproduce these data in a way that is quasi-equivalent to Eq. (4) of Ref. 27.

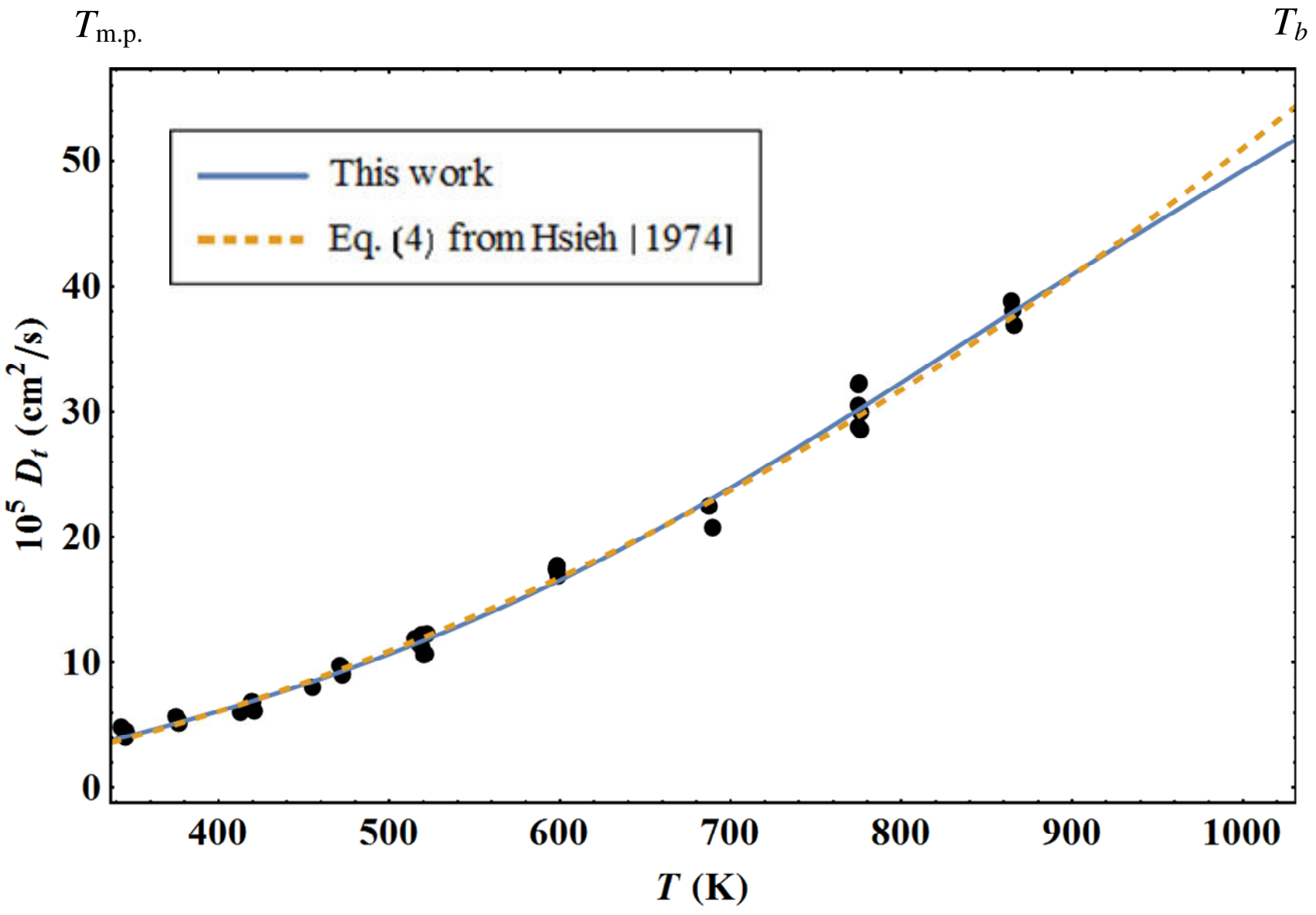

Fig. 25. Self-diffusion coefficient of potassium versus temperature along the atmospheric isobar. The black points represents the data from Hsieh *et al.* (Ref. 27). The orange dashed curve represents Eq. (4) from Ref. 27 and the blue curve corresponds to the present modeling with $C_d = 5$, $C_N = 0.0540234$, $C_{q_c} = 0.793522$ and $C_{K^*} = 1.21108$. $T_{\text{m.p.}}$ represents the melting temperature and $T_b$ the boiling temperature for the atmospheric pressure.

More recently, Novikov *et al.* (Ref. 28) determined values for the self-diffusion coefficient of liquid potassium from inelastic neutron scattering experiment. Fig. 26 shows not only that the data of Novikov *et al.* have much lower values than those of Hsieh *et al.* but also that the variation with temperature seems very different *a priori*. The values of Novikov *et al.* are consistent, for example, with those obtained by Bove *et al.* in water (Ref. 29), which are lower than the NMR data of Krynicki *et al.* (Ref. 30). As mentioned in Ref. 4 about inelastic neutron scattering, the $D_t$ values determined by this method are dependent on the modeling done to extract them from the spectra and of the energy resolution of the spectrometer used.

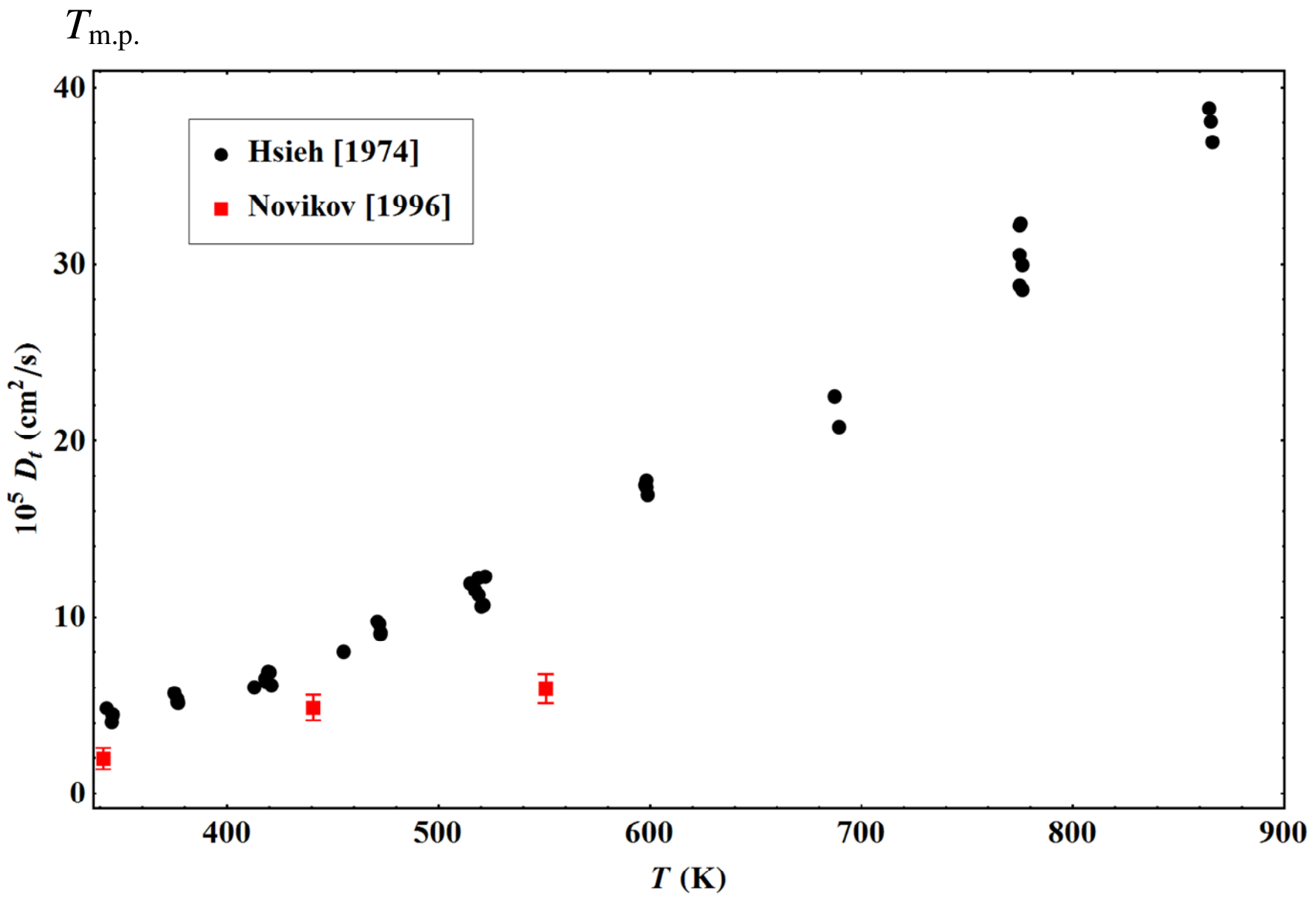

Fig. 26. Experimental self-diffusion coefficients of potassium from Hsieh *et al.* (Ref. 27) and Novikov *et al.* (Ref. 28) along the atmospheric isobar. $T_{\text{m.p.}}$ represents the melting temperature at atmospheric pressure.

Having now fixed all the parameters of the present modeling with the data of Hsieh *et al.*, we can try to analyze temperature variations of Novikov *et al.*'s data. Taking into account the specific characteristics of this experiment, it must be considered here that $C_{q_c} = 1$ and $C_{K^*} = 1$. Therefore only the parameters $C_d$ and $C_N$ can be considered as "free" parameters of the model. Fig. 27 shows that the present modeling allows to reproduces the data of Novikov *et al.* by crossing the error bars so that the values of $d$ and $d_N$ are compatible with the sample size. Therefore, it is again observed that the difference in temperature variation of various datasets is simply due to different geometrical characteristics of the experimental devices.

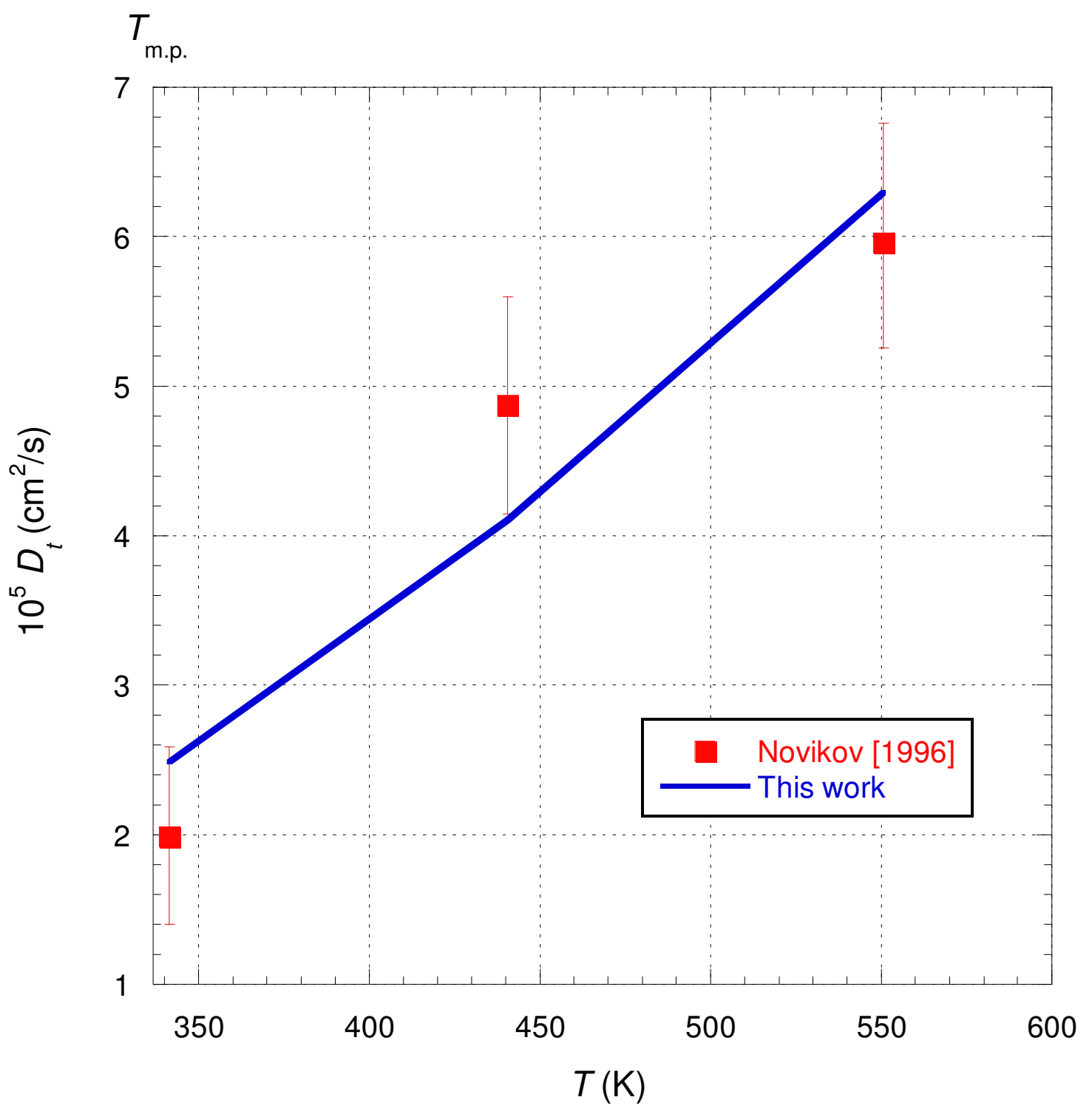

Fig. 27. Self-diffusion coefficient of potassium versus temperature along the atmospheric isobar. The red points represents the data from Novikov *et al.* (Ref. 28). The blue curve corresponds to the present modeling with $C_d$ = 0.46, $C_N$ = 0.0116. $T_{\text{m.p.}}$ represents the melting temperature at atmospheric pressure.

## 4 Application to thallium fluid

The properties for thallium are much more fragmentary than for potassium, therefore it is not possible to determine all the parameters of the theory. To determine the fundamental scaling of the model, it is first necessary to know the critical properties of thallium. According to Ref. 31 the value of the molar mass $M$ is between 204.382 g/mole and 204.385 g/mole. We have retained here the value in Table 5 which is the one most frequently found in the literature. The value of the critical temperature and density are taken from Ref. 32. These two values are given with an uncertainty of ±10% so the scaling parameters will have at best this uncertainty. It has been shown that at atmospheric pressure and temperatures close to $T_{\text{m.p.}}$ = 577 K, the solid thallium crystal structure is body-centred cubic (e.g. Ref. 33) which leads us to the $n_B$ value of Table 5.

| $M$ (g/mole) | $n_B$ | $T_c$ (K) | $P_c$ (MPa) | $\rho_c$ (g/cm$^3$) | $v_c$ (Å$^3$) | $z_c$ |
|---|---|---|---|---|---|---|
| 204.383 | 2 | 2260 | 1.659 | 3.5869 | 94.620 | 0.005 |

Table 5. Characteristic parameters of fluid thallium. The value of $M$ is from Ref. 31 and the critical parameters are from Ref. 32 except for $P_c$. Indeed, the value of $P_c$ is determined in the analysis below. All critical values are defined to within 10%. Below the melting line, solid thallium is b.c.c. with 2 atoms per unit cell (Ref. 7, Ref. 33) hence the value of $n_B$.

The following values are then deduced for the scalings $K_0$ = 19.5286 GPa and $q_{c0,\text{crit}} = 6.78915\times10^{7}\ \text{cm}^{-1}$.

As mentioned in the caption of Table 5, unlike potassium the value of the critical pressure of thallium is not given in the literature. However, there are different data sets for Saturated Vapor Pressure (SVP) versus temperature and knowing the SVP pressure $P_\sigma(T)$ is useful in particular to extrapolate a critical pressure value. For our analysis, we considered the four data sets from Refs. 34 to 37. These different data sets have a large dispersion between them (i.e. greater than the uncertainty of each data set) as mentioned for example by Aldred *et al.* (Ref. 34):

> "The present results appear to be about 40% higher than the earlier ones [i.e. those of Coleman *et al.*, Ref. 35]."

Therefore a deviation plot is not significant. Fig. 28 shows that an equation can be determined that represents an average of these data sets and can be extrapolated to the limits at the melting temperature $T_{\text{m.p.}}$ and at the critical temperature $T_c$. The equation thus determined is written in the following form:

$$\log_{10}\left(\frac{P_\sigma}{P_0}\right) = 5.30825 - \frac{9532.31}{T} + \frac{798956}{T^2} - \frac{3.08727\times10^{8}}{T^3} \qquad (21)$$

where $P_0 = 0.1\,\text{MPa}$ and $T$ is in Kelvin.

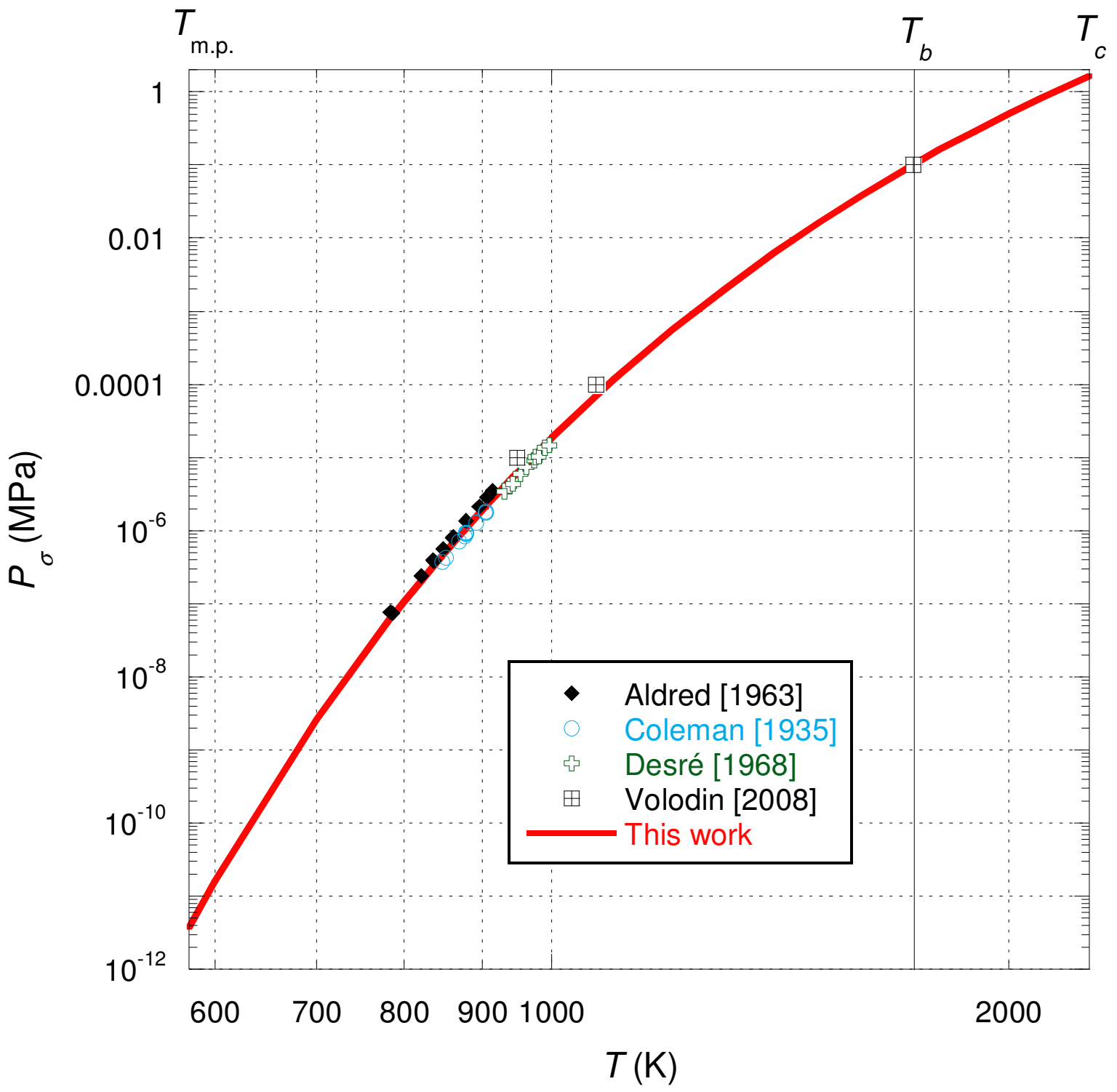

Fig. 28. Saturated vapor pressure of thallium versus temperature. The data points correspond to Refs. 34 to 37. The red curve corresponds to Eq. (21) of the present modeling. $T_{m.p.}$ = 577 K represents the melting point temperature and $T_b$ = 1730.05 K the boiling temperature for the atmospheric pressure.

From Eq. (21), we determine that $P_\sigma(600\,\text{K}) = 1.625 \times 10^{-5}$ Pa ; however it is a value very close to the one for example found in the periodic table of the Royal Society of Chemistry (Ref. 38). This shows that the extrapolation to $T_{m.p.}$ is reasonable.

Concerning the boiling temperature (i.e. at atmospheric pressure), values between 1730 K and 1746 K can be found in the literature. Eq. (21) gives the value of $T_b$ = 1730.05 K which is practically identical to that determined by Leitgebel (Ref. 39) and more recently by Volodin *et al.* (Ref. 37) to within 0.072%.

Given that Eq. (21) is consistent between $T_{m.p.}$ and $T_b$, one can consider its extrapolation to $T_c$, which leads to a value $P_c$ = 1.659 MPa. Given the uncertainty on $T_c$, it can be assumed that this value of $P_c$ is determined at best to within 10%. Table 5 shows that with the critical values thus determined, the critical compressibility factor $z_c$ is excessively small. We can note that the value of $z_c$ is very different from the one of potassium, but these two metal compounds do not belong to the same family of chemical elements in the Mendeleev table.

Using the Clausius-Clapeyron formula (instead of the Clapeyron formula because the gas density along SVP is not known), one can determine approximate values of the latent heat of vaporization $L_v(T)$ and its evolution as long as one remains sufficiently far from the critical point. Coleman *et al.* (Ref. 35) determined by their analysis a value of $L_v$(876.9 K) = 171 kJ/mole and a value of 170,657 kJ/mole is determined with the present modeling, which is therefore perfectly consistent.

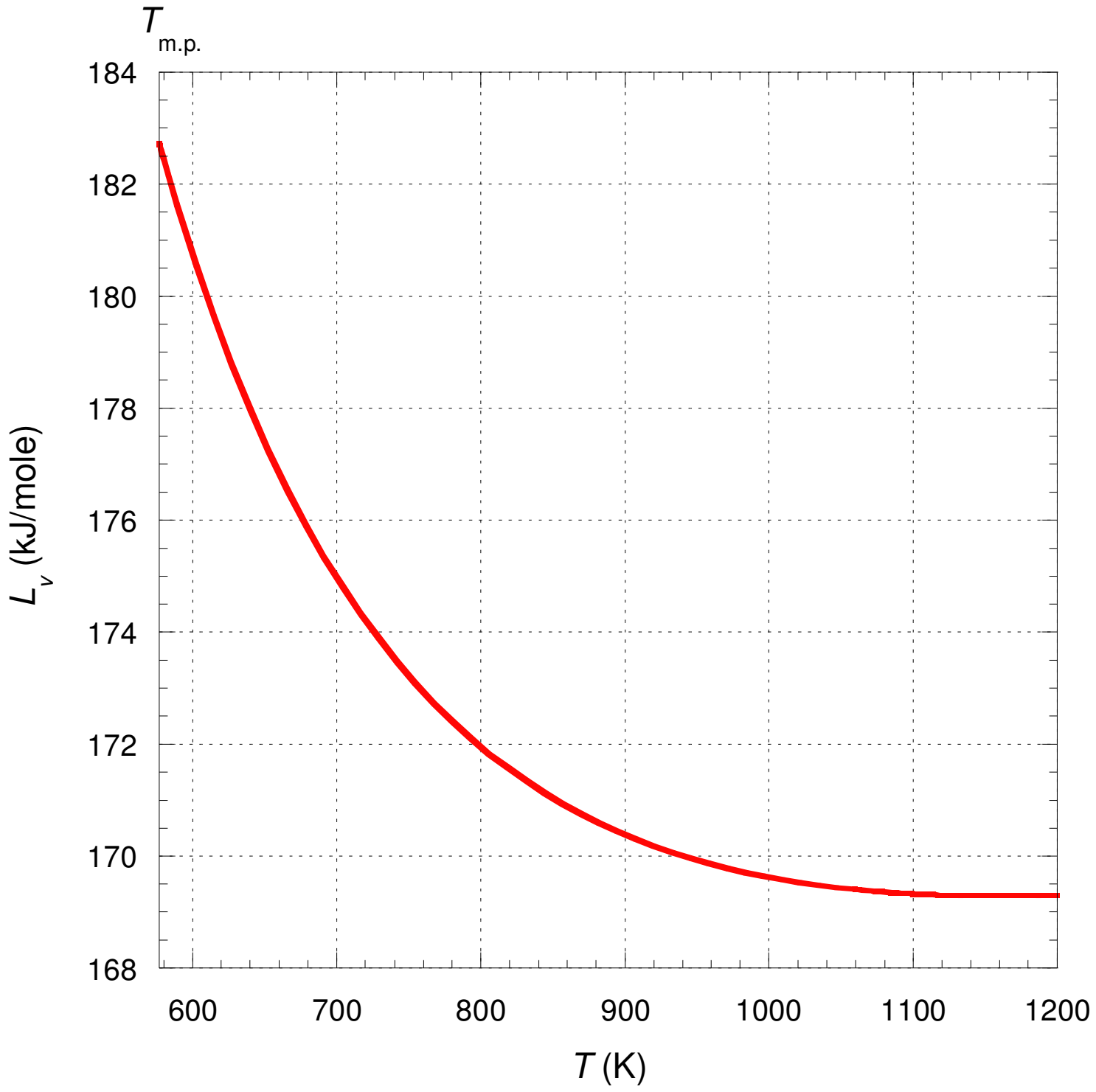

Fig. 29. Thallium latent heat of vaporization versus temperature deduced from the Clausius-Clapeyron formula. $T_{\mathrm{m.p.}}$ represents the melting point temperature at atmospheric pressure.

Now to put the value of $K_0$ in perspective with the latent heat of vaporization, we need to know the liquid density on the coexistence curve. This density is unknown. The only known density that can be used is the density along the atmospheric isobar. To do this we will use Eq. (1) determined by Assael *et al.* (Ref. 40) which is considered to be valid between $T_{\mathrm{m.p.}}$ and 1200 K. Fig. 30 shows that the value of $K_0$ is higher than the value $L_v(T_{\mathrm{m.p.}})$ as is the case for potassium and the other fluids considered in Appendix B of Ref. 4.

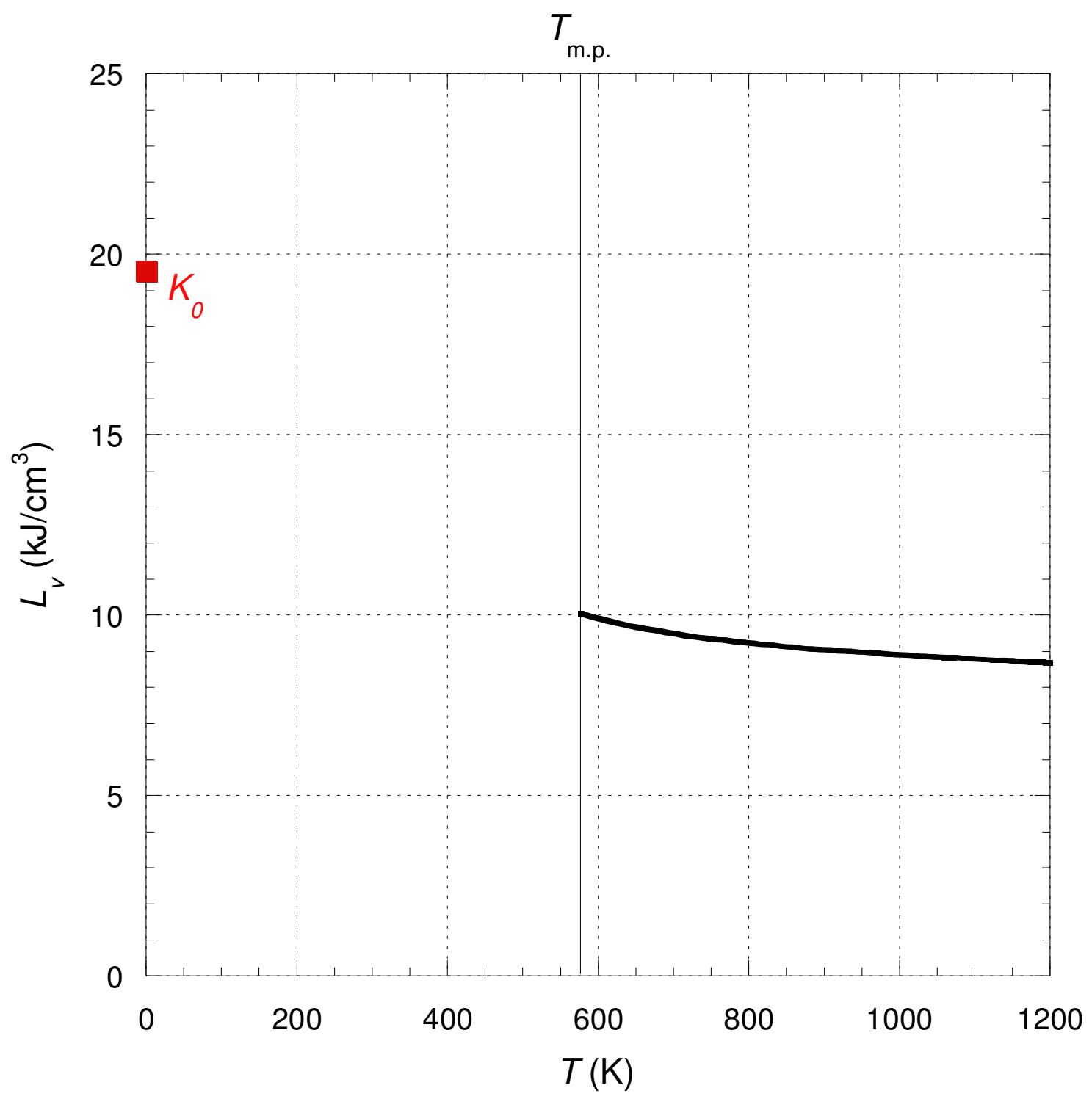

Fig. 30. Thallium latent heat of vaporization versus temperature when combining the Clausius-Clapeyron formula and the density along the atmospheric isobar from Ref. 40. $T_{\mathrm{m.p.}}$ represents the melting point temperature at atmospheric pressure.

### 4.1. Viscosity of thallium in the dilute-gas limit

To determine the other parameters of the model, viscosity data in the gas phase must be analyzed. There are no such data for thallium in the literature; therefore the dilute-gas limit of the liquid-like term $\eta_l$ can only be determined here. As recalled in section 3.4 for potassium, this limit has the following value $\eta_{l0} = \pi \hbar / \mathcal{V}_{\mathrm{mol}}$ with $\mathcal{V}_{\mathrm{mol}} = \sqrt{\mathcal{V}_{\mathrm{c}} \mathcal{V}_0}$ where $\mathcal{V}_{\mathrm{c}}$ and $\mathcal{V}_0$ are the critical molecular volume and the molecular volume at zero temperature, respectively. Here $\mathcal{V}_0$ can only be determined by extrapolating to zero temperature Eq. (1) of Assael *et al.* (Ref. 40) which describes the atmospheric isobar, i.e. $\mathcal{V}_0 = 28.46\,\mathrm{\mathring{A}}^3$. As a result, the low limit value of the liquid-like term is: $\eta_{l0} = \lim_{\rho \to 0} \eta_l = 6.384 \times 10^{-3}$ mPa.s.

Now, assuming as in the case of potassium that within the dilute-gas limit the shear elastic constant varies proportionally to the cube of the density such that $K^*_{\mathrm{lim}}(\rho) = c_{K0} \left( \dfrac{\rho}{\rho_c} \right)^3$ and the function $f_N$ varies proportionally to the square of the density such that $f_{N,\mathrm{lim}}(\rho) = c_{N0} \left( \dfrac{\rho}{\rho_c} \right)^2$, then one of the two constants $c_{K0}$ and $c_{N0}$ can be determined from the resulting expression of the dilute-gas limit of the liquid-like term, i.e. $\eta_{l0} = 2\pi \dfrac{\sqrt{c_{K0}}}{c_{N0}} \dfrac{\sqrt{K_0 \rho_c}}{q_{c0,\mathrm{crit}}}$. Contrary to the case of potassium, the value of $c_{N0}$ cannot be fixed to that of water because thallium is liquid at temperatures that almost correspond to supercritical water. Experimental devices for determining liquid thallium viscosity have geometrical dimensions comparable to those of liquid water and liquid potassium. In other words, for liquid thallium temperatures, $d_N$ values comparable to those obtained for liquid water and liquid potassium must be found. This will be verified in the following section, but for this purpose it must be assumed here that $c_{N0} = 12.5695$. It is then deduced that $c_{K0} = 1.0733\times10^{-4}$.

Finally, it is possible to calculate the liquid-like term $\eta_l(T,\rho)$ in the gaseous phase such that:

$$\eta_l(\rho \le \rho_{\sigma,\mathrm{Vap}}, T) = \frac{d}{H_N(v)}\sqrt{\rho\, K^*_{\lim}(\rho) K_0} \tag{22}$$

with $N-1 = f_{N,\lim}(\rho)\, d\, \dfrac{q_{c0,\mathrm{crit}}}{2\pi}$. But here it is not possible to determine the gas-like term $\eta_{Knu}$ precisely. Assuming, however, that Eq. (15) can be applied to thallium, an order of magnitude can be deduced for the viscosity in the vicinity of SVP. One must just take into account that the amount of gas released in thallium is greater than in potassium. Indeed, we will see in the next section that along the atmospheric isobar we find approximately a factor of 7.95 on the density $\tilde{\rho}_{Knu}(T, 1\,\mathrm{atm.})$ with respect to potassium. Taking this factor into account for the density calculation $\tilde{\rho}_{Knu,0}(\rho)$ and by scaling the value of viscosity with that given by the kinetic theory of gases at 1600 K, that is to say 0.0717 mPa.s when using a molecular diameter of 380 pm, one determines that for thallium, it is necessary to take $T_{Knu} = 927\,\mathrm{K}$. Fig. 31 then shows the order of magnitude of thallium viscosity and its expected evolution in the vicinity of SVP in the temperature range from 900 K to 1600 K.

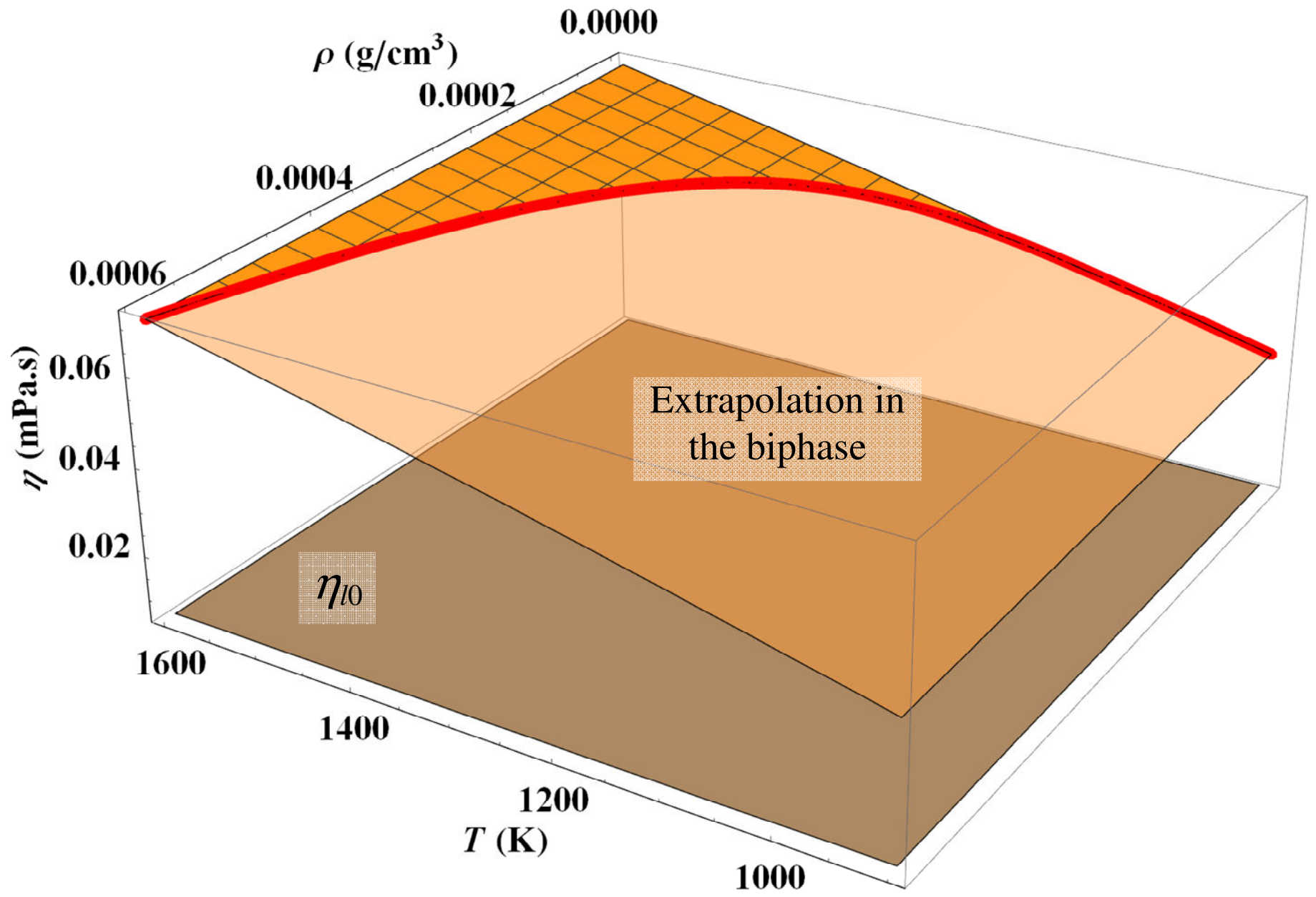

Fig. 31. Viscosity as function of density and temperature in the gaseous phase of thallium. The thick red curve represents the gaseous states along SVP from 900 K to 1600 K. The horizontal plane represents the dilute-gas limit value of the liquid-like term.

### 4.2. Viscosity and self-diffusion coefficient in the liquid phase of thallium

The different viscometry data sets in the thallium liquid phase have been compiled by Assael *et al.* (Ref. 40). All these sets correspond to data along the atmospheric isobar. Fig. 32 shows the three relevant data sets. Indeed, we have excluded from the list analyzed by Assael *et al.*, the 3 data points of Walsdorfer *et al.* (Ref. 41) corresponding to pure thallium in the studied mixtures because these do not represent experimental measurement points but only an interpolation of Crawley's data (Ref. 42).

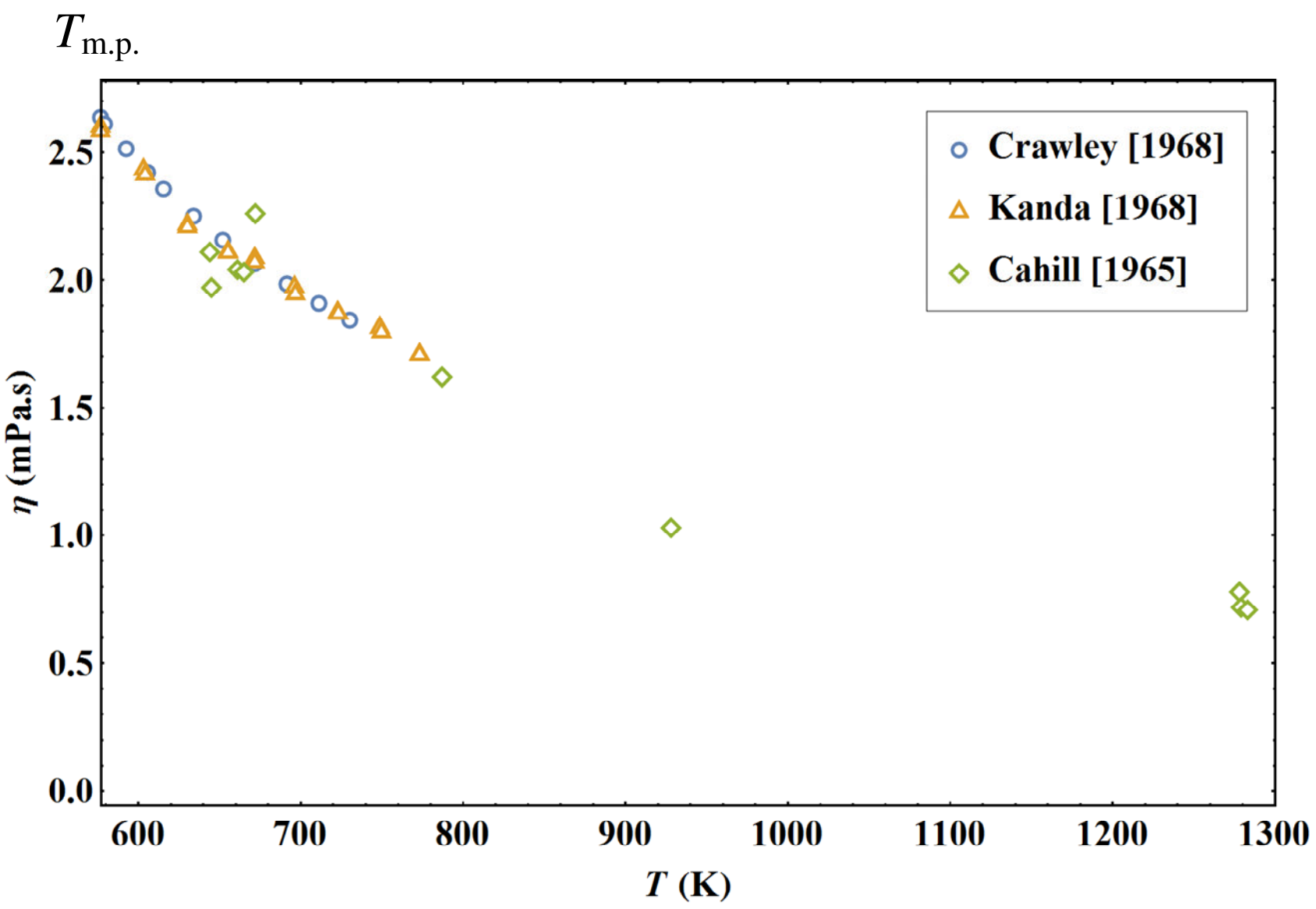

Fig. 32. Thallium viscosity data versus temperature along the atmospheric isobar from Crawley (Ref. 42), Kanda *et al.* (Ref. 43) and Cahill *et al.* (Ref. 44). $T_{\text{m.p.}}$ represents the melting point temperature at atmospheric pressure.

As in the case of potassium, the knowledge of the data along the atmospheric isobar does not allow to deduce a general law of $K^*(\rho) = K/K_0$ and $f_N(\rho,T)$ but only a particular approximation for this isobar. Consequently we will admit by simplification that $K^*$ and $f_N$ can be considered as constant along the atmospheric isobar for liquid thallium. Therefore we consider that the density of the released gas $\tilde{\rho}_{Knu}(T, 1\,\text{atm.})$ can still be described by Eq. (19). The analysis of the viscosity data in thallium leads to set the values given in Table 6. The parameters $\Delta T_{Knu}$ and $\gamma_{Knu,1\,\text{atm.}}$ have not been set to a precise value because we will see that this value depends on the analyzed data.

| $d$ (cm) | $K^*$ | $f_N$ | $\tilde{\rho}_{\text{m.p.}}$ (g/cm$^3$) | $\tilde{\rho}_\infty$ (g/cm$^3$) |
|---|---|---|---|---|
| 0.01 | 0.444 | 114.112 | 9.41 | 2.3 |

Table 6. Characteristic parameters to describe the viscosity of thallium along the atmospheric isobar.

The three experiments that we will analyze below have been performed with an oscillating-cup viscometer. The 3 cups were cylindrical graphite crucibles with comparable diameters of 3 cm to 3.5 cm. The most accurate data are those of Crawley (Ref. 42) and therefore deserve special attention. Indeed according to the author:

"The experimental results are considered to be accurate within ±0.5%."

Fig. 33 shows that the present modeling makes it possible to represent all the data with a deviation of ±0.3% which is well below the experimental uncertainty. On the other hand, it can be seen that the reference equation proposed by Assael *et al.* is not in agreement with the experimental uncertainty and is just at the limit at temperatures above 650 K. Moreover, at low temperatures Eq. (2) of Assael *et al.* does not have a variation with temperature in agreement with the experimental data.
Note that the present modeling requires that $C_d = 1$ and $C_N = 1.00778$ which leads to $d_N = 1.15$ cm. This value of $d_N$ is slightly smaller than the radius of the crucible. In other words these two coefficients are consistent with some geometrical characteristics of the experimental devices but also these values are consistent with what was found for the set-up with oscillating sphere in the case of water (see Fig. 19a) and potassium (see Fig. 20).

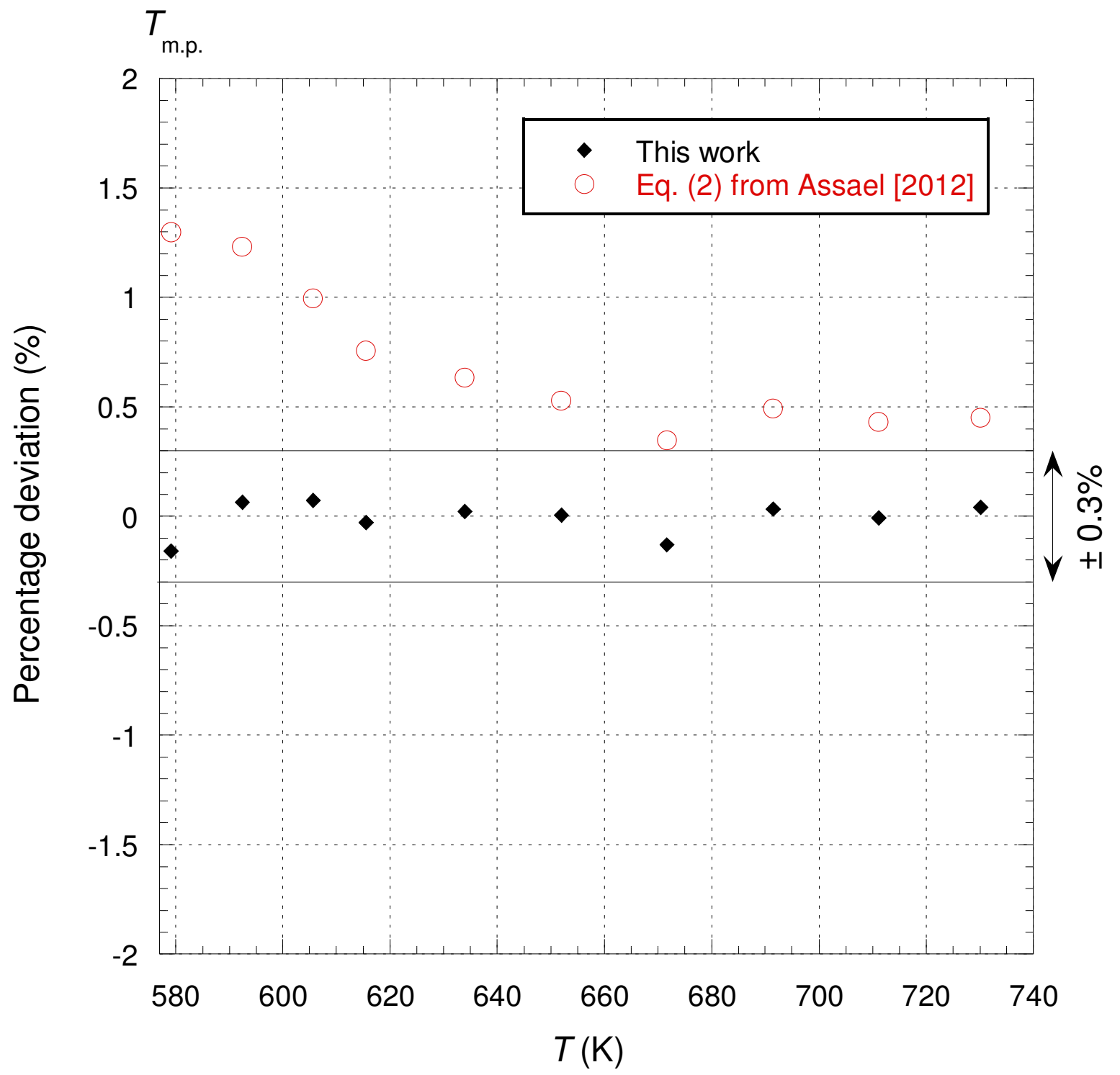

Fig. 33. Deviation plot for the viscosity data of liquid thallium along the atmospheric isobar, i.e. $100(\eta_{\mathrm{exp}} - \eta_{\mathrm{calc}})/\eta_{\mathrm{calc}}$ where $\eta_{\mathrm{exp}}$ are from Table 1 of Ref. 42. $\eta_{\mathrm{calc}}$ represents either Eq. (2) of Ref. 40 or the present modeling with $C_N = 1.00778$, $\Delta T_{Knu} = 413.609$ K and $\gamma_{Knu,1\,\mathrm{atm.}} = 1.194$ . $T_{\mathrm{m.p.}}$ represents the melting point temperature for the atmospheric pressure.

The experimental design of Kanda *et al.* (Ref. 43) is very similar to that of Crawley. However, the authors wrote that:

> "An estimation of the error inherent in the viscometer was made from the limits of error of the above parameters. This amounted to approximately 1% in the viscosity."

Fig. 34 shows that the present modeling and Eq. (2) of Assael *et al.* are quite close with respect to the uncertainty obtained with the experimental data. In both cases it does not seem possible to represent the experimental data within the experimental uncertainty announced by the authors. We note that it is the difference between the experimental data for a given temperature which is of the order of 1% according to what is written by the authors:

> "The results indicate a reproducibility of viscosity values of the order of 1%, which concurs with the error analysis mentioned in Section 2 [i.e. see the previous citation]."

Kanda *et al.* suggest that the viscosity variation can be represented in such a way that:

> "The viscosity-temperature behavior of all the samples studied appeared to follow the Andrade equation $\eta = A\ \exp(E/RT)$. This was inferred because the plots of log $\eta$ *versus* $1/T$(K) were observed to be linear within the limits of sensitivity of the viscometer."

Now this representation is exactly that of Eq. (2) of Assael *et al.* except that the values of the parameters $A$ and $E$ were determined by Assael *et al.* to have a single representation for all the data sets. But if we determine the values of the parameters $A$ and $E$ specific only to the data of Kanda *et al.* the green square points of Fig. 34 are obtained. It is then observed that the deviation associated with this representation is no better than ±2%. This suggests that the present modeling is the one that best accounts for the data.

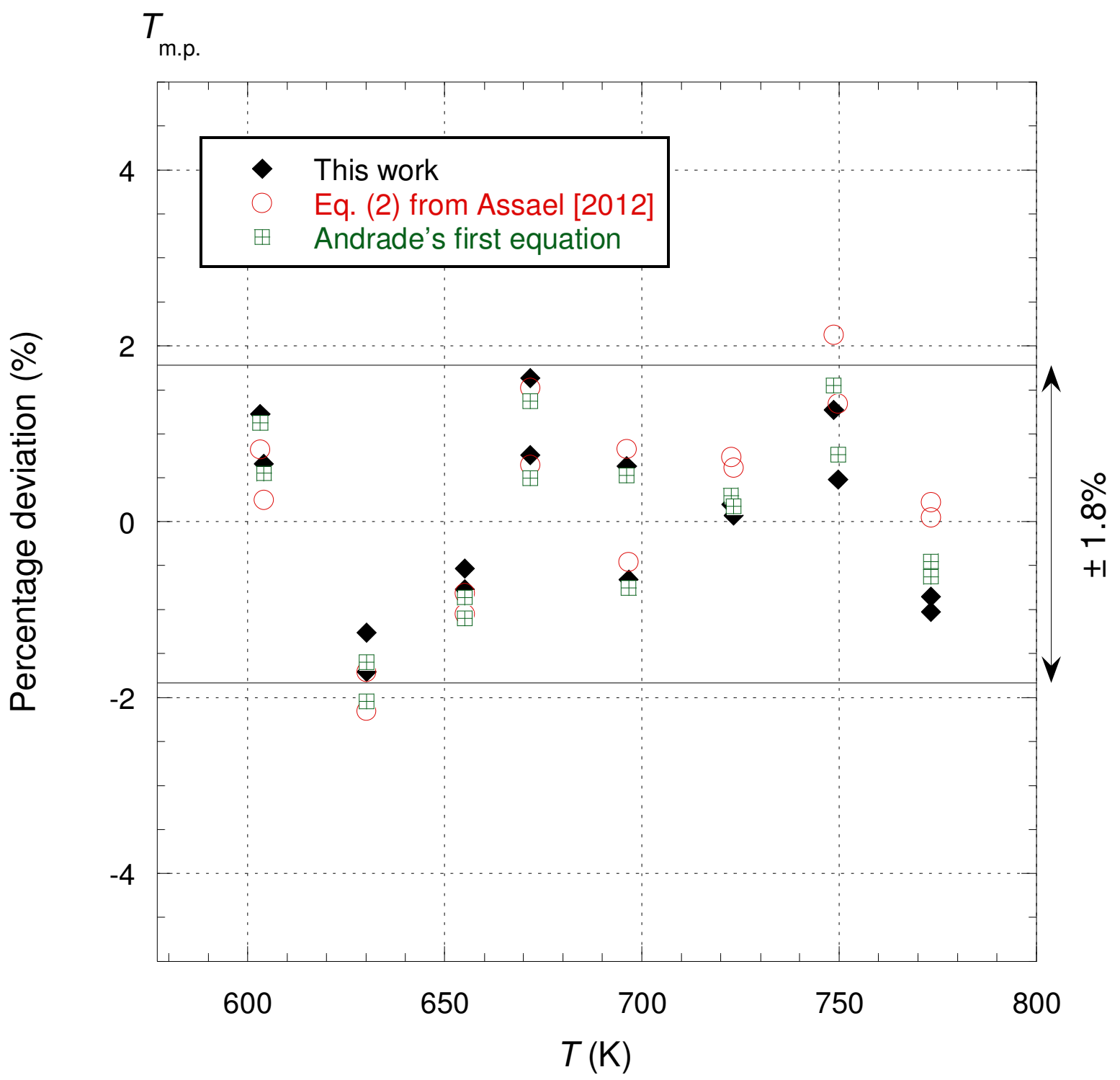

Fig. 34. Deviation plot for the viscosity data of liquid thallium along the atmospheric isobar, i.e. $100(\eta_{\text{exp}} - \eta_{\text{calc}})/\eta_{\text{calc}}$ where $\eta_{\text{exp}}$ are from Table 1 of Ref. 43. $\eta_{\text{calc}}$ represents either Andrade's first equation $\eta_{\text{calc}}(\text{mPa.s}) = 0.520412\exp(923.699/T)$ or Eq. (2) of Ref. 40 or even the present modeling with $C_N$ = 1.04748, $\Delta T_{Knu}$ = 460.878 K and $\gamma_{Knu,1\,\text{atm.}} = 1.169$. $T_{\text{m.p.}}$ represents the melting point temperature for the atmospheric pressure.

We end the analysis of the viscosity data with the data from Cahill *et al.* (Ref. 44). The crucible dimensions are slightly larger than in the previous experiments, but above all the thallium purity used here is less good, i.e. 99.99% instead of 99.999%. More marked differences are therefore expected with the two previous experiments. The data is also less accurate. Indeed the authors wrote:

> “Uncertainties in corrections for thermal expansion, inductive electrical effects, and ambient gas viscosity limit the accuracy of this determination to an error probability of ±0.05 cp. [i.e ~2.36%] at the melting point and ±0.1 cp. [i.e. ~14%] at the boiling point.”

Although less precise, however, these data are interesting since they cover the widest temperature range. Fig. 35 shows that the present modeling allows the data to be reproduced within a deviation of ±11% as well as Eq. (3) of Cahill *et al.* with the exception of one point. On the other hand, it is observed that the reference equation of Assael *et al.* does not correspond to the evolution law of the data beyond 800 K. It is probably for this reason that Assael *et al.* limited themselves to representing only the data below 800 K.

Given the large dispersion of the data for a given temperature, it does not seem possible to do better than a deviation of ±11%. This deviation is however consistent with the probable uncertainty inferred by the authors.

It is noted that the $C_N$ coefficient is greater than for the two previous experiments, which is consistent with the fact that the crucible has a larger diameter and height here. The coefficients $\Delta T_{Knu}$ and $\gamma_{Knu,1\,\mathrm{atm.}}$ are quite different from the two previous experiments which clearly indicates that the liquid thallium is of different composition.

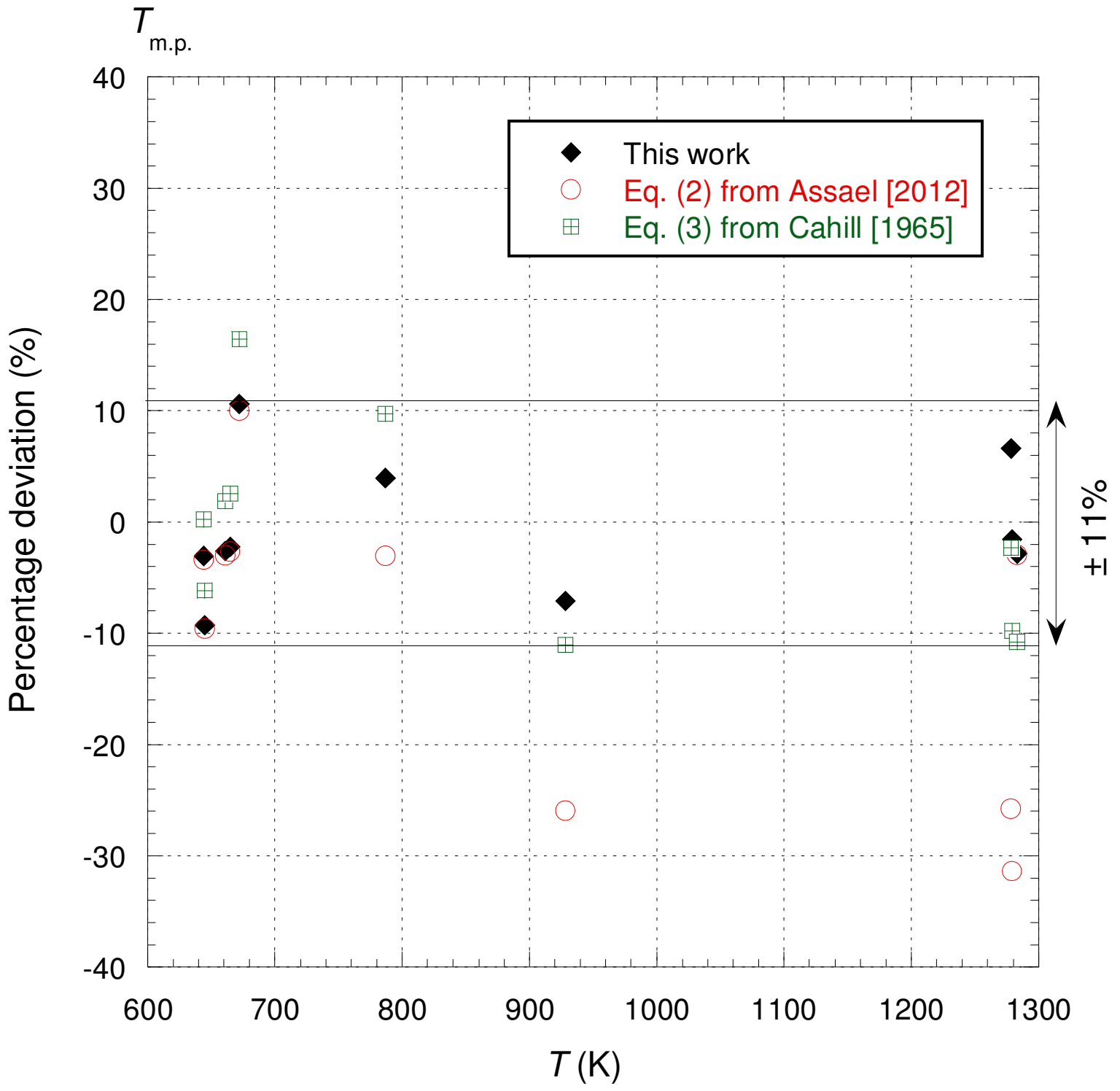

Fig. 35. Deviation plot for the viscosity data of liquid thallium along the atmospheric isobar, i.e. $100\left(\eta_{\mathrm{exp}}-\eta_{\mathrm{calc}}\right)/\eta_{\mathrm{calc}}$ where $\eta_{\mathrm{exp}}$ are from Table I of Ref. 44. $\eta_{\mathrm{calc}}$ represents either Eq. (3) of Ref. 44 or Eq. (2) of Ref. 40 or even the present modeling with $C_N$ = 1.08537, $\Delta T_{Knu}$ = 294.915 K and $\gamma_{Knu,1\,\mathrm{atm.}} = 1.717$ . $T_{\mathrm{m.p.}}$ represents the melting point temperature for the atmospheric pressure.

As for liquid potassium, it appears that in order to analyze the data of the different authors, the density of the released gas must be assumed to be slightly different in each of the experiments as shown in Fig. 36. It should be remembered that the density of this gas does not depend on the experimental device but only on the temperature and the density. However the variations according to these parameters depend on the purity of the studied medium. It is this purity difference which is apparent in the different variations of Fig. 36.
Fig. 36 shows also that the values of $\tilde{\rho}_{Knu}(T, 1\,\text{atm.})$ are seven times higher than in potassium on the atmospheric isobar but if we determine the "real" value of the density then we find that $\rho_{Knu}(T, 1\,\text{atm.})$ varies between 8.7×10$^{-5}$ g/cm$^3$ and 2.3×10$^{-5}$ g/cm$^3$ considering $\delta$=$d$, which corresponds to higher densities than in water and potassium, but the order of magnitude is still quite similar.

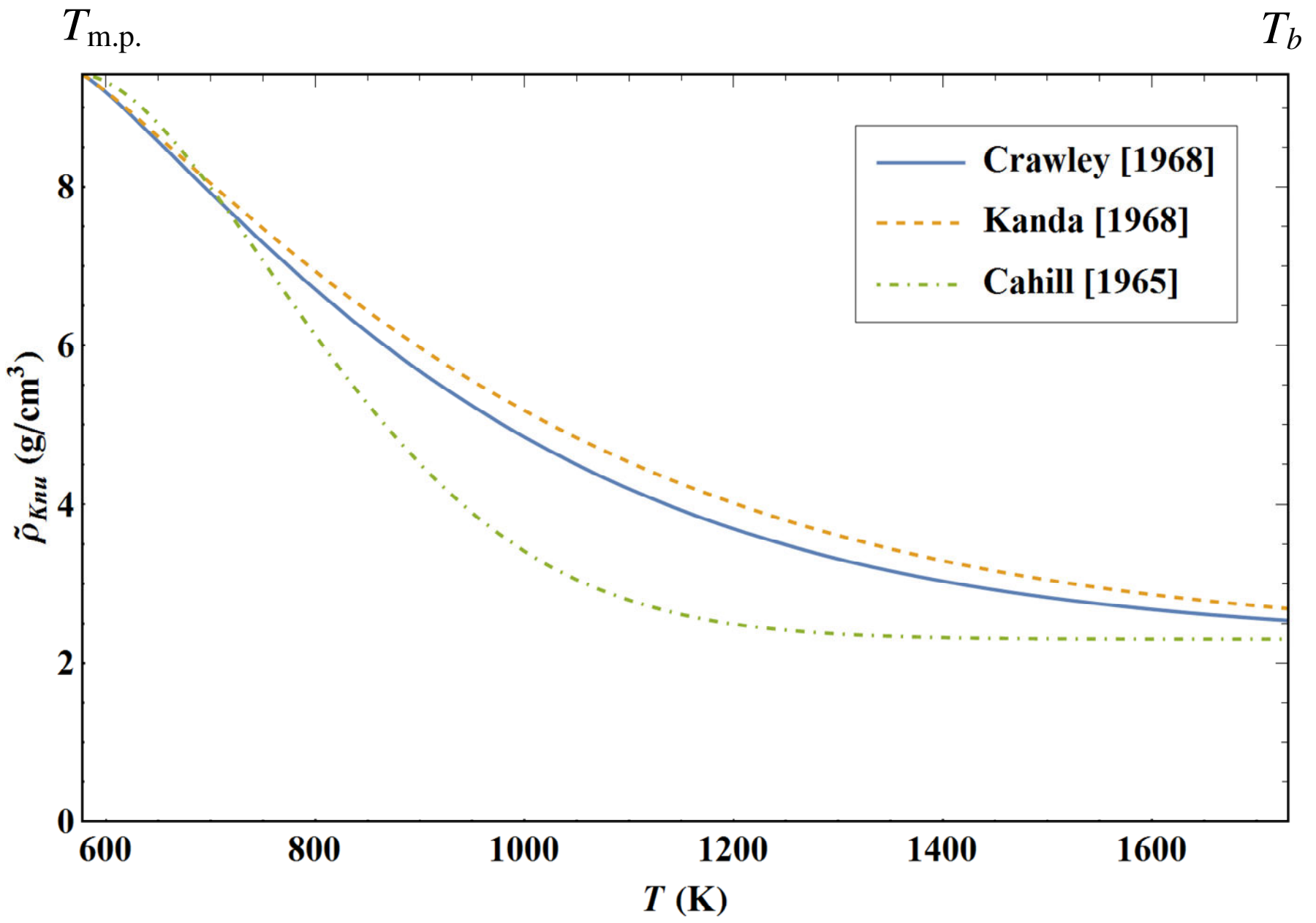

Fig. 36. Representation of Eq. (19) with the parameters used in the present modeling corresponding to the different experiments analyzed (Refs. 42 to 44) for liquid thallium. $T_{\text{m.p.}}$ represents the melting point temperature and $T_b$ the boiling temperature for the atmospheric pressure.

Finally, Eq. (9) allows to calculate the viscosity of thallium along the atmospheric isobar while Eq. (22) gives an order of magnitude (within a factor of 2) of the viscosity of thallium in the gas phase. It is not possible to define a general law of evolution of the functions $K^*(\rho)$ and $f_N(\rho)$, but a partial picture of this evolution can be obtained as we can see on Fig. 37: as for water and potassium, it can be seen that the deviation of the reduced elastic shear constant $K^*(\rho)$ from its limiting law is quite large, while $f_N(\rho)$ deviates little from its limiting law as the density increases.

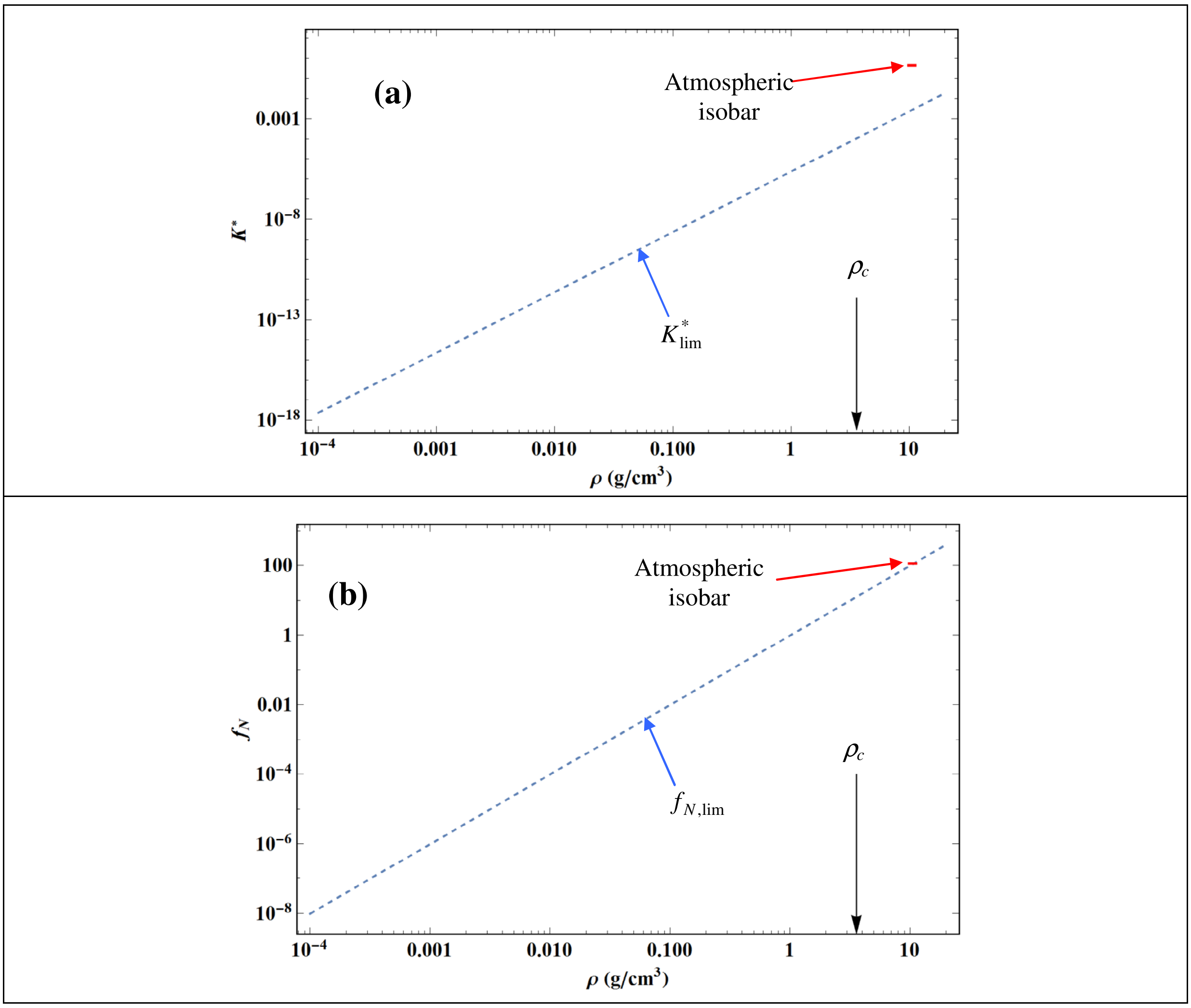

Fig. 37. Logarithmic plot of an overview for fluid thallium of the evolution versus density of **(a)** the reduced shear elastic constant and **(b)** the reduced fluctuative distance.

Contrary to the case of potassium, it is rather risky to extrapolate relations to the boiling point. Only an order of magnitude can be obtained. We have thus grouped few orders of magnitude in Table 7.

| $T_b$ (K) | $\rho_{\sigma,\rm Liq}$ (g/cm$^3$) | $\rho_{\sigma,\rm Vap}$ (g/cm$^3$) | $\eta_{\sigma,\rm Liq}$ (mPa.s) | $\eta_{\sigma,\rm Vap}$ (mPa.s) |
|---|---|---|---|---|
| 1730.05 | ~ 9.84 | ~ 0.0014 | ~ 0.80 | ~ 0.077 |

Table 7. Order of magnitude of some fluid thallium properties at the boiling point (i.e. for a pressure of 1 atm.). The viscosity value of $\eta_{\sigma,\rm Liq}$ is determined by considering Crawley's experimental parameters (i.e. $C_N$ = 1.00778, $\Delta T_{Knu}$ = 413.609 K and $\gamma_{Knu,1\,\rm atm.} = 1.194$). $\rho_{\sigma,\rm Vap}$ is determined by using the perfect gas equation of state.

On the other hand, the parameters at the melting temperature $T_{\rm m.p.}$ can be determined more precisely. Table 8 groups the values of different parameters at the melting temperature $T_{\rm m.p.}$.

| $T_{\text{m.p.}}$ (K) | $\rho_{\text{m.p.,Liq}}$ (g/cm$^3$) | $\eta_{\text{m.p.,Liq}}$ (mPa.s) | $P_\sigma$ (Pa) |
|---|---|---|---|
| 577 | 11.2326 | 2.631 | $3.805\times10^{-6}$ |

Table 8. Characteristic parameters of liquid thallium at the melting temperature. The viscosity value of $\eta_{\sigma,\text{Liq}}$ is determined by considering Crawley's experimental parameters (i.e. $C_N$ = 1.00778, $\Delta T_{Knu}$ = 413.609 K and $\gamma_{Knu,1\,\text{atm.}} = 1.194$). The values of these parameters are defined with an uncertainty of ±1%.

To complete this section we will analyze the self-diffusion coefficient data determined by Barras *et al.* (Ref. 45) along the atmospheric isobar. These data were obtained by measuring the diffusion of a tracer which is the radioactive isotope $^{204}$Tl.

The temperature range covered by the data here is small compared to the extent of the atmospheric isobar, therefore we stay in the region where $f_{q_c}(T,\rho_{1\text{atm.}})=1$ (i.e. $q_c = q_{c0} = \left(\frac{6\pi^2\rho\,\mathfrak{N}_a}{M\,n_B}\right)^{1/3}$). It is therefore not possible to determine here the evolution law of the function $f_{q_c}(T,\rho_{1\text{atm.}})$ unlike in the case of potassium. The analysis of the data shows in fact that to reproduce them correctly one must decrease $K^*$ by a factor $C_{K^*} = 0.7019$ in accordance with what can be expected with the tracer used.

By imposing values for $C_d$ and $C_N$ that correspond to the measuring tubes dimensions used by Barras *et al.* (i.e. 1.59 mm in diameter for ~3.1 cm in length), Fig. 38 shows that the present modeling can finally reproduce these data in a way that is quasi-equivalent to an Arrhenius law. We observe that the two curves do not satisfy the error bars indicated by the authors, but the dispersion of the points around 650 K clearly indicates that these error bars are underestimated. The dispersion of the experimental points can also be explained by the fact that the lengths of the tubes are slightly variable from one data to another.

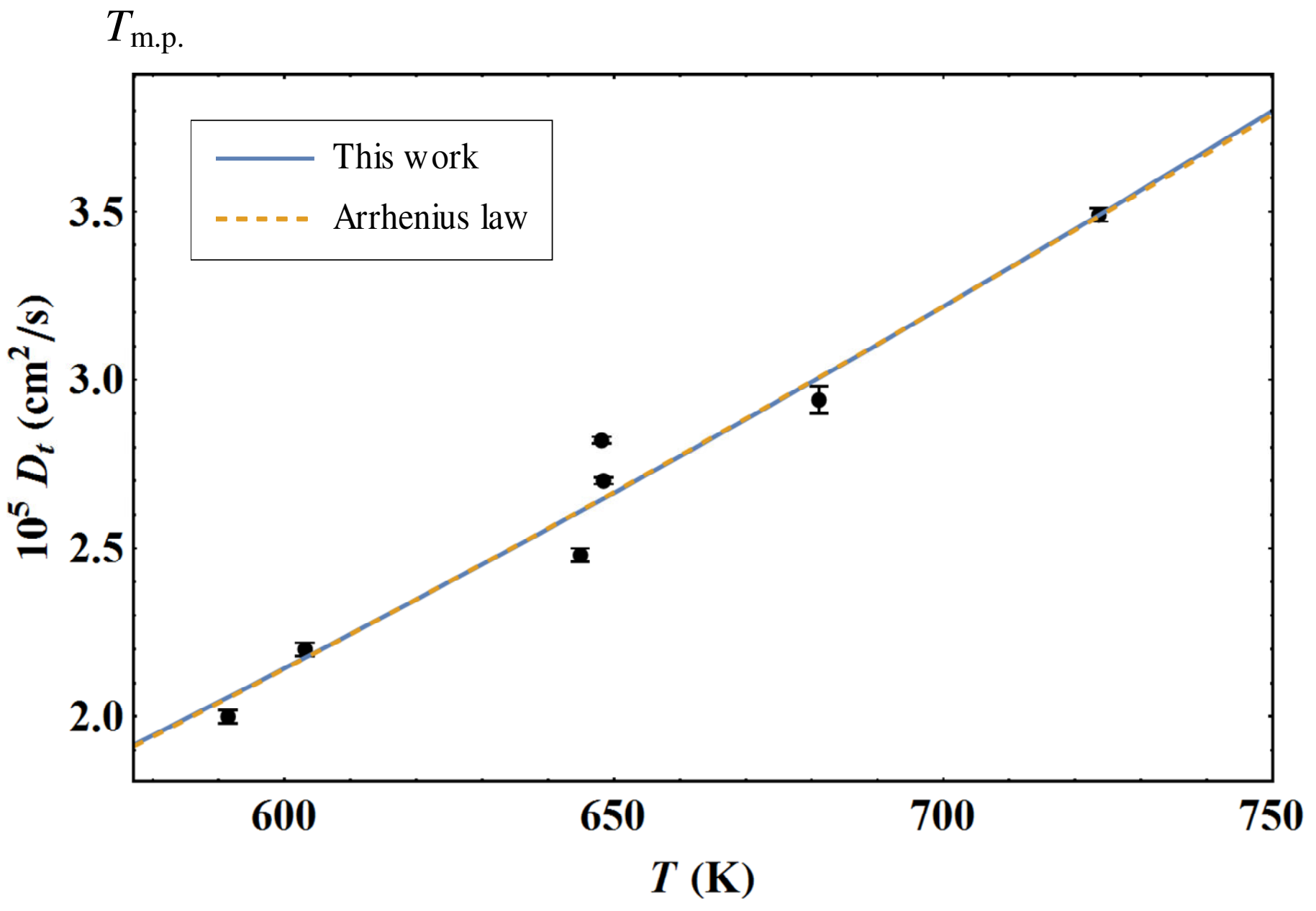

Fig. 38. Self-diffusion coefficient of thallium versus temperature along the atmospheric isobar. The black points represent the data from Barras *et al.* (Ref. 45). The orange dashed curve represents the following Arrhenius law $D_t\left(\text{cm}^2/\text{s}\right) = 3.7057\times10^{-4}\,\exp\left(-\frac{14220.4}{R_g T}\right)$ and he blue curve corresponds to the present modeling with $C_d$ =

7.95, $C_N =$ 0.05748 and $C_{K^*} = 0.7019$ . $T_{\mathrm{m.p.}}$ represents the melting point temperature for the atmospheric pressure.

It is important to note that the value of the critical temperature of thallium is poorly determined and this can vary between the value in Table 5 and 3219 K (Ref. 46). Taking this last value of critical temperature, we determine by extrapolation of Eq. (21) a value of $P_c$ = 25.99 MPa from which a new value of $z_c$ = 0.055. This value, although still rather low, seems more "realistic". It should be emphasized that changing the value of $T_c$ does not modify the analysis we have just done but only modifies some numerical values of few parameters. In particular, this has the effect of increasing the value of the scaling $K_0$.
In conclusion of this thallium section, we can say that the potassium transport properties analysis was a necessary support for the analysis of the fragmentary thallium data and thus a great coherence of the elastic mode theory could emerge.

## 5 Conclusion

First of all, it appears that the whole of the analysis carried out makes it possible to propose rather precise values of various thermodynamic parameters at the melting temperature $T_{\mathrm{m.p.}}$ and at the boiling temperature $T_b$ at atmospheric pressure.

Secondly, it has been shown that the translational elastic mode theory developed in Ref. 4 makes it possible to account for the experimental data on viscosity and self-diffusion coefficient of potassium and thallium in addition to water, but above all it makes it possible to account for the different variations observed between the authors, differences that are generally greater than the uncertainties of each dataset.

In particular, we have shown that the present modeling leads to a much better representation of viscosity data in thallium than the so-called reference equation proposed by Assael *et al.* (Ref. 40) since the latter combined data obtained with set-ups having different geometric characteristics and different purities of the samples.

Finally, we have been able to show that the variations as function of density and temperature of the different parameters of the elastic mode theory for potassium and thallium are perfectly analogous to those found in water in the same conditions. In particular, it has been shown that the dilute-gas limit laws relating different parameters which were found for water also apply in the case of potassium and thallium. Therefore, these limit laws seem to have a universal character that should be explored in other fluids. In this dilute gas limit, the fact that the Planck's constant appears in connection with the essential parameters of the model, without explicitly invoking quantum concepts, suggests the existence of a deeper physical meaning that will be explored in our subsequent publications.

## 6 REFERENCES

[1] R.C. Miller and P. Kusch, Phys. Rev. **99**, 1314 (1955).
[2] B.B. Bevard and G.L. Yoder, in *AIP Conf. Proc.* (AIP, 2003), pp. 629–634.
[3] Bariselli, F., Boccelli, S., Magin, T., Frezzotti, A., & Hubin, A. In *2018 Joint Thermophysics and Heat Transfer Conference,* p. 4180 (2018).
[4] F. Aitken and F. Volino, Phys. Fluids **33**, 117112 (2021).
[5] David R. Lide, *CRC Handbook of Chemistry and Physics*, CRC Press Inc, 2009, 90e ed., 2804 p.

[6] I.G. Dillon, P.A. Nelson, and B.S. Swanson, *Critical Temperatures and Densities of the Alkali Metals,* ANL-7025 (1965); I.G. Dillon, P.A. Nelson, and B.S. Swanson, J. Chem. Phys. **44**, 4229 (1966).

[7] Young, D. A. *Phase diagrams of the elements*. http://www.osti.gov/servlets/purl/4010212/ (1975) doi:10.2172/4010212.

[8] F. Aitken and F. Volino, Mathematica application for potassium viscosity and self-diffusion coefficient (2021). http://mathematica.g2elab.grenoble-inp.fr/newpotassiumthallium.html

[9] C. T. Ewing, J.P. Stone, J.R. Spann, E.W. Steinkuller, D. D. Williams and R. R. Miller, *High-temperature properties of potassium*, NRL Report 6233 (1965).

[10] C.T. Ewing, J.P. Stone, J.R. Spann, and R.R. Miller, J. Chem. Eng. Data **11**, 460 (1966).

[11] N.G. Vargaftik, A.N. Nikitin, V.G. Stepanov, and A.I. Abakumov, J. Eng. Phys. **60**, 371 (1991).

[12] R.T. Caldwell and D.M. Walley, *Physical and Thermodynamic Properties of Potassium.* GARRETT CORP PHOENIX AZ AIRESEARCH MFG DIV (1966).

[13] W.T. Hicks, J. Chem. Phys. **38**, 1873 (1963).

[14] W.F. Freyland and F. Hensel, Berichte Der Bunsengesellschaft Für Phys. Chemie **76**, 16 (1972).

[15] E.B. Hagen, Ann. Phys. **255**, 436 (1883).

[16] M.E. Rinck, C. R. Hebd. Séances Acad. Sci. **189**, 39 (1929).

[17] R.H. Stokes, J. Phys. Chem. Solids **27**, 51 (1966).

[18] C.T. Ewing, J.A. Grand, and R.R. Miller, J. Am. Chem. Soc. **73**, 1168 (1951).

[19] D.I. Lee and C.F. Bonilla, Nucl. Eng. Des. **7**, 455 (1968).

[20] B.I. Stefanov, D.L. Timrot, E.E. Totskii, and W. Chu, High Temp. **4**, 131 (1966).

[21] D.E. Briggs, *Thermal Conductivity of Potassium Vapor,* Ph.D. thesis (1968).

[22] N. Gerasimov, B. Stefanov, and L. Zarkova, J. Phys. D. Appl. Phys. **13**, 1841 (1980).

[23] Y.S. Chiong, Proc. R. Soc. London. Ser. A - Math. Phys. Sci. **157**, 264 (1936).

[24] C.T. Ewing, J.A. Grand, and R.R. Miller, J. Phys. Chem. **58**, 1086 (1954).

[25] A.W. Lemmon, H.W. Deem, E.H. Hall, and J.F. Walling, in *Proc. 1963 High-Temperature Liq. Heat Transf. Technol. Meet.,* ORNL-3605 (1964), pp. 88–115.

[26] E.N. da C. Andrade and Y.S. Chiong, Proc. Phys. Soc. **48**, 247 (1936).

[27] M. Hsieh and R.A. Swalin, Acta Metall. **22**, 219 (1974).

[28] A.G. Novikov, M.N. Ivanovskii, V. V. Savostin, A.L. Shimkevich, O. V. Sobolev, and M. V. Zaezjev, J. Phys. Condens. Matter **8**, 3525 (1996).

[29] L.E. Bove, S. Klotz, T. Strässle, M. Koza, J. Teixeira, and a. M. Saitta, Phys. Rev. Lett. **111**, 1 (2013).

[30] K. Krynicki, C.D. Green, and D.W. Sawyer, Faraday Discuss. Chem. Soc. **66**, 199 (1978).

[31] J. S. Coursey, D. J. Schwab, J. J. Tsai, and R. A. Dragoset, *Atomic Weights and Isotopic Compositions (version 4.1)*, 2015, National Institute of Standards and Technology, Gaithersburg, MD, accessed November 2016.

[32] R.W. Ohse, *Handbook of Thermodynamic and Transport Properties of Alkali Metals* (1985).

[33] A. Jayaraman, W. Klement, R.C. Newton, and G.C. Kennedy, J. Phys. Chem. Solids **24**, 7 (1963).

[34] A.T. Aldred and J.N. Pratt, J. Chem. Eng. Data **8**, 429 (1963).

[35] F.F. Coleman and A. Egerton, Phil. Trans. Roy. Soc. London A **234**, 177 (1935).

[36] P.J. Desré, D.T. Hawkins, and R. Hultgren, *Vapor Pressure of Thallium and Activity Measurements on Liquid Silver-Thallium Alloys by the Torsion Effusion Method* (Berkeley, California, 1968), LBNL Report #: UCRL-17828.

[37] V.N. Volodin, V.E. Khrapunov, and R.A. Isakova, Russ. J. Phys. Chem. A **82**, 1075 (2008).
[38] Periodic Table of Royal Society of Chemistry. *https://www.rsc.org/periodic-table/element/81/thallium*.
[39] W. Leitgebel, Z. anorg. u. allg. Chem. Bd. **202**, 305 (1931).
[40] M.J. Assael, I.J. Armyra, J. Brillo, S.V. Stankus, J. Wu, and W.A. Wakeham, J. Phys. Chem. Ref. Data **41**, (2012).
[41] H. Walsdorfer, I. Arpshofen, and B. Predel, Z. Metallkd. **79**, 654 (1988).
[42] A.F. Crawley, Trans. Met. Soc. AIME **242**, 2309 (1968).
[43] F.A. Kanda and J.A. Domingue, J. Less-Common Met. **64**, 135 (1979).
[44] J.A. Cahill and A. V. Grosse, J. Phys. Chem. **69**, 518 (1965).
[45] R.E. Barras, H.A. Walls, and A.L. Hines, Metall. Trans. B **6**, 347 (1975).
[46] D.S. Gates and G. Thodos, A.I.Ch.E. Journal **6**(1), 50 (1960).